\documentclass[12pt]{article}

\usepackage[dvipdfmx]{graphicx}
\usepackage{amsmath}        
\usepackage{amssymb}        
\usepackage{cite}
\usepackage{cancel}
\usepackage{fancybox}
\usepackage{ulem}
\usepackage{longtable}
\usepackage{color}

\def\mydate{26 February 2018}
\def\ignore#1{{}}

\def\go{\rightarrow}
\def\dd{\partial}
\def\ep{{\varepsilon}}

\def\checkpoint{{\color{red} \bf (check)}}
\def\eff{{\rm eff}}

\def\KK{{\rm KK}}
\def\onehalf{\hbox{$\frac{1}{2}$}}

\def\diag{{\rm diag}}
\def\mfrac#1#2{\displaystyle \frac{#1}{#2}}

\newsavebox{\BCmatrix}
\savebox{\BCmatrix}{$\begin{pmatrix} P_2 & P_3 \cr P_0 &P_1 \end{pmatrix}$}

\makeatletter
 
 \@addtoreset{equation}{section}
\makeatother

\begin{document}

\thispagestyle{empty}

{\small \noindent \mydate    \hfill OU-HET 943}

{\small      \hfill MISC-2017-08}

\vskip 3.5cm

\baselineskip=30pt plus 1pt minus 1pt

\begin{center}
{\Large \bf  Electroweak symmetry breaking and mass spectra}

{\Large \bf  in six-dimensional gauge-Higgs grand unification}

\end{center}



\baselineskip=22pt plus 1pt minus 1pt

\vskip 1.5cm

\begin{center}
{\bf
Yutaka Hosotani$^1$ and Naoki Yamatsu$^2$
}

\vskip 5pt
{\small \it $^1$Department of Physics, Osaka University, 
Toyonaka, Osaka 560-0043, Japan} \\
{\small \it $^2$Maskawa Institute for Science and Culture, 
Kyoto Sangyo University, Kyoto 603-8555, Japan} \\

\end{center}



\vskip 3.5cm
\baselineskip=16pt plus 1pt minus 1pt

\begin{abstract}
The mass spectra of the standard model particles 
are reproduced in the  $SO(11)$ gauge-Higgs grand  unification 
in the six-dimensional warped space without introducing exotic light fermions.
Light neutrino masses are explained by the gauge-Higgs seesaw mechanism.
We evaluate the effective potential of the 4d Higgs boson  appearing as
a fluctuation mode of the Aharonov-Bohm phase $\theta_H$ in the extra-dimensioal space,
and show that the dynamical electroweak symmetry breaking takes place
with the Higgs boson mass $m_H \sim 125\,$GeV and $\theta_H \sim 0.1$.
The Kaluza-Klein mass scale in the fifth dimension is approximately given by 
{$m_\KK \sim 1.230\,{\rm TeV}/\sin \theta_H$}.  
\end{abstract}

\newpage

\baselineskip=20pt plus 1pt minus 1pt
\parskip=0pt

\section{Introduction}

The discovery of the last piece of the standard model (SM) particle, the
Higgs boson, seems to imply that the non-vanishing vacuum expectation
value (VEV) of the Higgs field spontaneously breaks the electroweak
(EW) gauge symmetry $SU(2)_L\times U(1)_Y$ to the electromagnetic gauge
symmetry $U(1)_{\rm EM}$. 
Almost all observational data at low-energy experiments are
consistent with the SM.  The SM is a good low-energy effective theory.
However, it is not clear whether or not 
the Higgs boson is  a genuine fundamental scalar field,
which usually suffers from the so-called gauge hierarchy problem.

There are several proposals to overcome the problem by making use of symmetries. 
One of them is the gauge-Higgs unification. In this theory, the Higgs boson is identified 
with a part of the extra dimensional component of gauge fields in higher dimensional spacetime
\cite{Hosotani:1983xw,Hosotani:1988bm,Davies:1987ei,Davies:1988wt,Hatanaka:1998yp}.
It is described as a four-dimensional (4D) fluctuation mode of the
Aharonov-Bohm (AB) phase $\theta_H$ along the extra-dimensional space.

The $SU(2)_L\times U(1)$ EW unification of the SM is formulated as the
$SO(5)\times U(1)$ gauge-Higgs unification in the five-dimensional (5D)
Randall-Sundrum (RS) warped space
\cite{Agashe:2004rs,Medina:2007hz,Hosotani:2007qw,Hosotani:2008tx,Hosotani:2009qf,Funatsu:2013ni,Funatsu:2014fda,Funatsu:2014tka,Funatsu:2015xba,Funatsu:2016uvi,Funatsu:2017nfm}.
According to
Refs.~\cite{Funatsu:2013ni,Funatsu:2014fda,Funatsu:2014tka,Funatsu:2015xba},
its phenomenology at low energies under the mass scale of the first
Kaluza-Klein (KK) modes is almost the same as in the
SM for the AB phase $\theta_H  \lesssim 0.1$.
$Z'$ bosons, which are the first KK modes of $\gamma$, $Z$, and
$Z_R$, are predicted around $7 \sim 9\,$TeV range for $\theta_H=0.1 \sim 0.07$.  
$Z'$ bosons can be produced  at 14 TeV LHC.\cite{Funatsu:2016uvi}  
At electron-positron linear colliders at the energies of $250\,$GeV $\sim  1\,$TeV,  
the interference effects among $\gamma$, $Z$ and $Z'$ bosons
give distinct signals of the gauge-Higgs unification.\cite{Funatsu:2017nfm}

To incorporate the $SU(3)_C$ gauge symmetry in higher dimensional gauge
theories and gauge-Higgs unification scenario, grand unified theories
(GUTs) based on a GUT gauge group
$G_{\rm GUT}(\supset G_{\rm SM}:=SU(3)_C\times SU(2)_L\times U(1))$ 
have been discussed in
Refs.~\cite{Kawamura:1999nj,Kawamura:2000ir,Kawamura:2000ev,Frigerio:2011zg,Yamamoto:2013oja,Yamatsu:2017sgu,Yamatsu:2017ssg,Burdman:2002se,Haba:2002py,Haba:2003ux,Haba:2004qf,Lim:2007jv,Kojima:2011ad,Kojima:2016fvv,Kojima:2017qbt,Hosotani:2015hoa,Hosotani:2015wmb,Yamatsu:2015rge,Furui:2016owe,Hosotani:2016njs,Hosotani:2017krs}.
The $SO(11)$ gauge-Higgs grand unified theory (GHGUT) is proposed in the
5D Randall-Sundrum (RS) warped space in Ref.~\cite{Hosotani:2015hoa}.
The EW Higgs boson is identified with a part of the 5th dimensional
component of the $SO(11)$ gauge bosons. 
The $SO(11)$ gauge symmetry is reduced first by two different orbifold 
boundary conditions (BCs) on the UV and IR branes in the RS space.
It is reduced to $SO(10)$ by the BC on the UV brane, and to 
$SO(4) \times SO(7)$ by the BC on the IR brane, 
the resultant symmetry being  $SO(4) \times SO(6)$.
Secondly the brane scalar on the UV brane, which is an $SO(10)$ spinor ${\bf 16}$,
spontaneously breaks $SO(10)$ to $SU(5)$ by the Higgs mechanism on the UV brane.
As a net result the $SO(11)$ gauge symmetry is reduced to the SM gauge symmetry 
$G_{\rm SM}$, which is dynamically broken to $SU(3)_C \times U(1)_{\rm EM}$ by the 
Hosotani mechanism. 

Zero modes of 5D $SO(11)$ ${\bf 32}$ fermions in the bulk are
identified with each generation of quarks and leptons.
In Ref.~\cite{Furui:2016owe} it is found that the observed mass spectra of the quarks 
and leptons can be reproduced, while there appear additional exotic particles having 
unacceptably small masses.

Recently, to avoid these  light exotic particles, 
$SO(11)$ GHGUT in six-dimensional (6D) hybrid warped space has been proposed
in Ref.~\cite{Hosotani:2017ghg}. 
For neutrinos a new  seesaw mechanism in 6D hybrid
warped space has been formulated 
by using a 5D symplectic Majorana fermion \cite{Mirabelli:1997aj}, 
which generalizes  the well-known 4D seesaw mechanism \cite{Minkowski:1977sc}. 

This paper is organized as follows. 
In Sec.~\ref{Sec:6D-SO(11)-GHGUT}, 6D $SO(11)$ GHGUT is introduced.
The matter content is specified and the action is given 
that contains both 6D bulk and 5D brane terms.
In Sec.~\ref{Sec:Spectrum-Bosons},  a  summary is given for
the mass spectrum of the 4D gauge and scalar bosons originating from 
the 6D $SO(11)$ gauge bosons. 
In Sec.~\ref{Sec:Spectrum-Fermions}, 
we derive the mass spectrum of  4D SM fermions. 
By using these mass spectra, we evaluate the 
effective potential $V_{\rm eff} (\theta_H)$ in
Sec.~\ref{Sec:Effective-potential} to show that the dynamical EW symmetry breaking
takes place and the 4D Higgs boson mass $m_H = 125.1\,$GeV is obtained.
Section~\ref{Sec:Summary-Discussion} is devoted to a summary and discussions.
In the Appendix basics for the KK expansion in 6D warped space are explained.

\section{6D $SO(11)$ GHGUT}
\label{Sec:6D-SO(11)-GHGUT}

We construct an $SO(11)$ gauge-Higgs grand unified model on 
the six dimensional hybrid warped space introduced in  Ref.~\cite{Hosotani:2017ghg}.
The metric of generalized Randall-Sundrum (RS) space
\cite{Randall:1999ee,Hosotani:2017ghg} is given by
\begin{align}
ds^2=e^{-2\sigma(y)}
\left(\eta_{\mu\nu}dx^\mu dx^\nu+dv^2\right)+dy^2,
\label{Eq:6D-metric}
\end{align}
where 
$e^{-2\sigma(y)}$ is a warped factor and $\eta_{\mu\nu}=\mbox{diag}(-1,+1,+1,+1)$.
$\sigma(y)$ satisfies 
 $\sigma(y)=\sigma(-y)=\sigma(y+2L_5)$ and $\sigma(y)=k|y|\ \mbox{for}\ |y|\leq L_5$.
The fifth dimension with the coordinate $y$ behaves as the EW dimension, whereas 
the sixth dimension with the coordinate $v$ is  $S^1$ $(v\sim v+2\pi R_6)$ and
behaves as the GUT dimension.
Two spacetime points $(x^\mu, y , v)$ and $(x^\mu, -y, -v)$ are identified
by the $\mathbb{Z}_2$ transformation.
As a result the spacetime has the same topology as the orbifold 
$M^4\times T^2/\mathbb{Z}_2$.
The spacetime  \eqref{Eq:6D-metric} solves the Einstein equations
with brane tensions at $y=0$ and $y=L_5$.
The bulk region is 
the anti-de-Sitter space with a negative cosmological constant 
$\Lambda=-10k^2$. 
The five-dimensional branes at $y=0$ and $y=L_5$ have the same 
topology as  $M^4 \times S^1$.

The extra dimensional space  has four fixed points under $\mathbb{Z}_2$: 
$(y_0,v_0)=(0,0)$, $(y_1,v_1)=(L_5,0)$, 
$(y_2,v_2)=(0,\pi R_6)$, and $(y_3,v_3)=(L_5,\pi R_6)$.
In terms of the conformal coordinate $z=e^{ky} \, (1\leq z\leq z_L=e^{kL_5})$
in the region $0\leq y \leq L_5$, the metric becomes
\begin{align}
ds^2=\frac{1}{z^2}\left(
\eta_{\mu\nu}dx^\mu dx^\nu+dv^2+\frac{dz^2}{k^2}\right).
\end{align}
The fifth and sixth dimensional Kaluza-Klein (KK) mass scales are given
by $m_{\rm KK_5}=\pi k/(e^{kL_5}-1)$ and $m_{\rm KK_6}=R_6^{-1}$.
The warp factor is supposed to be large; $z_L = e^{kL_5} \gg 1$.
$m_{\rm KK_6}$ is  expected to be a GUT scale while $m_{\rm KK_5}$ is
$O(10) \,$TeV so that $m_{\rm KK_6}\gg m_{\rm KK_5}$. 

Parity transformations $P_j$ $(j=0,1,2,3)$ around the four  fixed 
points are defined as $(x^\mu,y_j+y,v_j+v) \go (x^\mu,y_j-y,v_j-v)$.
Only three of the four parity transformations $P_j$ are independent.
They satisfy the relation $P_3=P_2P_0P_1=P_1P_0P_2$.

We adopt  orbifold boundary conditions (BCs)  such that 
$P_{0}=P_{1}$ and $P_{2}=P_{3}$, which enables us to avoid the problem
of unwanted light exotic fermions mentioned in the Introduction.
More specifically, we choose BCs such that 
$P_{0}=P_{1}$ and $P_{2}=P_{3}$
break  $SO(11)$ to $SO(4)\times SO(7)$ and $SO(10)$, respectively.
The orbifold BCs reduce $SO(11)$ symmetry to the Pati-Salam symmetry
$SO(4) \times SO(6) \simeq SU(2)_L\times SU(2)_R\times SU(4)_C=:G_{\rm PS}$.
We note that the 5th and 6th dimensional loop translations 
$U_5:(x^\mu,y,v)\to(x^\mu,y+2L_5,v)$ and
$U_6:(x^\mu,y,v)\to(x^\mu,y,v+2\pi R_6)$ are related to $P_j$'s by 
$U_5=P_1P_0=P_3P_2$ and $U_6=P_2P_0=P_3P_1$.

\subsection{6D bulk and 5D brane fields and orbifold boundary conditions }

The matter content in the $SO(11)$ GHGUT consists of 6D $SO(11)$ gauge
bosons $A_M$, 
6D $SO(11)$ ${\bf 32}$ Weyl fermions $\Psi_{\bf 32}^{\alpha}(x,y,v)$
$(\alpha=1,2,3,4)$,
6D $SO(11)$ ${\bf 11}$ Dirac fermions $\Psi_{\bf 11}^{\beta}(x,y,v)$
and $\Psi_{\bf 11}^{\prime\beta}(x,y,v)$ $(\beta=1,2,3)$ 
in the six-dimensional bulk space, and
a 5D $SO(11)$ ${\bf 32}$ brane scalar boson $\Phi_{\bf 32}(x,v)$ and
5D $SO(11)$ ${\bf 1}$ brane symplectic Majorana fermions 
$\chi_{\bf 1}^\beta (x,v)$ $(\beta=1,2,3)$
on the UV brane $y=0$ \cite{Hosotani:2017ghg}. 
Their orbifold BCs are listed below.
\begin{itemize}
\item For the 6D $SO(11)$ gauge boson $A_M$, the orbifold BCs are given by
\begin{align}
&\left(
\begin{array}{c}
A_\mu\\
A_{y}\\
A_{v}\\
\end{array}
\right)
(x,y_j-y,v_j-v)
=P_{j}
\left(
\begin{array}{c}
A_\mu\\
-A_{y}\\
-A_{v}\\
\end{array}
\right)
(x,y_j+y,v_j+v)P_{j}^{-1},
\label{Eq:BC-SO(11)-gauge}
\end{align}
where in the $SO(11)$ vector representation, we take the orbifold BCs
$P_{0}=P_{1}$ and $P_{2}=P_{3}$ as
\begin{align}
P_{0}^{\rm vec}=P_{1}^{\rm vec}=\mbox{diag}(I_4,-I_7),\ \ \
P_{2}^{\rm vec}=P_{3}^{\rm vec}=\mbox{diag}(I_{10},-I_1).
\label{Eq:BC-SO(11)-11}
\end{align}
By using $(P_0=P_1,P_2=P_3)$, the parity assignment of 
$A_\mu$, $A_y$, and $A_v$ are summarized in Table~\ref{Tab:BC-gauge}.

\item For the four 6D $SO(11)$ spinor {\bf 32} bulk Weyl fermions
$\Psi_{\bf 32}^{\alpha}(x,y,v)$ $(\alpha=1,2,3,4)$, the orbifold BCs are
given by 
\begin{align}
\Psi_{{\bf 32}}^{\alpha}(x,y_j-y,v_j-v)
=\eta_j^\alpha\overline{\gamma}P_j^{\rm sp}
\Psi_{\bf 32}^{\alpha}(x,y_j+y,v_j+v),
\end{align}
where $\eta_j^{\alpha}=\pm 1$.  
6D Dirac matrices $\gamma^a$ $(a=1,2,\cdots,6)$ satisfy 
$\{\gamma^a,\gamma^b\}=2\eta^{ab}$
$(\eta^{ab}=\mbox{diag}(-I_1,I_5))$, and 
$\overline{\gamma}:=-i\gamma^5\gamma^6
=\gamma_{6D}^{7}\gamma_{4D}^{5}=\gamma_{4D}^{5}\gamma_{6D}^{7}$,
$\gamma_{4D}^{5}=I_2\otimes\sigma^3\otimes I_2$,
$\gamma_{6D}^{7}=I_4\otimes\sigma^3$.
$P_j^{\rm sp}$'s are
\begin{align}
&P_{0}^{\rm sp}=P_{1}^{\rm sp}=I_2\otimes \sigma^3\otimes I_8,\ \ \
 P_{2}^{\rm sp}=P_{3}^{\rm sp}=I_{16}\otimes\sigma^3.
\label{Eq:BC-SO(11)-32}
\end{align}
To ensure the 6D $SO(11)$ chiral anomaly cancellation, we assign
$\gamma_{6D}^7=+1$ for $\alpha=1,2$;
$\gamma_{6D}^7=-1$ for $\alpha=3,4$.
We take $\eta_j^{1,2}=-1$; $\eta_j^3=1$, and 
$\eta_{0,2}^4=-\eta_{1,3}^4=1$.
The parity assignment 
\begin{align}
\left(
\begin{array}{cc}
\eta_2^\alpha \overline{\gamma}P_2^{\rm sp}&
\eta_3^\alpha \overline{\gamma}P_3^{\rm sp}\\
\eta_0^\alpha \overline{\gamma}P_0^{\rm sp}&
\eta_1^\alpha \overline{\gamma}P_1^{\rm sp}\\
\end{array}
\right)
\label{parityassign32}
\end{align}
of 4D left- and right-handed component of $\Psi_{\bf 32}^{\alpha}$ are
summarized in Table~\ref{Tab:BC-fermion-spinor}.
We find that $\Psi_{\bf 32}^{\alpha}$ $(\alpha=1,2,3)$ has zero modes,
corresponding to one generation of quarks and leptons for each $\alpha$.
The corresponding names adopted in Ref.~\cite{Furui:2016owe} are also
listed in Table~\ref{Tab:BC-fermion-spinor} for $\alpha=1,2,3$.

\item For the three 6D $SO(11)$ vector {\bf 11} bulk Dirac
fermions $\Psi_{\bf 11}^{\beta}(x,y,v)$ and 
$\Psi_{\bf 11}^{\prime\beta}(x,y,v)$ $(\beta=1,2,3)$, 
the orbifold BCs are given by
\begin{align}
\Psi_{{\bf 11}}^{(\prime)\beta}(x,y_j-y,v_j-v)
=\eta_j^{(\prime)\beta}\overline{\gamma}P_j^{\rm vec}
\Psi_{\bf 11}^{(\prime)\beta}(x,y_j+y,v_j+v),
\end{align}
where $\eta_{0,1}^{\beta}=-\eta_{2,3}^{\beta}=-1$ 
for $\Psi_{\bf 11}^{\beta}$;
$\eta_{0,1}^{\prime\beta}=\eta_{2,3}^{\prime\beta}=-1$ 
for $\Psi_{\bf 11}^{\prime\beta}$.
The parity assignment 
$(\eta_{0}^{\prime\beta}\overline{\gamma}P_0^{\rm vec}=
\eta_{1}^{\prime\beta}\overline{\gamma}P_1^{\rm vec},
\eta_{2}^{\prime\beta}\overline{\gamma}P_2^{\rm vec}=
\eta_{3}^{\prime\beta}\overline{\gamma}P_3^{\rm vec})$
of 4D left- and right-handed component of 
$\Psi_{\bf 11}^{(\prime)\beta}$ are summarized in
Table~\ref{Tab:BC-fermion-vector}. 

\item For the 5D $SO(11)$ spinor ${\bf 32}$ brane scalar field on the UV
brane at $y=0$, $\Phi_{\bf 32}(x,v)$, the orbifold BCs are given by
\begin{align}
\Phi_{\bf 32}(x,v_j-v)=\eta_j P_j^{\rm sp}\Phi_{\bf 32}(x,v_j+v),
\end{align}
where $j=0,2$, $\eta_0=-\eta_2=-1$.
The components of the $G_{\rm PS}$ $({\bf 1,2,\overline{4}})$ have zero
modes, one of which corresponds to the $SU(5)$ ${\bf 1}$ of
$SO(10)$ ${\bf 16}$. It is responsible for reducing $SO(11)$ to $SU(5)$
at the UV brane $y=0$. The BCs of $\Phi_{\bf 32}$ are summarized in
Table~\ref{Tab:BC-boson-spinor}. 

\item For the 5D $SO(11)$ singlet fermions 
$\chi^\beta (x,v) = \chi_{\bf 1}^\beta (x,v)$ ($\beta =1,2,3$) at $y=0$,
     their orbifold BCs are given by
\begin{align}
\chi^\beta (x,v_j-v)=\overline{\gamma} \chi^\beta (x,v_j+v)\ \ (j=0,2).
\end{align}
      These brane fields also satisfy the 5D symplectic Majorana condition
      $\chi^C=\widetilde{\chi}$, 
      $\widetilde{\chi}:=i\overline{\Gamma}\chi$, 
      where $\overline{\Gamma}:=\gamma_{4D}^5\gamma^6$:
\begin{align}
\chi=
\left(
\begin{array}{c}
\xi_+\\
\eta_+\\
\xi_-\\
\eta_-\\
\end{array}
\right),\ \ \
\chi^C=
\left(
\begin{array}{c}
+\eta_+^C\\
-\xi_+^C\\
-\eta_-^C\\
+\xi_-^C\\
\end{array}
\right)
=e^{i\delta_C}
\left(
\begin{array}{c}
-\sigma^2\eta_+^*\\
-\sigma^2\xi_+^*\\
+\sigma^2\eta_-^*\\
+\sigma^2\xi_-^*\\
\end{array}
\right)
=\left(
\begin{array}{c}
+\xi_-\\
-\eta_-\\
-\xi_+\\
+\eta_+\\
\end{array}
\right)=
\widetilde{\chi}.
\label{sympMajorana1}
\end{align}
Here the generation index $\beta$ has been suppressed.
\end{itemize}

\begin{table}[tbh]
{\small
\renewcommand{\arraystretch}{1.05}
\begin{center}
\begin{tabular}{cccc}\hline
$G_{\rm PS}$  &$A_\mu$&$A_y$  &$A_v$\\\hline
({\bf 3,1,1}) &$(+,+)$&$(-,-)$&$(-,-)$\\
({\bf 1,3,1}) &$(+,+)$&$(-,-)$&$(-,-)$\\
({\bf 1,1,15})&$(+,+)$&$(-,-)$&$(-,-)$\\
({\bf 2,2,6}) &$(-,+)$&$(+,-)$&$(+,-)$\\
({\bf 2,2,1}) &$(-,-)$&$(+,+)$&$(+,+)$\\
({\bf 1,1,6}) &$(+,-)$&$(-,+)$&$(-,+)$\\\hline
\end{tabular}
\caption{Parity assignment $(P_0, P_2)=(P_1, P_3)$ of $A_\mu$, $A_y$, and $A_v$ in 
$G_{\rm PS}=SU(2)_L\times SU(2)_R\times SU(4)_C$.}
\label{Tab:BC-gauge}
\end{center}
}
\end{table}

\begin{table}[t]
{\small
\renewcommand{\arraystretch}{1.05}
\begin{center}
\begin{tabular}{c}
$\Psi_{\bf 32}^{\alpha=1,2,3}$\\
\begin{tabular}{cccc}\hline
$G_{\rm PS}$&Left&Right &name\\  \hline
$({\bf 2,1,4})$&
$\left(
\begin{array}{cc}
+&+\\
+&+\\
\end{array}
\right)$&
$\left(
\begin{array}{cc}
-&-\\
-&-\\
\end{array}
\right)$
&$\begin{matrix} \nu \cr e\end{matrix}$~~$\begin{matrix} u_j \cr d_j\end{matrix}$\\
$({\bf 1,2,\overline{4}})$&
$\left(
\begin{array}{cc}
+&+\\
-&-\\
\end{array}
\right)$&
$\left(
\begin{array}{cc}
-&-\\
+&+\\
\end{array}
\right)$
&$\begin{matrix} \hat e \cr \hat \nu \end{matrix}$~~$\begin{matrix} \hat d_j \cr \hat u_j\end{matrix}$\\
$({\bf 2,1,\overline{4}})$&
$\left(
\begin{array}{cc}
-&-\\
+&+\\
\end{array}
\right)$&
$\left(
\begin{array}{cc}
+&+\\
-&-\\
\end{array}
\right)$
&$\begin{matrix} \hat e' \cr \hat \nu' \end{matrix}$~~$\begin{matrix} \hat d_j' \cr \hat u_j'\end{matrix}$\\
$({\bf 1,2,4})$&
$\left(
\begin{array}{cc}
-&-\\
-&-\\
\end{array}
\right)$&
$\left(
\begin{array}{cc}
+&+\\
+&+\\
\end{array}
\right)$
&$\begin{matrix} \nu' \cr e'\end{matrix}$~~$\begin{matrix} u_j' \cr d_j'\end{matrix}$\\
\hline
\end{tabular}
\end{tabular}
\begin{tabular}{c}
$\Psi_{\bf 32}^{\alpha=4}$\\
\begin{tabular}{ccc}\hline
$G_{\rm PS}$&Left&Right\\\hline
$({\bf 2,1,4})$&
$\left(
\begin{array}{cc}
+&-\\
+&-\\
\end{array}
\right)$&
$\left(
\begin{array}{cc}
-&+\\
-&+\\
\end{array}
\right)$\\
$({\bf 1,2,\overline{4}})$&
$\left(
\begin{array}{cc}
+&-\\
-&+\\
\end{array}
\right)$&
$\left(
\begin{array}{cc}
-&+\\
+&-\\
\end{array}
\right)$\\
$({\bf 2,1,\overline{4}})$&
$\left(
\begin{array}{cc}
-&+\\
+&-\\
\end{array}
\right)$&
$\left(
\begin{array}{cc}
+&-\\
-&+\\
\end{array}
\right)$\\
$({\bf 1,2,4})$&
$\left(
\begin{array}{cc}
-&+\\
-&+\\
\end{array}
\right)$&
$\left(
\begin{array}{cc}
+&-\\
+&-\\
\end{array}
\right)$\\
\hline
\end{tabular}
\end{tabular}
\caption{Parity assignment \usebox{\BCmatrix} in (\ref{parityassign32}) 
of $\Psi_{\bf 32}^{\alpha}$
$(\alpha=1,2,3,4)$ in $G_{\rm PS}$.} 
\label{Tab:BC-fermion-spinor}
\end{center}
}
\end{table}

\begin{table}[tbh]
{\small
\renewcommand{\arraystretch}{1.1}
\begin{center}
\begin{tabular}{c}
$\Psi_{\bf 11}^{\beta}$\\
\begin{tabular}{cccccc}\hline
$G_{\rm PS}$&Left${}_+$&Right${}_+$&Left${}_-$&Right${}_-$&name \\
\hline
$({\bf 2,2,1})$&
$\left(+,-\right)$&$\left(-,+\right)$&
$\left(-,+\right)$&$\left(+,-\right)$&$\begin{matrix} N & \hat E \cr E & \hat N \end{matrix}$\\
$({\bf 1,1,6})$&
$\left(-,-\right)$&$\left(+,+\right)$&
$\left(+,+\right)$&$\left(-,-\right)$&$D_j, \hat D_j$\\
$({\bf 1,1,1})$&
$\left(-,+\right)$&$\left(+,-\right)$&
$\left(+,-\right)$&$\left(-,+\right)$&$S$ \\
\hline
\end{tabular}
\end{tabular}
\vskip 5pt
\begin{tabular}{c}
$\Psi_{\bf 11}^{\prime\beta}$\\
\begin{tabular}{cccccc}\hline
$G_{\rm PS}$&Left${}_+$&Right${}_+$&Left${}_-$&Right${}_-$&name \\
\hline
$({\bf 2,2,1})$&
$\left(+,+\right)$&$\left(-,-\right)$&
$\left(-,-\right)$&$\left(+,+\right)$
&$\begin{matrix} N' & \hat E' \cr E' & \hat N' \end{matrix}$\\
$({\bf 1,1,6})$&
$\left(-,+\right)$&$\left(+,-\right)$&
$\left(+,-\right)$&$\left(-,+\right)$& $D_j' , \hat D_j'$\\
$({\bf 1,1,1})$&
$\left(-,-\right)$&$\left(+,+\right)$&
$\left(+,+\right)$&$\left(-,-\right)$&$S'$ \\
\hline
\end{tabular}
\end{tabular}
\caption{Parity assignment $(P_0, P_2)=(P_1, P_3)$ for $\Psi_{\bf 11}^{(\prime)\beta}$
$(\beta=1,2,3)$ in $G_{\rm PS}$.} 
\label{Tab:BC-fermion-vector}
\end{center}
}
\end{table}

\begin{table}[tbh]
{\small
\renewcommand{\arraystretch}{1.1}
\begin{center}
\begin{tabular}{c}
$\Phi_{\bf 32}$\\
\begin{tabular}{cc}\hline
$G_{\rm PS}$&BCs\\\hline
$({\bf 2,1,4})$           &$(-,+)$\\
$({\bf 1,2,\overline{4}})$&$(+,+)$\\
$({\bf 2,1,\overline{4}})$&$(-,-)$\\
$({\bf 1,2,4})$           &$(+,-)$\\
\hline
\end{tabular}
\end{tabular}
\caption{Parity assignment  $(P_0, P_2)$ for $\Phi_{\bf 32}$
in $G_{\rm PS}$.}
\label{Tab:BC-boson-spinor}
\end{center}
}
\end{table}

\subsection{Action}
\label{Sec:action}

The action consists of the 6D bulk  and 5D brane terms.

\subsubsection{Bulk terms}
\label{Sec:bulk-action}


The bulk part of the action is given by 
\begin{align}
S_{\rm bulk}=&
S_{\rm bulk}^{\rm gauge}+S_{\rm bulk}^{\rm fermion},
\label{Eq:Action-bulk}
\end{align}
where $S_{\rm bulk}^{\rm gauge}$ and $S_{\rm bulk}^{\rm fermion}$ are
bulk actions of gauge and fermion fields, respectively.
The action of $SO(11)$ gauge field {\bf 55} $A_M(x,y,v)$ is
\begin{align}
S_{\rm bulk}^{\rm gauge}=&
\int d^6x\sqrt{-\mbox{det}G}\bigg[-\mbox{tr}\left(
\frac{1}{4}F_{}^{MN}F_{MN}
+\frac{1}{2\xi}(f_{\rm gf})^2+{\cal L}_{\rm gh}\right)\bigg],
\label{Eq:Action-bulk-gauge}
\end{align}
where $\sqrt{-\mbox{det}G}=1/k z^6$, $z=e^{ky}$, $M,N=0,1,2,3,5,6$, 
and the field strength $F_{MN}$ is
defined by  
\begin{align}
F_{MN}&:=
\partial_MA_N-\partial_NA_M-i g[A_M,A_N].
\end{align}
We take the following gauge fixing and ghost terms 
\begin{align}
f_{\rm gf}&=
z^2\left\{\eta^{\mu\nu}{\cal D}_\mu^c A_\nu^q
+{\cal D}_6^c A_6^q
+{\xi k^2 z^2 {\cal D}_z^c\Big(\frac{A_z^q}{z^2}\Big)} \right\},\cr
{\cal L}_{\rm gh}&=
\bar{c}\left\{
\eta^{\mu\nu}{\cal D}_\mu^c{\cal D}_\nu^{c+q}
+{\cal D}_6^c{\cal D}_6^{c+q}
+{ \xi k^2 z^2{\cal D}_z^c\frac{1}{z^2}{\cal D}_z^{c+q}} \right\}c,
\end{align}
where $\mu,\nu=0,1,2,3$, 
$\eta^{\mu\nu}=\eta_{\mu\nu}=\mbox{diag}(-1,1,1,1)$, and
$A_M=A_M^c+A_M^q$.
${\cal D}_M^{c}B=\partial_M B-ig[A_M^c,B]$ and
${\cal D}_M^{c+q}B=\partial_M B-ig[A_M,B]$ 
where $B=A_\mu^q, {A_z^q/z^2}, A_6^q$ and $c$.

The action of fermions in $SO(11)$ spinor and vector
representations {\bf 32} and {\bf 11}, 
$\Psi_{\bf 32}^{\alpha}(x,y,v)$, $\Psi_{\bf 11}^{\beta}(x,y,v)$ and
$\Psi_{\bf 11}^{\prime\beta}(x,y,v)$,  is given by
\begin{align}
S_{\rm bulk}^{\rm fermion}=&
\int d^6x\sqrt{-\mbox{det}G} \, 
\bigg\{
\sum_{\alpha=1}^4\overline{\Psi_{\bf 32}^{\alpha}}{\cal D}
(c_{\Psi_{\bf 32}^{\alpha}})
\Psi_{\bf 32}^{\alpha} \cr
\noalign{\kern 5pt}
&\hskip 2.cm
+\sum_{\beta=1}^3
\overline{\Psi_{\bf 11}^{\beta}}{\cal D}
(c_{\Psi_{\bf 11}^{\beta}})
\Psi_{\bf 11}^{\beta}
+\sum_{\beta=1}^3
\overline{\Psi_{\bf 11}^{\prime\beta}}{\cal D}
(c_{\Psi_{\bf 11}^{\prime\beta}})
\Psi_{\bf 11}^{\prime\beta}
\bigg\} , 
\label{Eq:Action-bulk-fermion}
\end{align}
where $\overline{\Psi_{\bf 32}^{\alpha}}:=
i{\Psi_{\bf 32}^{\alpha\dag}}\gamma^0$.  
$c_{\Psi_{\bf 32}^{\alpha}}$, $c_{\Psi_{\bf 11}^{\beta}}$ and
$c_{\Psi_{\bf 11}^{\prime\beta}}$ are bulk mass parameters,
and ${\cal D}(c)$ is a covariant derivative given by
\begin{align}
{\cal D}(c)=z\left\{
\gamma^\mu D_\mu+\gamma^6 D_v-\frac{5}{2}\frac{\sigma'}{z}\gamma^5
+\sigma'\gamma^5 D_z+ic\frac{\sigma'}{z}\gamma^6\right\},
\end{align}
where $\sigma'(y):=d\sigma(y)/dy$ and $\sigma'(y) =k$ for $0< y < L_5$.

Here, we introduce $\check{\Psi}$ defined by 
\begin{align}
\check{\Psi}:=\frac{1}{z^{5/2}}\Psi,\ \ 
\left(D_z-\frac{5}{2}\frac{1}{z}\right)\Psi
=z^{5/2}D_z\check{\Psi}, 
\end{align}
which turns out convenient to discuss mass spectra for fermions in
Sec.~\ref{Sec:Spectrum-Fermions}. 
The bulk part of the bulk fermion action becomes 
\begin{align}
S=&\int d^4x\int_0^{2\pi R_6}dv\int_1^{z_L}\frac{dz}{k}
\bigg[
 \overline{\check{\Psi}_{\bf 32}^{\alpha}}
\left(\gamma^\mu D_\mu+\gamma^6 D_v
+\sigma'\gamma^5 D_z+ic_{\Psi_{\bf 32}}\frac{\sigma'}{z}\gamma^6\right)
\check{\Psi}_{\bf 32}^{\alpha}\nonumber\\
\noalign{\kern 5pt}
&\hspace{10em}
+\overline{\check{\Psi}_{\bf 11}^{\beta}}
\left(\gamma^\mu D_\mu+\gamma^6 D_v
+\sigma'\gamma^5 D_z+ic_{\Psi_{\bf 11}}\frac{\sigma'}{z}\gamma^6\right)
\check{\Psi}_{\bf 11}^{\beta}
\nonumber\\
\noalign{\kern 5pt}
&\hspace{10em}
+\overline{\check{\Psi}_{\bf 11}^{\prime\beta}}
\left(\gamma^\mu D_\mu+\gamma^6 D_v
+\sigma'\gamma^5 D_z+ic_{\Psi_{\bf 11}'}\frac{\sigma'}{z}\gamma^6\right)
\check{\Psi}_{\bf 11}^{\prime\beta}
\bigg].
\label{Eq:Action-bulk-fermion-check}
\end{align}

\subsubsection{Action for the brane scalar $\Phi_{\bf 32}$ and the Higgs mechanism}

We consider the 5D brane terms for a brane scalar $\Phi_{\bf 32}(x,v)$:
\begin{align}
&S_{\rm 5D\ brane}^{\rm brane}=
\int d^6x\sqrt{-\mbox{det}G} \, \delta(y)
{\cal L}_{\Phi_{\bf 32}} ~, \cr
\noalign{\kern 5pt}
&{\cal L}_{\Phi_{\bf 32}}:=
-(D_\mu\Phi_{\bf 32})^{\dag}D^\mu\Phi_{\bf 32}
-(D_v\Phi_{\bf 32})^{\dag}D^v\Phi_{\bf 32}
-\lambda_{\Phi_{\bf 32}}
\big( \Phi_{\bf 32}^\dag\Phi_{\bf 32}-|w|^2 \big)^2 ~,
\label{Eq:Action-brane-scalar}
\end{align}
where $\lambda_{\Phi_{\bf 32}}$ and $w$ are constants, and
\begin{align}
D_\mu\Phi_{\bf 32}=&
\left(\partial_\mu-ig A_\mu^{SO(11)}\right)\Phi_{\bf 32}=
\bigg\{\partial_\mu-\frac{ig}{\sqrt{2}}
\sum_{j<k}^{11}A_\mu^{(jk)}T_{jk}^{\rm sp}\bigg\}\Phi_{\bf 32} ~,\cr
D_v\Phi_{\bf 32}=&
\left(\partial_v-ig A_v^{SO(11)}\right)\Phi_{\bf 32}=
\bigg\{\partial_v-\frac{ig}{\sqrt{2}}
\sum_{j<k}^{11}A_v^{(jk)}T_{jk}^{\rm sp}\bigg\}\Phi_{\bf 32} ~.
\end{align}

We denote the 5D brane scalar field $\Phi_{\bf 32}$ as 
\begin{align}
\Phi_{\bf 32}:=
\left(
\begin{array}{c}
\Phi_{\bf 16}\\
\Phi_{\overline{\bf 16}}\\
\end{array}
\right)
\end{align}
and  assume that the component  $({\bf 1,2,\overline{4}})$ of 
$G_{\rm PS}(=SU(2)_L\times SU(2)_R\times SU(4)_C)$ for $\Phi_{\bf 32}$
develops the nonvanishing VEV: 
\begin{align}
\langle\Phi_{\bf 16}\rangle=
\left(
\begin{array}{c}
0_4\\
0_4\\
v_4\\
0_4\\
\end{array}
\right),\ \ \
v_4=
\left(
\begin{array}{c}
0\\
0\\
0\\
w\\
\end{array}
\right),\ \ 
\langle\Phi_{\overline{\bf 16}}\rangle=0,
\end{align}
where one component of $({\bf 1,2,\overline{4}})$ of 
$G_{\rm PS}(\subset SO(10))$
corresponds to ${\bf 1}_{+5}$ of 
$SU(5)\times U(1)_Z(\subset SO(10))$.
The VEV breaks $SO(11)$ to $SU(5)$.
Note that branching rules of $SO(11)$ ${\bf 32}$ and ${\bf 55}$
representations for $SO(11)\supset SO(10)\supset SU(5)$ are given by  
${\bf 32}=({\bf 16})\oplus({\bf \overline{16}})=
({\bf 10}\oplus{\bf \overline{5}}\oplus{\bf 1})\oplus
({\bf \overline{10}}\oplus{\bf 5}\oplus{\bf 1})$
and
${\bf 55}=({\bf 45})\oplus({\bf 10})=
({\bf 24}\oplus{\bf 10}\oplus{\bf \overline{10}}\oplus{\bf 1})
\oplus({\bf 5}\oplus{\bf \overline{5}})$, respectively.
(For further information, see e.g.,
Refs.~\cite{McKay:1981,Slansky:1981yr,Yamatsu:2015gut}.) 

We also use the 5D brane scalar field $\widetilde{\Phi}_{\bf 32}$ 
related to $\Phi_{\bf 32}$ by  
\begin{align}
\widetilde{\Phi}_{\bf 32}=
\left(
\begin{array}{c}
-\widehat{R}\Phi_{\overline{\bf 16}}^*\\
\widehat{R}\Phi_{\bf 16}^*\\
\end{array}
\right),\ \ \
\widehat{R}=\sigma^2\otimes\sigma^3\otimes\sigma^2\otimes\sigma^3.
\end{align}
For $\langle\Phi_{\bf 32}\rangle\not=0$, 
\begin{align}
\widehat{R}\langle\Phi_{\bf 16}^* \rangle=
\left(
\begin{array}{c}
0_4\\
0_4\\
0_4\\
\tilde{v}_4\\
\end{array}
\right),\ \ \
\tilde{v}_4=
\left(
\begin{array}{c}
0\\
0\\
w^*\\
0\\
\end{array}
\right).
\end{align}
The combination of the nonvanishing VEV $\langle\Phi_{\bf 32}\rangle$ on
the UV brane (at $y=0$) and the orbifold BCs $P_j (j=0,1,2,3)$
reduces $SO(11)$ to the SM gauge group 
$G_{\rm SM}(=SU(3)_C\times SU(2)_L\times U(1)_Y)$.

\subsubsection{Action for the singlet brane fermion $\chi_{\bf 1}$}

The action for $\chi_{\bf 1}^\beta (x,v)$, which satisfies the symplectic Majorana 
condition (\ref{sympMajorana1}), is
\begin{align}
&S_{\rm 5D\ brane}^{\rm brane}=
\int d^6x\sqrt{-\mbox{det}G} \, \delta(y)
{\cal L}_{\chi_{\bf 1}} ~, \cr
&{\cal L}_{\chi_{\bf 1}}:=
\frac{1}{2}\overline{\chi}_{\bf 1}^\beta
\left(\gamma^\mu\partial_\mu+\gamma^6\partial_v\right)\chi_{\bf 1}^\beta
- \frac{1}{2} M^{\beta \beta'} \overline{\chi}_{\bf 1}^\beta \chi_{\bf 1}^{\beta'} ,
\label{Eq:Action-brane-fermion-chi}
\end{align}
where $M^{\beta \beta'}$ is a constant matrix.

\subsubsection{Brane masses and interactions}

On the UV brane there can be $SO(11)$-invariant brane interactions
among the bulk fermions, the singlet brane fermion, and the brane scalar.
We consider  
\begin{align}
S_{\rm 5D\ brane}^{\rm fermion}=&
\int d^6x\sqrt{-\mbox{det}G}\delta(y)
\Big({\cal L}_1+{\cal L}_2+{\cal L}_3+{\cal L}_4+{\cal L}_5
+{\cal L}_6+{\cal L}_7 + {\cal L}_8 \Big) ~, \cr
\noalign{\kern 5pt}
{\cal L}_1:=&
-\kappa^{\alpha\beta}
\overline{\Psi_{\bf 32}^{\alpha}}\Gamma^a\Phi_{\bf 32}
\cdot(\Psi_{\bf 11}^{\beta})_a
-\kappa^{\alpha\beta*}
(\overline{\Psi_{\bf 11}^{\beta}})_{a}\Phi_{\bf 32}^{\dag}
\Gamma^a(\Psi_{\bf 32}^{\alpha})_a,\cr
{\cal L}_2:=&
-\widetilde{\kappa}^{\alpha\beta}
\overline{\Psi_{\bf 32}^{\alpha}}\Gamma^a
\widetilde{\Phi}_{\bf 32}\cdot(\Psi_{\bf 11}^{\beta})_a
-\widetilde{\kappa}^{\alpha\beta*}
(\overline{\Psi_{\bf 11}^{\beta}})_{a}
\widetilde{\Phi}_{\bf 32}^{\dag}\Gamma^a(\Psi_{\bf 32}^{\alpha})_a,\cr
{\cal L}_3:=&
-2\mu_{\bf 11}^{\beta\beta'}
\overline{\Psi_{\bf 11}^{\beta}}\Psi_{\bf 11}^{\beta'},
\allowdisplaybreaks[1]\cr
{\cal L}_4:=&
-\kappa^{\prime\alpha\beta}
\overline{\Psi_{\bf 32}^{\alpha}}\Gamma^a\Phi_{\bf 32}
\cdot(\Psi_{\bf 11}^{\prime\beta})_a
-\kappa^{\prime\alpha\beta*}
(\overline{\Psi_{\bf 11}^{\prime\beta}})_{a}\Phi_{\bf 32}^{\dag}
\Gamma^a(\Psi_{\bf 32}^{\alpha})_a,\cr
{\cal L}_5:=&
-\widetilde{\kappa}^{\prime\alpha\beta}
\overline{\Psi_{\bf 32}^{\alpha}}\Gamma^a
\widetilde{\Phi}_{\bf 32}\cdot(\Psi_{\bf 11}^{\prime\beta})_a
-\widetilde{\kappa}^{\prime\alpha\beta*}
(\overline{\Psi_{\bf 11}^{\prime\beta}})_{a}
\widetilde{\Phi}_{\bf 32}^{\dag}\Gamma^a(\Psi_{\bf 32}^{\alpha})_a,\cr
{\cal L}_6:=&
-2\mu_{\bf 11}^{\prime\beta\beta'}
\overline{\Psi_{\bf 11}^{\prime\beta}}\Psi_{\bf 11}^{\prime\beta'},
\allowdisplaybreaks[1]\cr
{\cal L}_7:=&
-\widetilde{\kappa}_{\bf 1}^{\alpha \beta}
\overline{\chi_{\bf 1}^\beta}
\widetilde{\Phi}_{\bf 32}^\dagger    \Psi_{\bf 32}^{\alpha}
-\widetilde{\kappa}_{\bf 1}^{\alpha \beta *}
\overline{\Psi_{\bf 32}^{\alpha}}
\widetilde{\Phi}_{\bf 32}^{}\chi_{\bf 1}^\beta ,  \cr
{\cal L}_8:=&
-{\kappa}_{\bf 1}^{\alpha \beta}
\overline{\chi_{\bf 1}^\beta}
{\Phi}_{\bf 32}^\dagger    \Psi_{\bf 32}^{\alpha}
-{\kappa}_{\bf 1}^{\alpha \beta *}
\overline{\Psi_{\bf 32}^{\alpha}}
{\Phi}_{\bf 32}^{}\chi_{\bf 1}^\beta  ,  
\label{Eq:Action-brane-fermion}
\end{align}
where $\kappa$'s and $\mu$'s are coupling constants.
Note that 
$\overline{\Psi_{\bf 11}^{\alpha}}\Psi_{\bf 11}^{\prime\beta}$ and 
$\overline{\Psi_{\bf 11}^{\prime\beta}}\Psi_{\bf 11}^{\alpha}$
are forbidden by the parity assignment of 
$\overline{\Psi_{\bf 11}^{\alpha}}$ and $\Psi_{\bf 11}^{\prime\beta}$
shown in Table~\ref{Tab:BC-fermion-vector}.

\subsubsection{Mass terms for fermions on the UV brane}

From the action in Eqs.~(\ref{Eq:Action-brane-fermion-chi}) and
(\ref{Eq:Action-brane-fermion}), the quadratic mass terms on the UV brane 
for the bulk and brane fermions with $\langle\Phi_{\rm 32}\rangle\not=0$
are given by 
\begin{align}
S_{\rm brane\ mass}^{\rm fermion}=&
\int d^4x\int_0^{2\pi R_6}dv
\bigg(
\sum_{j=1}^8 {\cal L}_j^m
+{\cal L}_{\chi_{\bf 1}}^m\bigg),
\label{Eq:Action-brane-fermion-mass}
\end{align}
where  
\begin{align}
{\cal L}_1^m=&
-\mu_1^{\alpha\beta}
\left\{-i2
(\overline{\check{\hat{e}}^{\prime\alpha}}\check{\hat{E}}^{\beta}
+\overline{\check{\hat{\nu}}^{\prime\alpha}}\check{\hat{N}}^{\beta})
- 2\overline{\check{d}^{\prime\alpha}}\check{D}^{\beta} 
+\sqrt{2}\overline{\check{\hat{\nu}}^{\alpha}}\check{S}^{\beta}\right\}
+\mbox{h.c.},
\allowdisplaybreaks[1]\cr
{\cal L}_2^m=&
-\widetilde{\mu}_1^{\alpha\beta}
\left\{i2
(\overline{\check{e}^{\alpha}}\check{E}^{\beta}
+\overline{\check{\nu}^{\alpha}}\check{N}^{\beta})
-2\overline{\check{\hat{d}}^{\alpha}}\check{\hat{D}}^{\beta}
+\sqrt{2}\overline{\check{\nu}^{\prime\alpha}}\check{S}^{\beta}\right\}
+\mbox{h.c},
\allowdisplaybreaks[1]\cr
{\cal L}_3^m=&
-2 \mu_{\bf 11}^{\beta\beta'} 
\bigg\{
 \overline{\check{D}^{\beta}}\check{D}^{\beta'}
+\overline{\check{\hat{D}}^{\beta}}\check{\hat{D}}^{\beta'}
+\overline{\check{E}^{\beta}}\check{E}^{\beta'}
+\overline{\check{\hat{E}}^{\beta}}\check{\hat{E}}^{\beta'}  \cr
&\hskip 5.cm
+\overline{\check{N}^{\beta}}\check{N}^{\beta'}
+\overline{\check{\hat{N}}^{\beta}}\check{\hat{N}}^{\beta'}
+\overline{\check{S}^{\beta}}\check{S}^{\beta'}
\bigg\},
\allowdisplaybreaks[1]\cr
{\cal L}_4^m=&
-\mu_2^{\alpha\beta}
\left\{-i
2(\overline{\check{\hat{e}}^{\prime\alpha}}\check{\hat{E}}^{\prime\beta}
+\overline{\check{\hat{\nu}}^{\prime\alpha}}\check{\hat{N}}^{\prime\beta})
-2\overline{\check{d}^{\prime\alpha}} {\check{D}^{\prime\beta}}
+\sqrt{2}\overline{\check{\hat{\nu}}^{\alpha}}\check{S}^{\prime\beta}
\right\}
+\mbox{h.c.},
\allowdisplaybreaks[1]\cr
{\cal L}_5^m=&
-\widetilde{\mu}_2^{\alpha\beta}
\left\{i2
(\overline{\check{e}^{\alpha}}\check{E}^{\prime\beta}
+\overline{\check{\nu}^{\alpha}}\check{N}^{\prime\beta})
-2\overline{\check{\hat{d}}^{\alpha}}\check{\hat{D}}^{\prime\beta}
+\sqrt{2}\overline{\check{\nu}^{\prime\alpha}}\check{S}^{\prime\beta}
\right\}
+\mbox{h.c},
\allowdisplaybreaks[1]\cr
{\cal L}_6^m=&
-2 \mu_{\bf 11}^{\prime\beta\beta'} 
\bigg\{
 \overline{\check{D}^{\prime\beta}}\check{D}^{\prime\beta'}
+\overline{\check{\hat{D}}^{\prime\beta}}\check{\hat{D}}^{\prime\beta'}
+\overline{\check{E}^{\prime\beta}}\check{E}^{\prime\beta'}
+\overline{\check{\hat{E}}^{\prime\beta}}\check{\hat{E}}^{\prime\beta'}  \cr
&\hskip 5.cm
+\overline{\check{N}^{\prime\beta}}\check{N}^{\prime\beta'}
+\overline{\check{\hat{N}}^{\prime\beta}}\check{\hat{N}}^{\prime\beta'}
+\overline{\check{S}^{\prime\beta}}\check{S}^{\prime\beta'}
\bigg\},
\allowdisplaybreaks[1]\cr
{\cal L}_7^m=&
- \frac{m_B^{\alpha \beta}}{\sqrt{k}} \,
( \overline{\chi^\beta}  \,  \check{\nu}^{\prime\alpha} + 
\overline{\check{\nu}^{\prime\alpha}} \, \chi^\beta )    ~,     \cr
\noalign{\kern 5pt}
{\cal L}_8^m=&
-  \frac{\hat m_B^{\alpha \beta} }{\sqrt{k}} \,
( \overline{\chi^\beta} \, \check{\hat \nu}^{\alpha} + 
\overline{\check{\hat \nu}^{\alpha}} \, \chi^\beta )    ~ ,   \cr
\noalign{\kern 5pt}
{\cal L}_{\chi_{\bf 1}}^m=&
-\frac{1}{2}  M^{\beta \beta'} \, \overline{\chi^\beta} \, \chi^{\beta '} ~.
\label{braneFmass1}
\end{align}
Here 
$2\mu_{1}^{\alpha\beta}:=\sqrt{2}\kappa^{\alpha\beta}w$,
$2\widetilde{\mu}_{1}^{\alpha\beta}:=
\sqrt{2}\widetilde{\kappa}^{\alpha\beta}w$,
$2\mu_{2}^{\alpha\beta}:=\sqrt{2}\kappa^{\prime\alpha\beta}w$,
$2\widetilde{\mu}_{2}^{\alpha\beta}:=
\sqrt{2}\widetilde{\kappa}^{\prime\alpha\beta}w$,
$m_B^{\alpha \beta}/\sqrt{k} :=\widetilde{\kappa}_{\bf 1}^{\alpha \beta}w$ and
$\hat m_B^{\alpha \beta }/ \sqrt{k} := {\kappa}_{\bf 1}^{\alpha \beta}w$.
For the sake of simplicity $\chi_{\bf 1}^\beta$ has been denoted as $\chi^\beta$.
All $\mu$ parameters are dimensionless quantities, whereas 
$m_B^{\alpha \beta}$,  $\hat m_B^{\alpha \beta}$ and $M^{\beta \beta'}$
have the dimension of mass.

\subsubsection{Mass terms for gauge bosons on the UV brane}

By replacing $\Phi_{\bf 32}$ to its VEV $\langle\Phi_{\bf 32}\rangle$ in
Eq.~(\ref{Eq:Action-brane-scalar}), the mass terms for the 4D components
of the $SO(11)$ gauge fields $A_\mu$ can be read off as 
\begin{align}
{\cal L}_{\rm brane\ mass}^{\rm 4D\ gauge}=&
-\frac{g^2w^2}{8}
\bigg\{
\left(A_\mu^{15}-A_\mu^{26}\right)^2
+\left(A_\mu^{16}+A_\mu^{25}\right)^2
\nonumber\\
&
+\left(A_\mu^{17}-A_\mu^{28}\right)^2
+\left(A_\mu^{18}+A_\mu^{27}\right)^2
+\left(A_\mu^{19}-A_\mu^{2,10}\right)^2
+\left(A_\mu^{1,10}+A_\mu^{29}\right)^2
\nonumber\\
&
+\left(A_\mu^{35}+A_\mu^{46}\right)^2
+\left(A_\mu^{36}-A_\mu^{45}\right)^2
+\left(A_\mu^{37}+A_\mu^{48}\right)^2
+\left(A_\mu^{38}-A_\mu^{47}\right)^2
\nonumber\\
&
+\left(A_\mu^{39}+A_\mu^{4,10}\right)^2
+\left(A_\mu^{3,10}-A_\mu^{49}\right)^2
+\left(A_\mu^{57}-A_\mu^{68}\right)^2
+\left(A_\mu^{58}+A_\mu^{67}\right)^2
\nonumber\\
&
+\left(A_\mu^{59}-A_\mu^{6,10}\right)^2
+\left(A_\mu^{5,10}+A_\mu^{69}\right)^2
+\left(A_\mu^{79}-A_\mu^{8,10}\right)^2
+\left(A_\mu^{7,10}+A_\mu^{89}\right)^2
\nonumber\\
&
+\left(A_\mu^{23}-A_\mu^{14}\right)^2
+\left(A_\mu^{31}-A_\mu^{24}\right)^2
+\left(A_\mu^{12}-A_\mu^{34}+A_\mu^{56}+A_\mu^{78}+A_\mu^{9,10}\right)^2
\nonumber\\
&
+\left(A_\mu^{1,11}\right)^2
+\left(A_\mu^{2,11}\right)^2
+\left(A_\mu^{3,11}\right)^2
+\left(A_\mu^{4,11}\right)^2
+\left(A_\mu^{5,11}\right)^2
\nonumber\\
&
+\left(A_\mu^{6,11}\right)^2
+\left(A_\mu^{7,11}\right)^2
+\left(A_\mu^{8,11}\right)^2
+\left(A_\mu^{9,11}\right)^2
+\left(A_\mu^{10,11}\right)^2
\bigg\},
\label{Eq:Brane-mass-term-gauge-field-4D}
\end{align}
where the notation is the same as one in  Ref.~\cite{Furui:2016owe}. We
omit them here. 
All the 31 components of $SO(11)/SU(5)$ obtain large brane masses via 
$\langle\Phi_{\bf 32}\rangle\not=0$. 
Each mass term changes the 5th dimensional BCs on the UV brane for the
corresponding fields.
Similarly, from the action (\ref{Eq:Action-brane-scalar}), we  find 
the mass terms for the 6th dimensional component of the $A_v$:
\begin{align}
{\cal L}_{\rm brane\ mass}^{\rm 6th\ dim\ gauge}=&
-\frac{g^2w^2}{8}
\bigg\{
\left(A_v^{15}-A_v^{26}\right)^2
+\left(A_v^{16}+A_v^{25}\right)^2
\nonumber\\
&
+\left(A_v^{17}-A_v^{28}\right)^2
+\left(A_v^{18}+A_v^{27}\right)^2
+\left(A_v^{19}-A_v^{2,10}\right)^2
+\left(A_v^{1,10}+A_v^{29}\right)^2
\nonumber\\
&
+\left(A_v^{35}+A_v^{46}\right)^2
+\left(A_v^{36}-A_v^{45}\right)^2
+\left(A_v^{37}+A_v^{48}\right)^2
+\left(A_v^{38}-A_v^{47}\right)^2
\nonumber\\
&
+\left(A_v^{39}+A_v^{4,10}\right)^2
+\left(A_v^{3,10}-A_v^{49}\right)^2
+\left(A_v^{57}-A_v^{68}\right)^2
+\left(A_v^{58}+A_v^{67}\right)^2
\nonumber\\
&
+\left(A_v^{59}-A_v^{6,10}\right)^2
+\left(A_v^{5,10}+A_v^{69}\right)^2
+\left(A_v^{79}-A_v^{8,10}\right)^2
+\left(A_v^{7,10}+A_v^{89}\right)^2
\nonumber\\
&
+\left(A_v^{23}-A_v^{14}\right)^2
+\left(A_v^{31}-A_v^{24}\right)^2
+\left(A_v^{12}-A_v^{34}+A_v^{56}+A_v^{78}+A_v^{9,10}\right)^2
\nonumber\\
&
+\left(A_v^{1,11}\right)^2
+\left(A_v^{2,11}\right)^2
+\left(A_v^{3,11}\right)^2
+\left(A_v^{4,11}\right)^2
+\left(A_v^{5,11}\right)^2
\nonumber\\
&
+\left(A_v^{6,11}\right)^2
+\left(A_v^{7,11}\right)^2
+\left(A_v^{8,11}\right)^2
+\left(A_v^{9,11}\right)^2
+\left(A_v^{10,11}\right)^2
\bigg\}.
\label{Eq:Brane-mass-term-gauge-field-6th}
\end{align}
These mass terms give masses for the corresponding fields by changing the BCs.
We shall see more details in Section 3.

\subsection{EW Higgs boson and twisted gauge}

The orbifold BCs break $SO(11)$ to $G_{\rm PS}$, and further the
nonvanishing VEV of the 5D brane scalar $\Phi_{\rm 32}$ reduces 
$G_{\rm PS}$ to $G_{\rm SM}$.
The bilinear terms of the action of gauge fields in (\ref{Eq:Action-bulk-gauge}) are 
written down as
\begin{align}
&\left(S_{\rm bulk}^{\rm gauge}\right)_{\rm quadratic}  \cr
\noalign{\kern 10pt}
&=
\int d^4xdv {\frac{dz}{kz^2}}  \, \sum_{j<k} \bigg[
\frac{1}{2}A_{\lambda}^{(jk)}
\bigg\{ \hat \eta^{\lambda \rho}  \big(\Box+k^2 {{\cal P}_5}  +\partial_v^2 \big)-
\Big( 1-\frac{1}{\xi} \Big) \partial^\lambda \partial^\rho \bigg\} A_\rho^{(jk)}   \cr
\noalign{\kern 5pt}
&\hskip 0.5cm
+\frac{1}{2}k^2A_z^{(jk)}
\left(\Box+\xi k^2{\cal P}_z+\partial_v^2\right)A_z^{(jk)}  
+\overline{c}^{(jk)}
\left(\Box+\xi k^2 {{\cal P}_5} +\partial_v^2\right)c^{(jk)}    \bigg],  \cr
\noalign{\kern 10pt}
&
\hat \eta_{\lambda \rho} = \hat \eta^{\lambda \rho} =  \diag (-1, +1, +1, +1, +1) ~, \cr
\noalign{\kern 5pt}
&
\Box =\eta^{\mu\nu}\partial_\mu\partial_\nu,\ 
\partial^\lambda = \hat \eta^{\lambda \rho} \partial_\rho, ~
{{\cal P}_5 =z^2 \frac{\partial}{\partial z}
\frac{1}{z^2}\frac{\partial}{\partial z}} ,\
{{\cal P}_z =\frac{\partial}{\partial z}
z^2 \frac{\partial}{\partial z}\frac{1}{z^2} }.
\label{gauge-free-action1}
\end{align}
Here $\lambda, \rho$ run over $0,1,2,3,6$ so that 
$A_\lambda \hat \eta^{\lambda \rho} B_\rho = A_\mu \eta^{\mu \nu} B_\nu + A_v B_v$. 
The 4D and 6th dimensional components $A_\mu$ and $A_v$
have additional bilinear terms that come from the 5D brane
scalar term ${\cal L}_{\Phi_{\bf 32}}$ in (\ref{Eq:Action-brane-scalar}) 
with $\langle\Phi_{\bf 32}\rangle\not=0$.
The parity assignments of $A_\mu$, $A_z$, and $A_v$ are summarized in 
Table~\ref{Tab:BC-gauge}.
The components $({\bf 2,2,1})$ of $A_z$ and $A_v$ have parity $(+,+)$.
The components $({\bf 2,2,1})$ of $A_v$ obtain masses through
the brane mass terms in Eq.~(\ref{Eq:Brane-mass-term-gauge-field-6th}),
so only the components $({\bf 2,2,1})$ of $A_z$, which is identified
with the SM Higgs scalar fields, have zero modes. 
Three of them are absorbed by $W$ and $Z$ bosons. 
The remaining neutral component, the observed Higgs boson, also
becomes massive by radiative corrections through the Hosotani mechanism
as shown below.
We note that the components $({\bf 2,2,1})$ of $A_v$   get
additional masses of order $g R_6^{-1}$ by radiative corrections at one loop.
The components $({\bf 3,1,1})$, $({\bf 1,3,1})$, and 
$({\bf 1,1,15})$ of $A_\mu$  have parity $(+,+)$.
$G_{\rm PS}$ is spontaneously broken to 
$G_{\rm SM}$ by the nonvanishing VEV of the component 
$({\bf 1,2,\overline{4}})$ of $\Phi_{\bf 32}$.

The zero modes of $A_z$ are physical degrees of freedom which cannot
be gauged away. By using the KK mode expansion discussed in
Appendix~\ref{Sec:6D-scalar-KK-modes},   
the 6th dimensional $n=0$ modes ($v$-independent modes) of $A_z^{a\, 11}$ 
are expanded, 
in terms of mode functions $\big\{h_n^{(++)}(z)\big\}$ for
the parity $(+,+)$ boundary condition, as  
\begin{align}
A_z^{a \, 11}=&
\frac{1}{\sqrt{2\pi R_6}} \bigg\{ 
\phi_H^{a (0)}(x)u_H(z)+
\sum_{n=1}^{\infty}\phi_H^{a(n)}(x)h_n^{(++)}(z) \bigg\}   \ \ 
(a=1,2,3,4),
\label{AzHiggs1}
\end{align}
{where the zero mode function  in the fifth dimension is given by
$u_H (z) = [3/k (z_L^3 -1) ]^{1/2} \, z^2$. }
The four-component real field $\phi_H^{a (0)}(x)$ is identified
with the EW Higgs doublet field in the SM. We will show later that the
$\phi_H:=\phi_H^{ 4 (0)}$ dynamically obtain the nonvanishing VEV 
$\langle\phi_H\rangle\not=0$. The VEV breaks $G_{\rm SM}$ to 
$SU(3)_C\times U(1)_{\rm EM}$. $\phi_H^{1,2,3 (0)}$ are absorbed by $W$
and $Z$ bosons.

Under a general gauge transformation 
$A_M'=\Omega A_M \Omega^{-1}+(i/g)\Omega\partial_M\Omega^{-1}$, 
new gauge potentials satisfy the following new BCs:
\begin{align}
&\begin{pmatrix} A_\mu'\\ A_y'\\ A_v'  \end{pmatrix} (x,y_j-y,v_j-v) =
P_j'
\begin{pmatrix} A_\mu'\\ -A_y'\\ -A_v'  \end{pmatrix}(x,y_j+y,v_j+v) P_j^{\prime -1}\cr
\noalign{\kern 5pt}
&\hskip 7.5cm
+\frac{i}{g} P_j'
\begin{pmatrix}  \partial_\mu\\ -\partial_y\\ -\partial_v \end{pmatrix} P_j^{\prime -1}, \cr
\noalign{\kern 5pt}
&\hskip .5cm
P_j' =\Omega(x,y_j-y,v_j-v)P_j\Omega(x,y_j+y,v_j+v)^{-1} ~.
\end{align}
It is convenient to  work in the twisted gauge discussed,  e.g.,  in Ref.~\cite{Furui:2016owe}. 
With the following large gauge transformation $\Omega(y;\alpha)$, 
$\hat{\theta}_H$ is transformed to $\hat{\theta}_H'=0$:
\begin{align} 
\Omega(y)=&\mbox{exp}
\left\{i\frac{g\theta_H f_H}{ { \sqrt{4 \pi R_6}} }\int_y^{L}dy
\tilde{u}_H(y)T_{4,11}\right\}
=\mbox{exp}
\big\{ i\theta(z)T_{4,11}\big\}, \cr
\noalign{\kern 5pt}
\theta(z)=& \theta_H { \frac{z_L^3- z^3 }{z_L^3 -1} } \ \ 
\mbox{for}\ 1\leq z\leq z_L.
\label{largeGT1}
\end{align}
The new BC matrices $P_j'$ are given by
\begin{align}
\tilde{P}_j=&\Omega(0)^{2}P_j=e^{2i\theta_H T_{4,11}}P_j\ \ (j=0,2),\cr
\tilde{P}_k=&P_k\ \ (k=1,3).
\end{align}
For fields in the $SO(11)$ vector and spinor representations, the BC matrices are
given by
\begin{align}
\tilde{P}_0^{\rm vec}=&
\begin{cases}
\begin{pmatrix} \cos\theta_H&-\sin\theta_H\cr -\sin\theta_H&-\cos\theta_H  \end{pmatrix}
&\mbox{in the $4-11$ subspace},\cr
\noalign{\kern 5pt}
I_3&\mbox{in the $1,2,3$ subspace}, \cr
\noalign{\kern 5pt}
- I_6&\mbox{in the $5, \cdots, 10$ subspace},
\end{cases}  \cr
\noalign{\kern 10pt}
\tilde{P}_2^{\rm vec}=&
\begin{cases}
\begin{pmatrix} \cos\theta_H&-\sin\theta_H\cr -\sin\theta_H&-\cos\theta_H  \end{pmatrix}
&\mbox{in the $4-11$ subspace},\cr
\noalign{\kern 5pt}
I_9&\mbox{otherwise},
\end{cases}  \cr
\noalign{\kern 10pt}
\tilde{P}_0^{\rm sp}=&
\begin{pmatrix}
\pm \cos\frac{\theta_H}{2}& - i \sin\frac{\theta_H}{2}\\
 i\sin\frac{\theta_H}{2}& \mp\cos\frac{\theta_H}{2}  \end{pmatrix}
\quad
\mbox{for}\
\bigg\{ \,
\begin{matrix} \psi \cr \hat \psi \end{matrix} ~ ,   \cr
\noalign{\kern 10pt}
\tilde{P}_2^{\rm sp}=&
\begin{pmatrix}
\cos\frac{\theta_H}{2}&\mp i \sin\frac{\theta_H}{2}\\
\pm i\sin\frac{\theta_H}{2}&-\cos\frac{\theta_H}{2}  \end{pmatrix}
\quad
\mbox{for}\
\bigg\{ \,
\begin{matrix} \psi \cr \hat \psi \end{matrix} ~ ,   
\end{align}
where  
\begin{align}
{\psi}=&
\left(
\begin{array}{c}
{\nu}\\
{\nu}'\\
\end{array}
\right),\ 
\left(
\begin{array}{c}
{e}\\
{e}'\\
\end{array}
\right),\ 
\left(
\begin{array}{c}
{u}_j\\
{u}_j'\\
\end{array}
\right),\ 
\left(
\begin{array}{c}
{d}_j\\
{d}_j'\\
\end{array}
\right),\nonumber\\
\hat{\psi}=&
\left(
\begin{array}{c}
\hat{\nu}\\
\hat{\nu}'\\
\end{array}
\right),\ 
\left(
\begin{array}{c}
\hat{e}\\
\hat{e}'\\
\end{array}
\right),\ 
\left(
\begin{array}{c}
\hat{u}_j\\
\hat{u}_j'\\
\end{array}
\right),\ 
\left(
\begin{array}{c}
\hat{d}_j\\
\hat{d}_j'\\
\end{array}
\right).
\label{chihatchi}
\end{align}


\section{Mass spectrum of bosons}
\label{Sec:Spectrum-Bosons}

In this section we examine the spectrum of gauge fields, particularly  for 
6th-dimensional $n=0$ KK modes ($v$-independent modes) 
because we are interested in the mass
spectrum of the SM particles and the effective potential $V_\eff (\theta_H)$.
In the following we omit  the modes which are odd under the 6th dimensional
loop translation $U_6 = P_2 P_0 = P_3 P_1 = -1$.
Those modes have masses  $\ge \onehalf m_{\KK_6}$.
\begin{table}
\begin{center}
\ovalbox{
\begin{tabular}{l}
\\[-1.em]
$SO(11)$\\[0.75em]
\ (5),(6)\\[0.5em]
\end{tabular}
\ovalbox{
\begin{tabular}{l}
\\[-1.0em]
$SO(10)$\\[0.75em]
\ \ (4)\\[0.5em]
\end{tabular}
\ovalbox{
\begin{tabular}{l}
\\[-1em]
$SU(5)$\ \ \ \ \ \ \ \  $G_{SM}$\\[0.5em]
\ \ (2)\ \ \ \ \ \ \ \ \ \ (1) \\[0.5em]
\end{tabular}
}
\hspace{-6em}
\ovalbox{
\begin{tabular}{l}
\\[-1.25em]
\hspace{5em}
$G_{PS}$\\[0.5em]
\hspace{5.3em}(3) \\[0.4em]
\end{tabular}
}}}
\end{center}
\begin{center}
\renewcommand{\arraystretch}{1.1}
\begin{tabular}{cccccc}\hline
&&$\begin{matrix}{\rm No.~of}\cr{\rm generators} \end{matrix}$
&$A_\mu$&$A_z$&$A_v$\\
\hline 
(1)&$G_{\rm SM}$ &12 
&$\begin{pmatrix} N&N\cr N&N \end{pmatrix}$
&$\begin{pmatrix} D&D\cr D&D \end{pmatrix}$
&$\begin{pmatrix} D&D\cr D&D \end{pmatrix}$ \cr
(2)&$SU(5)/G_{\rm SM}$&12  
&$\begin{pmatrix} N&N\cr D&D \end{pmatrix}$
&$\begin{pmatrix} D&D\cr N&N \end{pmatrix}$
&$\begin{pmatrix} D&D\cr N&N \end{pmatrix}$ \cr
(3)&$G_{\rm PS}/G_{\rm SM}$&9 
&$\begin{pmatrix} D_{\rm eff} &N\cr D_{\rm eff} &N \end{pmatrix}$
&$\begin{pmatrix} D&D\cr D&D \end{pmatrix}$
&$\begin{pmatrix} D&D\cr D&D \end{pmatrix}$ \cr
(4)&$SO(10)/(SU(5)\cup G_{\rm PS})$&12
&$\begin{pmatrix} D_{\rm eff} &N\cr D&D \end{pmatrix}$
&$\begin{pmatrix} D&D\cr N&N \end{pmatrix}$
&$\begin{pmatrix} D&D\cr D_{\rm eff} &N \end{pmatrix}$ \cr
(5)&$SO(5)/SO(4)$&4
&$\begin{pmatrix} D&D\cr D&D \end{pmatrix}$
&$\begin{pmatrix} N&N\cr N&N \end{pmatrix}$
&$\begin{pmatrix} D_{\rm eff} &N\cr D_{\rm eff} &N \end{pmatrix}$ \cr
(6)&$SO(7)/SO(6)$&6
&$\begin{pmatrix} D&D\cr D_{\rm eff} &N \end{pmatrix}$
&$\begin{pmatrix} N&N\cr D&D \end{pmatrix}$
&$\begin{pmatrix} D_{\rm eff} &N\cr D&D \end{pmatrix}$ \cr
\hline
\end{tabular}
\end{center}
\caption{The Venn diagram for symmetry breaking pattern and 
the BCs \usebox{\BCmatrix} 
of the $SO(11)$ gauge field are summarized.
$N$ and $D$ stand for Neumann and Dirichlet conditions, respectively. 
$D_{\rm eff}$ represents the effective Dirichlet condition given by (\ref{effDirichletBC}).
}
\label{Table:Venn-diagram-Symmetry}
\end{table}

The BCs for gauge fields in the absence of  the brane terms are
given by 
\begin{align}
&\left\{
\begin{array}{lll}
N:\ &\frac{\partial}{\partial z}A_\mu=0\ &\mbox{for parity}=+,\cr
\noalign{\kern 5pt}
D:\ &A_\mu=0\ &\mbox{for parity}=-,
\end{array}
\right.,\cr
\noalign{\kern 5pt}
&\left\{
\begin{array}{lll}
N:\ &\frac{\partial}{\partial z}\left( {\frac{1}{z^2}} A_z\right)=0\ &
\mbox{for parity}=+, \cr
\noalign{\kern 5pt}
D:\ &A_z=0\ &\mbox{for parity}=-,
\end{array}
\right., \cr
\noalign{\kern 5pt}
&\left\{
\begin{array}{lll}
N:\ &\frac{\partial}{\partial z}A_v=0\ &\mbox{for parity}=+,  \cr
\noalign{\kern 5pt}
D:\ &A_v=0\ &\mbox{for parity}=-,
\end{array}
\right.
\end{align}
at $z=1 \, (y=0)$ and $z=z_L \, (y=L)$. 
In the presence of  the brane mass terms 
$| g A_\mu \langle\Phi_{\rm 32} \rangle |^2$ on the UV brane,
the Neumann BC $N$ is modified to an effective Dirichlet BC 
$D_{\rm eff}$ for the 5th dimensional zero mode of the $SO(11)/SU(5)$
components of $A_\mu$ and $A_v$.\cite{Furui:2016owe}
The equation of motion for $A_\mu^a$ in the $y$-coordinate is 
given in the form 
\begin{align}
\bigg( \Box + { e^{\sigma (y) } }\frac{\dd}{\dd y} { e^{- 3\sigma (y)}}  \frac{\dd}{\dd y}
+ \frac{\dd^2}{\dd v^2} - \frac{g^2 w^2}{2} \, \delta (y) \bigg) A_\mu^a
-
\bigg(1 - \frac{1}{\xi}  \bigg) \dd_\mu ( \dd^\nu A_\nu^a + \dd_v A_v^a) = \cdots
\label{gaugeequation1}
\end{align}
where the right-hand side involves interaction terms.
Suppose that $A_\mu^a$ is parity even, namely 
$A_\mu^a (x, -y, -v) = + A_\mu^a (x, y, v)$.  
Note that $A_v^a$ is parity odd so that $\dd_v A_v^a$ is parity even. 
By integrating the equation $\int_{-\epsilon}^\epsilon dy \cdots$ and taking the limit 
$\epsilon \go 0$, one finds 
$\dd A_\mu^a / \dd y |_{y=\epsilon} = (g^2 w^2/4) A_\mu^a |_{y=0}$.
In the $z$ coordinate the Neumann condition is changed to the effective Dirichlet condition
\begin{align}
D_\eff (\omega) : ~ \Big( \frac{\dd}{\dd z} - \omega \Big) \, A_\mu^a =0 , ~~
\omega = \frac{g^2 w^2}{4k} ~, \quad {\rm at}~z=1^+ .
\label{effDirichletBC}
\end{align}
Note that in the current six-dimensional model the mass dimensions of $g$ and $w$ are
$[g] = M^{-1}$, $[w] = M^{3/2}$ so that $\omega$ is dimensionless.
When $\omega \not=0$, $A_\mu^a$ develops a cusp at $y=0$.
The lowest mode of $A_\mu^a$ becomes massive, whose mass is $O(m_{\rm KK_5})$. 
The value of $\omega$ depends on the brane mass terms.
The BCs for the gauge fields 
are summarized
in Table~\ref{Table:Venn-diagram-Symmetry}.

In the  twisted gauge 
$\tilde{A}_M=\Omega(z)A_M\Omega(z)^{-1}
+\frac{i}{g}\Omega(z)\partial_M\Omega(z)^{-1}$ where $\Omega (z) = e^{i \theta (z) T_{4, 11}}$.
$\theta (z)$ is defined in (\ref{largeGT1}).
The $(k,4)$, $(k,11)$, and $(4,11)$ components of
the $SO(11)$ gauge fields are changed as
\begin{align}
A_M^{k4}=&
\cos\theta(z)\tilde{A}_M^{k4}-\sin\theta(z)\tilde{A}_{M}^{k,11}\ 
(k\not=4,11),\cr
\noalign{\kern 5pt}
A_M^{k,11}=&
\sin\theta(z)\tilde{A}_M^{k4}+\sin\theta(z)\tilde{A}_{M}^{k,11},\cr
\noalign{\kern 5pt}
A_z^{4,11}=&
\tilde{A}_z^{4,11}-\frac{\sqrt{2}}{g}\theta^{\prime}(z)
=\tilde{A}_z^{4,11}+ { \frac{3\sqrt{2}}{g} \frac{z^2}{z_L^3-1}}  \, \theta_H ,
\end{align}
while the other components remain unchanged.
As in Ref.~\cite{Furui:2016owe}, the BCs at $z=z_L$ determine 
wave functions for $\tilde{A}_\mu$, $\tilde{A}_z$, and $\tilde{A}_v$:
\begin{align}
&\tilde{A}_\mu\ \ N:\ C(z;\lambda),\  D:\ S(z;\lambda),\cr
&\tilde{A}_z\ \   N:\ S'(z;\lambda),\  D:\ C'(z;\lambda),\cr
&\tilde{A}_v\ \   N:\ S'(z;\lambda),\  D:\ C'(z;\lambda),
\end{align}
where $C(z;\lambda)$ and $S(z;\lambda)$ are defined in {(\ref{APgaugeF1}) and 
(\ref{APgaugeF2})}.

From the BCs for the gauge fields summarized in
Table~\ref{Table:Venn-diagram-Symmetry}, 
the components (2) $SU(5)/G_{\rm SM}$, (4) $SO(10)/(SU(5)\cup G_{\rm PS})$, and
(6) $SO(7)/SO(6)$ are parity odd under 
the 6th dimensional loop translation $U_6$, while
the components (1) $G_{\rm SM}$, (3) $G_{\rm PS}/G_{\rm SM}$, and (5) $SO(5)/SO(4)$
are parity even under  $U_6$.
Thus, we find that the low-lying
modes of the components (2), (4), (6)  of $A_M$ have masses of $O(m_{\rm KK_6})$, 
while the low-lying modes of the components (1), (3), and (5)of $A_M$
have masses less than or around $O(m_{\rm KK_5})$. 

We summarize the formulas determining  the mass spectrum of 6th dimensional
$n=0$ modes of $A_M$, 
for which the arguments in Ref.~\cite{Furui:2016owe} remain intact. 
We use  the same notation as in Ref.~\cite{Furui:2016owe}.
{
$A_M^{a_L}$ and $A_M^{a_R}$ ($a_L, a_R=1,2,3$) stand for $SU(2)_L$ and 
$SU(2)_R$ components of $A_M$, respectively. }
\begin{itemize}
\item $A_\mu$ components.
\begin{itemize}
\item[(i)]$(\tilde{A}_\mu^{a_L},\tilde{A}_\mu^{a_R},\tilde{A}_\mu^{a,11})$
$(a=1,2)$: For $W$ and $W_R$ towers, there are 6th dimensional
$n=0$ modes, and there are also 5th dimensional zero  modes in the absence of  brane terms. 
In the presence of the brane terms in (\ref{Eq:Brane-mass-term-gauge-field-4D}), 
their mass spectra of the 5th dimensional modes are given by
\begin{align}
2C'(SC'+\lambda\sin^2\theta_H)-\omega C(2SC'+\lambda\sin^2\theta_H)=0,
\end{align}
where $\omega:=g^2w^2/4k$.  
Let us denote the average magnitude of $C(1;\lambda)$ and $C'(1;\lambda)$ by
$\overline{C}$ and $\overline{C'}$, respectively. When $\omega \overline{C} \gg \overline{C'}$,
the spectrum is determined by
\begin{align}
W~\hbox{tower:}\ &2S(1;\lambda)C'(1;\lambda)+\lambda\sin^2\theta_H=0 ~, \cr
W_R ~\hbox{tower:}\ &C(1;\lambda)=0 ~.
\label{spectrumW}
\end{align}
Indeed, in typical cases examined below, 
$\overline{C}/\overline{C'} \sim {10^{6}}$  so that
the condition $\omega \overline{C} \gg \overline{C'}$ is well satisfied for $w \gg (m_{\KK_5})^{3/2}$.
      The mass of $W$ boson $m_W = m_{W^{(0)}}$ is given by
\begin{align}
m_W \simeq { \sqrt{\frac{3}{2}} \,  k z_L^{-3/2}\sin\theta_H 
= \frac{\sqrt{3} \sin\theta_H}{\sqrt{2} \pi\sqrt{z_L}} \, m_{\rm KK_5} }.
\label{Wmass}
\end{align}
\item[(ii)] $(\tilde{A}_\mu^{3_L},\tilde{C}_\mu ,\tilde{B}_\mu^Y,\tilde{A}_\mu^{3,11})$: 
	    For $\gamma$, $Z$ and $Z_R$ towers, there are 6th dimensional
	    $n=0$ modes, and there are also 5th dimensional zero
	    modes in the absence of  brane terms. 
Here
$C_\mu=\sqrt{2/5}A_\mu^{3_R}+\sqrt{3/5} A_\mu^{0_C}$ and
$B_\mu^Y =\sqrt{3/5}A_\mu^{3_R}-\sqrt{2/5} A_\mu^{0_C}$, where
$A_\mu^{0_C}=(A_\mu^{56}+A_\mu^{78}+A_\mu^{9\ 10})/\sqrt{3}$.
	    In the presence of  the brane terms in
(\ref{Eq:Brane-mass-term-gauge-field-4D}), 
	    the mass spectra of the 5th dimensional modes are given by
\begin{align}
C'\left\{2C'(SC'+\lambda\sin^2\theta_H)
-\omega C(5SC'+4\lambda\sin^2\theta_H)\right\}=0.
\end{align}
For $\omega \overline{C} \gg \overline{C'}$
\begin{align}
\gamma~\hbox{tower:}\ &C'(1;\lambda)=0,\cr
Z~\hbox{tower:}\ &5S(1;\lambda)C'(1;\lambda)+4\lambda\sin^2\theta_H=0,\cr
Z_R ~\hbox{tower:}\ &C(1;\lambda)=0.
\label{spectrumZ}
\end{align}
	    The mass of $Z$ boson $m_Z=m_{Z^{(0)}}$ is given by
\begin{align}
m_Z \simeq  { \sqrt{\frac{12}{5}} \,  k z_L^{-3/2}\sin\theta_H }
= \frac{m_W}{\cos \theta_W},\ \ \ \sin^2 \theta_W = \frac{3}{8} ~.
\label{Zmass}
\end{align}
The relation for the $Z$ tower in (\ref{spectrumZ}) can be written, 
by using $\cos^2 \theta_W = \frac{5}{8}$,  as
\begin{align}
Z~\hbox{tower:} ~  &2 S(1;\lambda)C'(1;\lambda)
+\frac{\lambda}{\cos^2 \theta_W} \sin^2\theta_H=0 ~.
\label{spectrumZ2}
\end{align}
The value $\sin^2 \theta_W = \frac{3}{8}$ is valid at the GUT scale.
The gauge coupling constants evolve as the energy scale, following the 
renormalization group equation (RGE).
It is not clear whether $\sin^2 \theta_W$ evolves to the observed value at low energies
as there are KK modes which do not respect $SU(5)$ below the GUT scale.
We leave solving the RGE  for future investigation.
When we evaluate the effective potential $V_\eff (\theta_H)$ at the EW scale 
in Section 5, we adopt the formula (\ref{spectrumZ2}) where
the observed value, $\sin^2 \theta_W \simeq 0.2312$,  at the EW scale is inserted.
\item[(iii)] $\tilde A_\mu^{4, 11}$: For $\hat A^{4}$ tower,
	     there is 6th dimensional $n=0$ modes, but there is no 5th
	     dimensional zero modes. The mass spectra of
	     the 5th dimensional modes is given by
\begin{align}
\hat{A}^{4}~\hbox{tower:}\ &S(1;\lambda)=0.
\end{align}
\item[(iv)] For $SU(3)_C$ gluons, there are 6th dimensional
	    $n=0$ modes, and there are also 5th dimensional zero
	    modes which correspond to 4d gluons.
	    The mass spectrum of the 5th dimensional massive modes is
	    given by
\begin{align}
\hbox{gluon tower:}\ &C'(1;\lambda)=0.
\end{align}
\item[(v)] For $X$-gluons, there are 6th dimensional
	   $n=0$ modes, and there are also 5th dimensional zero
	   modes in the absence of brane terms. 
	   In the presence of  the brane terms in
(\ref{Eq:Brane-mass-term-gauge-field-4D}), 
	   the mass spectrum of the 5th dimensional modes is given by
\begin{align}
X\hbox{-gluon tower:}\ &C'(1;\lambda)-\omega C(1;\lambda)=0.
\end{align}
For $\omega \overline{C} \gg \overline{C'}$
\begin{align}
X\hbox{-gluon tower:}\ &C(1;\lambda)\simeq 0.
\end{align}
\item[(vi)] For $X$-bosons, there are no 6th dimensional $n=0$ modes.
\item[(vii)] For $X'$-bosons, there are no 6th dimensional $n=0$ modes.
\item[(viii)] For $Y$, $Y'$-bosons, there are no 6th dimensional
	    $n=0$ modes.
\end{itemize}
\item $A_z$ components.
\begin{itemize}
\item[(i)] $A_z^{ab}\ (1\leq a< b\leq 3)$ and 
	   $A_z^{jk}\ (5\leq j<k\leq 10)$:
	   There are 6th dimensional $n=0$ modes, but there are no 5th
	   dimensional zero modes. Their mass spectrum of the 5th
	   dimensional modes is given by
\begin{align}
C'(1;\lambda)=0.
\end{align}
\item[(ii)] $A_z^{a4},A_z^{a\, 11}(a=1,2,3)$:
There are 6th dimensional $n=0$ modes for $A_z^{a4},A_z^{a\, 11}(a=1,2,3)$.
There are no 5th dimensional zero modes for $A_z^{a4}$ while there are 
 5th dimensional zero modes for $A_z^{a\, 11}$.
Their mass spectra of the 5th dimensional modes are given by
\begin{align}
S(1;\lambda)C'(1;\lambda)+\lambda\sin^2\theta_H=0.
\label{spectrumAz}
\end{align}
\item[(iii)] $A_z^{4,11}$: Higgs tower.  
	     There are 6th dimensional
	     $n=0$ modes, and there is also  a 5th dimensional zero mode. 
	     The mass spectrum of the 5th  dimensional modes is given by
\begin{align}
\hbox{Higgs tower:}\ &S(1;\lambda)=0.
\end{align}
\item[(iv)] $A_z^{ak} (a=1,2,3, k=5,..,10)$: 
	    There are no 6th dimensional $n=0$ modes.
\item[(v)] $A_z^{k4}$, $A_z^{k11}$ $(k= 5,...,10)$: 
	   There are no 6th dimensional $n=0$ modes.
\end{itemize}
\item $A_v$ components. 
\begin{itemize}
\item[(i)] $A_v^{ab}\ (1\leq a< b\leq 3)$ and 
	   $A_v^{jk}\ (5\leq j<k\leq 10)$: 
	   There are 6th dimensional $n=0$ modes, but there are no 5th
	   dimensional zero modes. Their mass spectra of the 5th
	   dimensional modes are given by
\begin{align}
C'(1;\lambda)=0.
\end{align}
\item[(ii)] $A_v^{a4},A_v^{a\ 11}(a=1,2,3)$: 
	    There are 6th dimensional $n=0$ modes for 
	    $A_v^{a4}$ and $A_v^{a\ 11}(a=1,2,3)$.
	    There are no 5th dimensional zero modes for $A_v^{a4}$ 
	    while there are 5th dimensional zero modes 
	    for $A_v^{a\ 11}(a=1,2,3)$ in the absence of  brane terms.
	    In the presence of  the brane terms in
(\ref{Eq:Brane-mass-term-gauge-field-6th}), 
	    one find that at $z=1$,
\begin{align}
\left(
\begin{array}{cc}
\cos\theta_H C'&-\sin\theta_H S'\\
\sin\theta_H (C-\omega C')&\cos\theta_H (S-\omega S')\\
\end{array}
\right)
\left(
\begin{array}{c}
\beta_{a4}^{v}\\
\beta_{a11}^{v}\\
\end{array}
\right)=0.
\end{align}
	    Thus, their mass spectra of the 5th dimensional modes are
	    given by 
\begin{align}
C'(1;\lambda)\left\{S(1;\lambda)-\omega S'(1;\lambda)\right\}
+\lambda\sin^2\theta_H=0.
\end{align}
The average maginitude  of $S(1;\lambda)$ ($=\overline{S}$) is much larger than
that of $S'(1;\lambda)$ ($=\overline{S'}$).  In typical cases $\overline{S}/\overline{S'} \sim {10^{4}}$.
Hence the spectrum is determined by
\begin{align}
C'(1;\lambda)S(1;\lambda) + \lambda\sin^2\theta_H=0. 
\end{align}
\item[(iii)] $A_v^{4,11}$: $A_v$-Higgs tower.
	     There are 6th dimensional $n=0$ modes, and there are also
	     5th dimensional zero modes in the absence of brane terms. 
	     In the presence of  the brane terms in
(\ref{Eq:Brane-mass-term-gauge-field-6th}), 
	     their mass spectra of the 5th dimensional modes are given by
\begin{align}
A_v\hbox{-Higgs tower:}\ &
S(1;\lambda)-\omega S'(1;\lambda)=0, 
\end{align}
{which is well approximated by}
\begin{align}
A_v\hbox{-Higgs tower:}\ &
{S(1;\lambda)=0.}
\end{align}
The components $A_v^{a \, 11}$ ($a=1 \sim 4$) 
acquire large 1 loop corrections of order $g_w M_{\KK_6}$ to their masses
so that their contributions to the $\theta_H$-dependent part of $V_\eff (\theta_H)$ 
become negligible and may be dropped.
\item[(iv)] $A_v^{ak} (a=1,2,3, k=5,..,10)$: 
	    There are no 6th dimensional $n=0$ modes. 
\item[(v)] $A_v^{k4}$, $A_v^{k11}$ $(k= 5,...,10)$: 
	   There are no 6th dimensional $n=0$ modes.
\end{itemize}
\end{itemize}

\ignore{ From the above mass spectrum formula, we find that for large
	$w$, the 6th dimension gauge fields $A_v^{ij}$ give no
	substantial contribution to the 5th dimensional effective potential 
	$V_{\rm eff}(\theta_H)$.  }

\section{Mass spectrum of fermions}
\label{Sec:Spectrum-Fermions}

In this section we determine the mass spectrum of quarks and leptons. For up-type
quarks, there are no brane interactions on the UV brane $(y=0)$ as
in the 5D $SO(11)$ GHGUT. For down-type quarks, charged leptons, and
neutrinos, both 6D bulk mass terms and  brane mass terms on the UV brane
are important to reproduce the observed mass spectra. 

In the following, we will calculate mass spectra for one set of 
a 6D $SO(11)$ ${\bf 32}$ Weyl fermion $\Psi_{\bf 32}^{\alpha}$ 
$(\alpha=1,2 ~{\rm or}~3)$, 
6D $SO(11)$ ${\bf 11}$ Dirac fermions 
$\Psi_{\bf 11}^{\beta}$ and $\Psi_{\bf 11}^{\prime\beta}$ $(\beta=\alpha)$,
and 5D $SO(11)$ singlet $\chi_{\bf 1}^\alpha$, 
which is identified with one generation of the SM quarks and leptons.
We will also discuss mass spectra for the 6D $SO(11)$ ${\bf 32}$ 
Weyl fermion $\Psi_{\bf 32}^{\alpha=4}$ denoted by 
$\Psi_{\bf 32}^{\prime}$.
We will omit the superscripts $\alpha$ and $\beta$. 
We are interested in EW scale physics and  we keep only 6th
dimensional $n=0$ KK modes because the masses of 6th dimensional
$n\not=0$ modes are much larger than $O(m_{\rm KK_5})$ as 
$m_{\rm KK_6}\gg m_{\rm KK_5}$. That is, we will discuss mass
spectra for the $G_{\rm PS}$ $({\bf 2,1,4})$ and $({\bf 1,2,4})$
components of $\Psi_{\bf 32}$,
the $G_{\rm PS}$ $({\bf 1,1,6})$ components of $\Psi_{\bf 11}$, 
the $G_{\rm PS}$ $({\bf 2,2,1})$ and $({\bf 1,1,1})$ 
components of $\Psi_{\bf 11}^{\prime}$, and
the $G_{\rm PS}$ $({\bf 2,1,4})$ and $({\bf 1,2,4})$
components of $\Psi_{\bf 32}^{\prime}$.
We will denote $c_{\Psi_{\bf 32}}^{\alpha\not=4}$, 
$c_{\Psi_{\bf 11}}^{\beta}$, $c_{\Psi_{\bf 11}}^{\prime\beta}$,
and $c_{\Psi_{\bf 32}}^{\alpha=4}$ 
as $c_0$, $c_1$, $c_2$, and $c_0^{\prime}$,  respectively.
In this paper we assume that {$c_0, c_1,  c_0^{\prime} \ge 0$ while $c_2$ can be negative}.
The calculation method employed in the following  is the same as the one 
employed in the 5D $SO(11)$ GHGUT discussed in Ref.~\cite{Furui:2016owe}.

\subsection{Up-type quark}

\subsubsection*{(i) $Q_{\rm EM}=+2/3$: $u_j,u_j'$ $(\Psi_{\bf 32})$}

From the action (\ref{Eq:Action-bulk-fermion-check}), 
we find the equations of motion for up-type quarks with $\gamma_{6D}^7 = +1$: 
\begin{align}
-i\delta
\left(
\begin{array}{c}
u_{+L}^{\dag}\\
u_{+L}^{\prime\dag}\\
\end{array}
\right):\ &
\left(-k\hat{D}_-(c_{0})+i\partial_v\right)
\left(
\begin{array}{c}
\check{u}_{+R}\\
\check{u}_{+R}^{\prime}\\
\end{array}
\right)
+\sigma^\mu\partial_\mu
\left(
\begin{array}{c}
\check{u}_{+L}\\
\check{u}_{+L}^{\prime}\\
\end{array}
\right)
=0,
\allowdisplaybreaks[1] \cr
i\delta
\left(
\begin{array}{c}
u_{+R}^{\dag}\\
u_{+R}^{\prime\dag}\\
\end{array}
\right):\ &
\overline{\sigma}^\mu\partial_\mu
\left(
\begin{array}{c}
\check{u}_{+R}\\
\check{u}_{+R}^{\prime}\\
\end{array}
\right)
+\left(-k\hat{D}_+(c_{0})+i\partial_v\right)
\left(
\begin{array}{c}
\check{u}_{+L}\\
\check{u}_{+L}^{\prime}\\
\end{array}
\right)
=0.
\end{align}
Here $\sigma^\mu = (I_2, \vec \sigma)$, 
$\overline{\sigma}^\mu = (-I_2, \vec \sigma)$, and
\begin{align}
\hat{D}_\pm (c) &= \pm \Big( \frac{\dd}{\dd z} + i \theta ' (z) T_{4,11} \Big) + \frac{c}{z} ~, \cr
\noalign{\kern 5pt}
T_{4,11} &= \begin{cases} ~ \onehalf \tau_1 &{\rm for~} \psi, \cr
- \onehalf \tau_1 &{\rm for~} \hat \psi, \end{cases} 
\label{Dhat1}
\end{align}
where $\psi$, $\hat \psi$ are the pairs defined in (\ref{chihatchi}).
For up-type quark with $\gamma_{6D}^7 = -1$, namely for the top quark, 
the equations become
\begin{align}
&
\left( k\hat{D}_+ (c_{0}) - i\partial_v\right)
\left(
\begin{array}{c}
\check{u}_{-R}\\
\check{u}_{-R}^{\prime}\\
\end{array}
\right)
+\sigma^\mu\partial_\mu
\left(
\begin{array}{c}
\check{u}_{-L}\\
\check{u}_{-L}^{\prime}\\
\end{array}
\right)
=0,
\allowdisplaybreaks[1] \cr
&
\overline{\sigma}^\mu\partial_\mu
\left(
\begin{array}{c}
\check{u}_{-R}\\
\check{u}_{-R}^{\prime}\\
\end{array}
\right)
+\left( k\hat{D}_-(c_{0}) - i\partial_v\right)
\left(
\begin{array}{c}
\check{u}_{-L}\\
\check{u}_{-L}^{\prime}\\
\end{array}
\right)
=0.
\end{align}

In the twisted gauge, the equations of motion become
\begin{align}
&\left\{
\begin{array}{c}
\overline{\sigma}^\mu\partial_\mu\tilde{u}_{+R}
=\left(+kD_+(c_0)-i\partial_v\right)\tilde{u}_{+L} \cr
\noalign{\kern 5pt}
{\sigma}^\mu\partial_\mu\tilde{u}_{+L}
=\left(+kD_-(c_0)-i\partial_v\right)\tilde{u}_{+R}\\
\end{array}
\right.,\ \
\left\{
\begin{array}{c}
\overline{\sigma}^\mu\partial_\mu\tilde{u}_{+R}'
=\left(+kD_+(c_0)-i\partial_v\right)\tilde{u}_{+L}' \cr
\noalign{\kern 5pt}
{\sigma}^\mu\partial_\mu\tilde{u}_{+L}'
=\left(+kD_-(c_0)-i\partial_v\right)\tilde{u}_{+R}'\\
\end{array}
\right.,\cr
\noalign{\kern 5pt}
&\left\{
\begin{array}{c}
\overline{\sigma}^\mu\partial_\mu\tilde{u}_{-R}
=\left(-kD_-(c_0)+i\partial_v\right)\tilde{u}_{-L} \cr
\noalign{\kern 5pt}
{\sigma}^\mu\partial_\mu\tilde{u}_{-L}
=\left(-kD_+(c_0)+i\partial_v\right)\tilde{u}_{-R}\\
\end{array}
\right.,\ \
\left\{
\begin{array}{c}
\overline{\sigma}^\mu\partial_\mu\tilde{u}_{-R}'
=\left(-kD_-(c_0)+i\partial_v\right)\tilde{u}_{-L}' \cr
\noalign{\kern 5pt}
{\sigma}^\mu\partial_\mu\tilde{u}_{-L}'
=\left(-kD_+(c_0)+i\partial_v\right)\tilde{u}_{-R}'\\
\end{array}
\right.,
\end{align}
where 
\begin{align}
D_\pm (c) = \pm \frac{\dd}{\dd z} + \frac{c}{z}
\label{Dpm}
\end{align}
and the relation between the twisted gauge and original one is given by 
\begin{align}
&\begin{pmatrix}  u\\ u' \end{pmatrix}
=\begin{pmatrix}
\cos \onehalf \theta(z) &-i\sin \onehalf \theta(z)  \cr
\noalign{\kern 5pt}
-i\sin \onehalf \theta(z) &\cos \onehalf \theta(z)   \end{pmatrix}
\begin{pmatrix} \tilde  u\\  \tilde u' \end{pmatrix}
=:\Omega(z)^{-1}
\begin{pmatrix} \tilde  u\\  \tilde u' \end{pmatrix} .
\label{twistedgauge2}
\end{align}

Let us check KK mode expansions for $u_L$ and $u_R$. Since $u_L$ has the
parity assignment 
$\begin{pmatrix} P_2&P_3 \cr  P_0&P_1 \end{pmatrix} =
\begin{pmatrix} +&+ \cr  +&+ \end{pmatrix}$,
we can expand it by using parity even combinations:
\begin{align}
u_L(x,y,v) &=
 \sum_{n=0}^{\infty}u_{nL}^C(x,y)f_n^C(v)
+\sum_{n=1}^{\infty}u_{nL}^S(x,y)f_n^S(v) \cr
\noalign{\kern 5pt}
&
=\frac{1}{\sqrt{2\pi R_6}}
\sum_{n=-\infty}^{\infty}u_{nL}(x,y)e^{inv/R_6},
\end{align}
where 
\begin{align}
&u_{nL}(x,y)=
\begin{cases}
\displaystyle \frac{1}{\sqrt{2}} \big\{ u_{nL}^C(x,y)-i u_{nL}^S(x,y) \big\} &(n>0) \cr
\noalign{\kern 5pt}
 u_{0L}^C(x,y)&(n=0) \cr
 \noalign{\kern 5pt}
\displaystyle   \frac{1}{\sqrt{2}} \big\{ u_{-n  L}^C(x,y) + i u_{-n  L}^S(x,y) \big\} &(n< 0)
\end{cases} \cr
\noalign{\kern 5pt}
&u_{nL}(x,-y)=u_{-nL}(x,y) ~, ~~u_{nL}(x,L-y)=u_{-nL}(x,L+y) ~. 
\end{align}
Note that $u_{nL}^C(x,y)$ and $u_{nL}^S(x,y)$ satisfy the same parity
property as $\widetilde \Phi_n^C (x,y)$ and $\widetilde \Phi_n^S (x,y)$
in (\ref{APPhi1}), 
and $f_n^{C} (v), f_n^{S} (v)$ are defined in (\ref{APfnCS}).
On the other hand, 
$u_R$ has the parity assignment 
$\begin{pmatrix} P_2&P_3 \cr  P_0&P_1 \end{pmatrix} =
\begin{pmatrix}-&- \cr  -&- \end{pmatrix}$
so that 
one can expand it by using parity odd combinations:
\begin{align}
u_R(x,y,v) &=
 \sum_{n=0}^{\infty}u_{nR}^S(x,y)f_n^C(v)
+\sum_{n=1}^{\infty}u_{nR}^C(x,y)f_n^S(v) \cr
\noalign{\kern 5pt}
&=\frac{1}{\sqrt{2\pi R_6}}
\sum_{n=-\infty}^{\infty}u_{nR}(x,y)e^{inv/R_6},
\end{align}
where 
\begin{align}
&u_{nR}(x,y)=
\begin{cases}
\mfrac{1}{\sqrt{2}} \big\{ u_{nR}^S(x,y)-i u_{nR}^C(x,y) \big\} &(n>0) \cr
\noalign{\kern 5pt}
 u_{0R}^S(x,y)&(n=0) \cr
 \noalign{\kern 5pt}
\mfrac{1}{\sqrt{2}} \big\{ u_{-n  R}^S(x,y) + i u_{-n  R}^C(x,y) \big\} &(n< 0)
\end{cases} \cr
\noalign{\kern 5pt}
&u_{nR}(x,-y)=-u_{-nR}(x,y) ~,~~
u_{nR}(x,L-y)=-u_{-nR}(x,L+y) ~.
\end{align}

To investigate physics at energies $\ll m_{\KK_6}$, 
we keep only 6th dimensional $n=0$ KK modes, 
and replace $u_{L/R}^{(\prime)}(x,y,v)$ by
$u_{0L/R}^{(\prime)}(x,y)$: 
\begin{align}
\left(
\begin{array}{c}
u_L(x,y,v)\\
u_R(x,y,v)\\
u_L'(x,y,v)\\
u_R'(x,y,v)\\
\end{array}
\right)
\ \Rightarrow\ 
\frac{1}{\sqrt{2\pi R_6}}
\left(
\begin{array}{c}
u_{0L}(x,y)\\
u_{0R}(x,y)\\
u_{0L}'(x,y)\\
u_{0R}'(x,y)\\
\end{array}
\right).
\end{align}
They have the following BCs at $(y_0,y_1)=(0,L_5)$;
\begin{align}
\left\{
\begin{array}{c}
u_{0L}(x,y_j-y)=+u_{0L}(x,y_j+y)\\
u_{0R}(x,y_j-y)=-u_{0R}(x,y_j+y)\\
u_{0L}'(x,y_j-y)=-u_{0L}'(x,y_j+y)\\
u_{0R}'(x,y_j-y)=+u_{0R}'(x,y_j+y)\\
\end{array}
\right.,
\end{align}
where $j=0,1$.
The equations of motion in the twisted gauge (\ref{twistedgauge2}) 
for the $\gamma_{6D}^7=+1$ fields become 
\begin{align}
&\left\{
\begin{array}{c}
\overline{\sigma}^\mu\partial_\mu\tilde{\check u}_{+0R}
=+kD_+(c_0)\tilde{\check u}_{+0L}\\
{\sigma}^\mu\partial_\mu\tilde{\check u}_{+0L}
=+kD_-(c_0)\tilde{\check u}_{+0R}\\
\end{array}
\right., \cr
\noalign{\kern 5pt}
&\left\{
\begin{array}{c}
\overline{\sigma}^\mu\partial_\mu\tilde{\check u}_{+0R}'
=+kD_+(c_0)\tilde{\check u}_{+0L}'\\
{\sigma}^\mu\partial_\mu\tilde{\check u}_{+0L}'
=+kD_-(c_0)\tilde{\check u}_{+0R}'\\
\end{array}
\right. .
\label{EoMup1}
\end{align}
It follows that 
$k^2 D_+ D_- \tilde{\check u}_{+0R} = \Box \tilde{\check u}_{+0R} 
= m^2 \tilde{\check u}_{+0R}$ etc.
The BCs at $z=z_L$ in the twisted gauge are the same as those in the original gauge:
\begin{align}
\tilde{\check u}_{+0R}=0,\  D_+\tilde{\check u}_{+0L}=0,\ 
 \tilde{\check u}_{+0L}'=0,\ D_-\tilde{\check u}_{+0R}'=0.
\end{align}

In terms of the basis functions $C_{R/L} (z; \lambda, c)$, $S_{R/L} (z; \lambda, c)$
given in Appendix B of Ref.~\cite{Furui:2016owe}, which satisfy
\begin{align}
&D_+ (c) \begin{pmatrix} C_L \cr S_L \end{pmatrix} 
= \lambda \begin{pmatrix} S_R \cr C_R \end{pmatrix} , ~~
D_- (c) \begin{pmatrix} C_R \cr S_R \end{pmatrix} 
= \lambda \begin{pmatrix} S_L \cr C_L \end{pmatrix} , \cr
\noalign{\kern 5pt}
&C_R = C_L = 1 , ~~ S_R=S_L = 0 \quad {\rm at~}z = z_L ~, 
\end{align}
one can write the mode functions for $\gamma_{6D}^7=+1$  as
\begin{align}
\begin{pmatrix} \tilde{\check u}_{+0R} \cr  \tilde{\check u}_{+0R}'   \end{pmatrix}
&=\begin{pmatrix}  \alpha_R^u S_R(z;\lambda,c_0)\cr
\alpha_R^{\prime u} C_R(z;\lambda,c_0)  \end{pmatrix}  f_R (x) ~, ~~
 \bar \sigma^\mu \dd_\mu f_R(x) = m f_L (x),\cr
\noalign{\kern 5pt}
\begin{pmatrix} \tilde{\check u}_{+0L}\\ \tilde{\check u}_{+0L}'  \end{pmatrix}
&=\begin{pmatrix} \alpha_L^u C_L(z;\lambda,c_0)\cr
 \alpha_L^{\prime u} S_L(z;\lambda,c_0)\end{pmatrix} f_L (x) ~, ~~
\sigma^\mu \dd_\mu f_L(x) = m f_R (x) ~. 
\label{EoMup2}
\end{align}
The equations of motion (\ref{EoMup1}) in the bulk $0<y<L$ $(1<z<z_L)$ yield
\begin{align}
\begin{pmatrix} -k\lambda&m \cr  m&-k\lambda  \end{pmatrix}
\begin{pmatrix} \alpha_R^{u} \cr  \alpha_L^{u}  \end{pmatrix} = 0 ~,  \mbox{etc.}
\end{align}
so that one finds
\begin{align}
m=k\lambda ~.
\end{align}
The BCs at $z=1$ in the twisted gauge follow from those in the original gauge:
\begin{align}
u_{+0R}=0\ \Rightarrow\ &
\tilde{u}_{+0R} \cos\frac{\theta_H}{2}-i\tilde{u}_{+0R}'\sin\frac{\theta_H}{2}=0,\\
D_-u_{+0R}'=0\ \Rightarrow\ &
-iD_-\tilde{u}_{+0R}\sin\frac{\theta_H}{2}
+D_-\tilde{u}_{+0R}'\cos\frac{\theta_H}{2}=0.
\end{align}
We  write the above equations in the matrix form:
\begin{align}
M_u
\begin{pmatrix} \alpha_R^u \cr \alpha_R^{\prime u}  \end{pmatrix}
:=
\begin{pmatrix}
\cos\frac{\theta_H}{2}S_R^0 &-i\sin\frac{\theta_H}{2}C_R^0 \cr
\noalign{\kern 5pt}
-i\sin\frac{\theta_H}{2}\lambda C_L^0& \cos\frac{\theta_H}{2}\lambda S_L^0  \end{pmatrix}
\begin{pmatrix} \alpha_R^u \cr \alpha_R^{\prime u}  \end{pmatrix}   =0 ~, 
\end{align}
where we have used short-hand notation such as $S_L^{0}$ standing for
$S_L(1;\lambda,c_0)$.   In the following we will use the same notation. 
From $\mbox{det} \, M_u=0$, we find the mass formula of the up-type quarks
\begin{align}
S_L^0 S_R^0+\sin^2\frac{\theta_H}{2}=0 ~.
\label{Up-type-quark-mass}
\end{align}
The same formula holds for $\Psi_{\bf 32}$ with $\gamma_{6D}^7=-1$.
The formula (\ref{Up-type-quark-mass})   is the same as  in the 5D $SO(11)$ 
GHGUT  in Ref.~\cite{Furui:2016owe}.
For $\lambda z_L\ll 1$, the up-type quark mass spectrum is given by
\begin{align}
m_u\simeq
\left\{
\begin{array}{ll}
k z_L^{-1}\sqrt{1-4c_0^2} \, \sin \onehalf \theta_H & \mbox{for}\ c_0<\onehalf \\
\noalign{\kern 10pt}
k z_L^{-1/2-c_0}\sqrt{4c_0^2-1} \, \sin \onehalf \theta_H & \mbox{for}\ c_0>\onehalf \\
\end{array}
\right..
\label{Eq:Up-type-quark-mass-approximate}
\end{align}

\subsection{Down-type quark}

\subsubsection*{(ii) $Q_{\rm EM}=- \frac{1}{3}$: 
$d_j,d_j',D_{j+},  D_{j-}$ $(\Psi_{\bf 32}, \Psi_{\bf 11})$}

Consider $\Psi_{\bf 32}$ with $\gamma_{6D}^7 = +1$. 
Parity even modes at $y=0$ with $(P_0, P_2) = (+,+)$ are 
$d_{+L}, d_{+R}', D_{+R}$ and $D_{-L}$.
From the action (\ref{Eq:Action-bulk-fermion-check}) and 
the ${\cal L}_1^m$ and ${\cal L}_3^m$ terms in (\ref{Eq:Action-brane-fermion-mass}),
one finds the equations of motion for down-type quarks: 
\begin{align}
-i\delta
\begin{pmatrix} d_{+L}^{\dag}\\ d_{+L}^{\prime\dag}  \end{pmatrix}
:\ &
\big(-k\hat{D}_-(c_{0})+i\partial_v\big)
\begin{pmatrix}\check{d}_{+R}\\ \check{d}_{+R}^{\prime} \end{pmatrix}
+\sigma^\mu\partial_\mu
\begin{pmatrix} \check{d}_{+L}\\ \check{d}_{+L}^{\prime} \end{pmatrix}
=0 ~, \cr
%
i\delta
\begin{pmatrix} d_{+R}^{\dag}\\ d_{+R}^{\prime\dag} \end{pmatrix}
:\ &
\overline{\sigma}^\mu\partial_\mu
\begin{pmatrix} \check{d}_{+R}\\ \check{d}_{+R}^{\prime} \end{pmatrix}
+\big(-k\hat{D}_+(c_{0})+i\partial_v\big)
\begin{pmatrix} \check{d}_{+L}\\ \check{d}_{+L}^{\prime} \end{pmatrix}
= 2 \mu_1 \,  \delta(y)
\begin{pmatrix} 0\\ \check{D}_{-L}  \end{pmatrix},\cr
-i\delta D_{+L}^{\dag}:\ &
\big(-k\hat{D}_-(c_{1})+i\partial_v\big)\check{D}_{+R}
+\sigma^\mu\partial_\mu\check{D}_{+L}
=0 ~, \cr
i\delta D_{+R}^{\dag}:\ &
\overline{\sigma}^\mu\partial_\mu\check{D}_{+R}
+\big(-k\hat{D}_+(c_{1})+i\partial_v\big)\check{D}_{+L}
=2\mu_{\bf 11} \, \delta(y) \, \check{D}_{-L} ~,\cr
-i\delta D_{-L}^{\dag}:\ &
\big(k\hat{D}_+(c_{1})-i\partial_v\big)\check{D}_{-R}
+\sigma^\mu\partial_\mu\check{D}_{-L}
=\delta(y)
\left\{2\mu_{\bf 11}\check{D}_{+R}+2\mu_{1}\check{d}_{+R}'\right\},\cr
i\delta D_{-R}^{\dag}:\ &
\overline{\sigma}^\mu\partial_\mu\check{D}_{-R}
+\big(k\hat{D}_-(c_{1})-i\partial_v\big)\check{D}_{-L}
= 0 ~. 
\end{align}
For 6th dimensional $n=0$ KK modes, the equations of motion reduce to
\begin{align}
\begin{matrix} (a)\\ (b)  \end{matrix}
:\ &
-k\hat{D}_-(c_{0})
\begin{pmatrix} \check{d}_{+R}\\ \check{d}_{+R}^{\prime}  \end{pmatrix}
+\sigma^\mu\partial_\mu
\begin{pmatrix} \check{d}_{+L}\\ \check{d}_{+L}^{\prime} \end{pmatrix}
=0 ~,\cr
\begin{matrix} (c)\\ (d)  \end{matrix}
:\ &
\overline{\sigma}^\mu\partial_\mu
\begin{pmatrix} \check{d}_{+R}\\ \check{d}_{+R}^{\prime}  \end{pmatrix}
-k\hat{D}_+(c_{0})
\begin{pmatrix} \check{d}_{+L}\\ \check{d}_{+L}^{\prime}  \end{pmatrix}
=2\mu_1\delta(y)
\begin{pmatrix} 0\\ \check{D}_{-L}  \end{pmatrix}, \cr
(e):\ &
-k\hat{D}_-(c_{1})\check{D}_{+R}
+\sigma^\mu\partial_\mu\check{D}_{+L}
=0,\cr
(f):\ &
\overline{\sigma}^\mu\partial_\mu\check{D}_{+R}
-k\hat{D}_+(c_{1})\check{D}_{+L}
=2\mu_{\bf 11}\delta(y)\check{D}_{-L},\cr
(g):\ &
k\hat{D}_+(c_{1})\check{D}_{-R}
+\sigma^\mu\partial_\mu\check{D}_{-L}
=2\mu_{\bf 11}\delta(y)\check{D}_{+R}
+2\mu_1\delta(y)\check{d}_{+R}^{\prime},\cr
(h):\ &
\overline{\sigma}^\mu\partial_\mu\check{D}_{-R}
+k\hat{D}_-(c_{1})\check{D}_{-L}
=0.
\end{align}

To obtain BCs at $y=0$, we integrate the above $(a),(d),(f),(g)$ in
the vicinity of $y=0$ for parity-odd fields:
\begin{align}
(a)\ \Rightarrow\ & 
2\check{d}_{+R}(x,\epsilon)=0,\cr
(d)\ \Rightarrow\ & 
-2\check{d}_{+L}^{\prime}(x,\epsilon)=
2\mu_1\check{D}_{-L}(x,0),\cr
(f)\ \Rightarrow\ & 
-2\check{D}_{+L}(x,\epsilon)=
2\mu_{\bf 11}\check{D}_{-L}(x,0),\cr
(g)\ \Rightarrow\ & 
2\check{D}_{-R}(x,\epsilon)=
2\mu_{\bf 11}\check{D}_{+R}(x,0)
+2\mu_1\check{d}_{+R}^{\prime}(x,0).
\end{align}
For parity-even fields, we evaluate the equations of motion at
$y=+\epsilon$ by using the above conditions:
\begin{align}
(c)\ \Rightarrow\ &
\hat{D}_+\check{d}_{+L}(x,\epsilon)=0,\cr
(b)\ \Rightarrow\ &
-k\hat{D}_-\check{d}_{+R}^{\prime}
+\mu_1 k D_{+}\check{D}_{-R}
=0,\cr
(e)\ \Rightarrow\ &
-k{D}_-\check{D}_{+R}(x,\epsilon)
+\mu_{\bf 11}kD_{+}\check{D}_{-R}=0,\cr
(h)\ \Rightarrow\ &
k{D}_-\check{D}_{-L}
+\mu_{\bf 11}kD_{+}\check{D}_{+L}
+\mu_1 k\hat{D}_{+}\check{d}_{+L}^{\prime}
=0,
\end{align}
where we used the equations of motion $(d)$ and $(g)$ at $y=+\epsilon$.

We recall that in the twisted gauge all fields obey free-field equations
in the bulk. Their eigenmodes are determined by the BCs
on the IR brane. The mass spectra can be fixed by the BCs at the UV
brane. In the twisted gauge the mode functions are given by 
\begin{align}
&\begin{pmatrix}
\tilde{\check{d}}_{+R}\cr \tilde{\check{d}}_{+R}^{\prime}\cr
\tilde{\check{D}}_{+R}\cr \tilde{\check{D}}_{-R}  \end{pmatrix}
= \begin{pmatrix}
\alpha_R^{d}S_R(z;\lambda,c_0) \cr
\noalign{\kern 3pt}
\alpha_R^{d'}C_R(z;\lambda,c_0) \cr
\noalign{\kern 3pt}
\alpha_R^{D_+}C_R(z;\lambda,c_1)\cr
\noalign{\kern 3pt}
\alpha_R^{D_-}S_L(z;\lambda,c_1)  \end{pmatrix} f_R(x) ~,\cr
\noalign{\kern 5pt}
&\begin{pmatrix}
\tilde{\check{d}}_{+L}\cr
\tilde{\check{d}}_{+L}^{\prime}\cr
\tilde{\check{D}}_{+L}\cr
\tilde{\check{D}}_{-L}  \end{pmatrix}
=\begin{pmatrix}
\alpha_L^{d}C_L(z;\lambda,c_0)\cr
\noalign{\kern 3pt}
\alpha_L^{d'}S_L(z;\lambda,c_0)\cr
\noalign{\kern 3pt}
\alpha_L^{D_+}S_L(z;\lambda,c_1)\cr
\noalign{\kern 3pt}
\alpha_L^{D_-}C_R(z;\lambda,c_1) \end{pmatrix} f_L(x) ~. 
\end{align}
where $f_R (x), f_L (x)$ satisfy the relations in (\ref{EoMup2}) and $m= k\lambda$.
The BCs at $z=1^+$ in the twisted gauge are converted to 
\begin{align}
&K\begin{pmatrix}
\alpha_R^{d}\cr
\alpha_R^{d'}\cr
\alpha_R^{D_+}\cr
\alpha_R^{D_-}  \end{pmatrix} = 0 ~, \cr
\noalign{\kern 5pt}
&K =
\begin{pmatrix}
\cos\frac{\theta_H}{2}S_R^0&-i\sin\frac{\theta_H}{2}C_R^0&0&0\cr
\noalign{\kern 3pt}
-i\sin\frac{\theta_H}{2}\lambda C_L^0&\cos\frac{\theta_H}{2}\lambda S_L^0&
0&-\mu_1\lambda C_R^1\cr
\noalign{\kern 3pt}
0&0&\lambda S_L^1&-\mu_{\bf 11}\lambda C_R^1\cr
\noalign{\kern 3pt}
i\mu_1\sin\frac{\theta_H}{2}S_R^0&-\mu_1\cos\frac{\theta_H}{2}C_R^0&
-\mu_{\bf 11}C_R^1&S_L^1 \end{pmatrix} .
\end{align}
From $\mbox{det}K=0$, we find the mass spectrum formula for the
down-type quarks:
\begin{align}
S_L^0 S_R^0+\sin^2\frac{\theta_H}{2}=
-\frac{\mu_1^2 S_R^0 C_R^0 S_L^1 C_R^1}
{\mu_{\bf 11}^2(C_R^1)^2-(S_L^1)^2}.
\label{Down-type-quark-mass}
\end{align}
The same formula is obtained for $\Psi_{\bf 32}$ with $\gamma_{6D}^7 = -1$.
For $\lambda z_L\ll 1$, the down-type quark mass spectrum is given by
\begin{align}
m_d \simeq
\begin{cases}
k z_L^{c_0-c_1-1}
\mfrac{\mu_{\bf 11}}{\mu_1}
\sqrt{(1-2c_0)(1+2c_1)} \, \sin \mfrac{\theta_H}{2}&
\mbox{for}\ c_0<\onehalf  ~, \cr
\noalign{\kern 5pt}
k z_L^{-c_1-1/2}
\mfrac{\mu_{\bf 11}}{\mu_1}
\sqrt{(2c_0-1)(2c_1+1)} \, \sin \mfrac{\theta_H}{2}&
\mbox{for}\ c_0> \onehalf ~.  \end{cases}
\label{Eq:Down-type-quark-mass-approximate}
\end{align}
Combining (\ref{Eq:Up-type-quark-mass-approximate}) and 
(\ref{Eq:Down-type-quark-mass-approximate}), one finds
\begin{align}
\frac{m_d}{m_u} = z_L^{c_0 - c_1} \, \frac{\mu_{\bf 11}}{\mu_1}
\sqrt{\frac{1 + 2 c_1}{1+ 2 c_0}} ~.
\label{DownUpmassratio}
\end{align}

\subsection{Charged lepton}
\label{Sec:Charged_lepton}

\subsubsection*{(iii) $Q_{\rm EM}=-1$: $e,e',E_+^{\prime},E_-^{\prime}$ 
$(\Psi_{\bf 32}, \Psi_{\bf 11}')$}

Parity even modes at $y=0$ with $(P_0, P_2) = (+,+)$ are 
$e_{+L}, e_{+R}', E_{+L}'$ and $E_{-R}'$.
From the action (\ref{Eq:Action-bulk-fermion-check}) and 
the ${\cal L}_5^m$ and ${\cal L}_6^m$ terms in (\ref{Eq:Action-brane-fermion-mass}),
one finds the equations of motion for charged leptons: 
\begin{align}
-i\delta
\begin{pmatrix} e_{+L}^{\dag}\\ e_{+L}^{\prime\dag}  \end{pmatrix}
:\ &
\big(-k\hat{D}_-(c_{0})+i\partial_v\big)
\begin{pmatrix}\check{e}_{+R}\\ \check{e}_{+R}^{\prime} \end{pmatrix}
+\sigma^\mu\partial_\mu
\begin{pmatrix} \check{e}_{+L}\\ \check{e}_{+L}^{\prime} \end{pmatrix}
= 2 i \tilde\mu_2 \,  \delta(y)
\begin{pmatrix} \check{E}_{-R}^{\prime} \\ 0  \end{pmatrix} ,  \cr
i\delta
\begin{pmatrix} e_{+R}^{\dag}\\ e_{+R}^{\prime\dag} \end{pmatrix}
:\ &
\overline{\sigma}^\mu\partial_\mu
\begin{pmatrix} \check{e}_{+R}\\ \check{e}_{+R}^{\prime} \end{pmatrix}
+\big(-k\hat{D}_+(c_{0})+i\partial_v\big)
\begin{pmatrix} \check{e}_{+L}\\ \check{e}_{+L}^{\prime} \end{pmatrix}
= 0 ~, \cr
-i\delta E_{+L}^{\prime \dag}:\ &
\big(-k\hat{D}_-(c_{2})+i\partial_v\big)\check{E}_{+R}'
+\sigma^\mu\partial_\mu\check{E}_{+L}'
= 2\mu_{\bf 11}^{\prime}\delta(y)\check{E}_{-R}^{\prime} ~, \cr
i\delta E_{+R}^{\prime \dag}:\ &
\overline{\sigma}^\mu\partial_\mu\check{E}_{+R}'
+\big(-k\hat{D}_+(c_{2})+i\partial_v\big)\check{E}_{+L}' = 0 ~,  \cr
-i\delta E_{-L}^{\prime \dag}:\ &
\big(k\hat{D}_+(c_{2})-i\partial_v\big)\check{E}_{-R}'
+\sigma^\mu\partial_\mu\check{E}_{-L}' = 0 ~, \cr
i\delta E_{-R}^{\prime \dag}:\ &
\overline{\sigma}^\mu\partial_\mu\check{E}_{-R}'
+\big(k\hat{D}_-(c_{2})-i\partial_v\big)\check{E}_{-L}'
=\delta(y)\left\{
-2i\widetilde{\mu}_2 \check{e}_{+L}
+2\mu_{\bf 11}^{\prime}\check{E}_{+L}^{\prime} \right\}.
\end{align}
For 6th dimensional $n=0$ KK modes, the equations of motion for charged
leptons reduce to 

\begin{align}
\begin{matrix} (a)\\ (b)  \end{matrix}
:\ &
-k\hat{D}_-(c_{0})
\begin{pmatrix}\check{e}_{+R}\\ \check{e}_{+R}^{\prime} \end{pmatrix}
+\sigma^\mu\partial_\mu
\begin{pmatrix} \check{e}_{+L}\\ \check{e}_{+L}^{\prime} \end{pmatrix}
= 2 i \tilde\mu_2 \,  \delta(y)
\begin{pmatrix} \check{E}_{-R}^{\prime} \\ 0  \end{pmatrix} ,  \cr
\begin{matrix} (c)\\ (d)  \end{matrix}
:\ &
\overline{\sigma}^\mu\partial_\mu
\begin{pmatrix} \check{e}_{+R}\\ \check{e}_{+R}^{\prime} \end{pmatrix}
-k\hat{D}_+(c_{0}) 
\begin{pmatrix} \check{e}_{+L}\\ \check{e}_{+L}^{\prime} \end{pmatrix}
= 0 ~, \cr
(e):\ &
-k\hat{D}_-(c_{2}) \check{E}_{+R}'
+\sigma^\mu\partial_\mu\check{E}_{+L}'
= 2\mu_{\bf 11}^{\prime}\delta(y)\check{E}_{-R}^{\prime} ~, \cr
(f):\ &
\overline{\sigma}^\mu\partial_\mu\check{E}_{+R}'
-k\hat{D}_+(c_{2}) \check{E}_{+L}' = 0 ~,  \cr
(g):\ &
k\hat{D}_+(c_{2}) \check{E}_{-R}'
+\sigma^\mu\partial_\mu\check{E}_{-L}' = 0 ~, \cr
(h):\ &
\overline{\sigma}^\mu\partial_\mu\check{E}_{-R}'
+ k\hat{D}_-(c_{2}) \check{E}_{-L}'
=\delta(y)\left\{
-2i\widetilde{\mu}_2 \check{e}_{+L}
+2\mu_{\bf 11}^{\prime}\check{E}_{+L}^{\prime} \right\}.
\end{align}

To obtain boundary conditions at $y=0$, we integrate the above
$(a),(d),(e),(h)$ in the vicinity of $y=0$ for parity-odd fields:
\begin{align}
(a)\ \Rightarrow\ & 
2\check{e}_{+R}(x,\epsilon)=2i\widetilde{\mu}_2\check{E}_{-R}^{\prime}(x,0),\cr
(d)\ \Rightarrow\ & 
-2\check{e}_{+L}^{\prime}(x,\epsilon)=
0,\cr
(e)\ \Rightarrow\ & 
2\check{E}_{+R}^{\prime}(x,\epsilon)=
2\mu_{\bf 11}^{\prime}\check{E}_{-R}^{\prime}(x,0),\cr
(h)\ \Rightarrow\ & 
-2\check{E}_{-L}^{\prime}(x,\epsilon)=
-i2\widetilde{\mu}_{2}\check{e}_{+L}(x,0)
+2\mu_{\bf 11}^{\prime}\check{E}_{+L}^{\prime}(x,0).
\end{align}
For parity-even fields, we calculate the equations of motion at
$y=+\epsilon$ by using the above conditions:
\begin{align}
(b)\ \Rightarrow\ &
\hat{D}_-\check{e}_{+R}^{\prime}(x,\epsilon)=
0,\cr
(c)\ \Rightarrow\ &
\hat{D}_+\check{e}_{+L}
+i\widetilde{\mu}_2D_{-}\check{E}_{-L}^{\prime}
=0,\cr
(f)\ \Rightarrow\ &
D_+\check{E}_{+L}^{\prime}(x,\epsilon)
+\mu_{\bf 11}^{\prime}D_{-}\check{E}_{-L}^{\prime}=0,\cr
(g)\ \Rightarrow\ &
{D}_+\check{E}_{-R}^{\prime}
+i\widetilde{\mu}_{2}\hat{D}_{-}\check{e}_{+R}
-\mu_{\bf 11}^{\prime}{D}_{-}\check{E}_{+R}^{\prime}
=0,
\end{align}
where we used the equations of motion $(d)$ and $(h)$ at $y=+\epsilon$.

By using the BCs on the IR brane, the mode functions of charged leptons
in the twisted gauge are given by  
\begin{align}
&\begin{pmatrix}
\tilde{\check{e}}_{+R}\cr \tilde{\check{e}}_{+R}^{\prime}\cr
\tilde{\check{E}}_{+R}'\cr \tilde{\check{E}}_{-R}'  \end{pmatrix}
= \begin{pmatrix}
\alpha_R^{e}S_R(z;\lambda,c_0) \cr
\noalign{\kern 3pt}
\alpha_R^{e'}C_R(z;\lambda,c_0) \cr
\noalign{\kern 3pt}
\alpha_R^{E_+'}S_R(z;\lambda,c_2)\cr
\noalign{\kern 3pt}
\alpha_R^{E_-'}C_L(z;\lambda,c_2)  \end{pmatrix} f_R(x) ~,\cr
\noalign{\kern 5pt}
&\begin{pmatrix}
\tilde{\check{e}}_{+L}\cr
\tilde{\check{e}}_{+L}^{\prime}\cr
\tilde{\check{E}}_{+L}'\cr
\tilde{\check{E}}_{-L}'  \end{pmatrix}
=\begin{pmatrix}
\alpha_L^{e}C_L(z;\lambda,c_0)\cr
\noalign{\kern 3pt}
\alpha_L^{e'}S_L(z;\lambda,c_0)\cr
\noalign{\kern 3pt}
\alpha_L^{E_+'}C_L(z;\lambda,c_2)\cr
\noalign{\kern 3pt}
\alpha_L^{E_-'}S_R(z;\lambda,c_2) \end{pmatrix} f_L(x) ~. 
\end{align}
As in the case of  down-type quarks, the BCs at $z=1^+$ in the twisted gauge
are converted to  
\begin{align}
&K \begin{pmatrix}
\alpha_R^{e} \cr
\noalign{\kern 3pt}
\alpha_R^{e'} \cr
\noalign{\kern 3pt}
\alpha_R^{E_+'}\cr
\noalign{\kern 3pt}
\alpha_R^{E_-'}  \end{pmatrix} = 0 ~, \cr
\noalign{\kern 5pt}
&K = \begin{pmatrix}
\cos\frac{\theta_H}{2}S_R^0 &-i\sin\frac{\theta_H}{2}C_R^0 &0 
&-i\widetilde{\mu}_2C_L^2\cr
\noalign{\kern 3pt}
-i\sin\frac{\theta_H}{2}\lambda C_L^0 &\cos\frac{\theta_H}{2}\lambda S_L^0
&0 &0 \cr
\noalign{\kern 3pt}
0 &0 &S_R^2 &-\mu_{\bf 11}^{\prime}C_L^2 \cr 
\noalign{\kern 3pt}
i\widetilde{\mu}_2\cos\frac{\theta_H}{2}\lambda C_L^0 &
\widetilde{\mu}_2\sin\frac{\theta_H}{2}\lambda S_L^0 &
-\mu_{\bf 11}^{\prime}\lambda C_L^2 &\lambda S_R^2 
\end{pmatrix} .
\end{align}
From $\mbox{det}K=0$, we find the mass spectrum formula for the
charged leptons:
\begin{align}
S_L^0 S_R^0+\sin^2\frac{\theta_H}{2}=
-\frac{\widetilde{\mu}_2^2 S_L^0 C_L^0 S_R^2 C_L^2}
{\mu_{\bf 11}^{\prime 2}(C_L^2)^2-(S_R^2)^2}.
\label{Charged-lepton-mass}
\end{align}
The  mass spectrum of charged leptons, for which $\lambda z_L\ll 1$,  is given by
\begin{align}
m_e\simeq
\begin{cases}
k z_L^{-1+c_2-c_0} \,
\mfrac{\mu_{\bf 11}^{\prime}}{\widetilde{\mu}_2} \,
\sqrt{(2c_0+1)(1-2c_2)} \,
\sin \mfrac{\theta_H}{2}
&\mbox{for}\  c_2<\onehalf , \cr
\noalign{\kern 5pt}
k z_L^{-\frac{1}{2}-c_0} \, 
\mfrac{\mu_{\bf 11}^{\prime}}{\widetilde{\mu}_2} \,
\sqrt{(2c_0+1)(2c_2-1)} \,
\sin \mfrac{\theta_H}{2}
&\mbox{for}\  c_2>\onehalf, 
\end{cases}
\label{Eq:Charged-lepton-mass-approximate}
\end{align}
Here we have made use of $(m_t/m_\tau)^2, (m_c/m_\mu)^2,  (m_u/m_e)^2 \gg 1$
which assures $|S_L^0 S_R^0 | \ll \sin^2 \onehalf \theta_H$,
and have assumed that {$\lambda^2 \ll {\mu_{\bf 11}'}^2$} for $c_2 > \onehalf$
and {$\lambda^2 \ll ({\mu_{\bf 11}'}  z_L^{2 c_2-1})^2$} for $c_2 < \onehalf$ so that
$\mu_{\bf 11}^{\prime 2}(C_L^2)^2 \gg (S_R^2)^2$.
Combining (\ref{Eq:Up-type-quark-mass-approximate}) and 
(\ref{Eq:Charged-lepton-mass-approximate}), one finds 
\begin{align}
&\frac{m_e}{m_u} =\begin{cases}
z_L^{c_2 - c_0}\,  \mfrac{\mu_{\bf 11}'}{\widetilde{\mu}_2}\, \sqrt{\frac{1 - 2 c_2}{1-2 c_0}}
&\mbox{for}\ c_0 <\onehalf,~  c_2 < \onehalf , \cr
\noalign{\kern 7pt}
z_L^{\frac{1}{2} - c_0}\,  \mfrac{\mu_{\bf 11}'}{\widetilde{\mu}_2}\, 
\sqrt{\frac{2 c_2 - 1}{1-2 c_0}}
&\mbox{for}\ c_0 <\onehalf,~ c_2 > \onehalf , \cr
\noalign{\kern 7pt}
z_L^{c_2 - \frac{1}{2}}\,  \mfrac{\mu_{\bf 11}'}{\widetilde{\mu}_2}\, 
\sqrt{\frac{1 - 2 c_2}{2 c_0 - 1}}
&\mbox{for}\ c_0 > \onehalf,~  c_2 < \onehalf , \cr
\noalign{\kern 7pt}
  \mfrac{\mu_{\bf 11}'}{\widetilde{\mu}_2}\, 
\sqrt{\frac{2 c_2 - 1}{2 c_0 - 1}}
&\mbox{for}\ c_0 >\onehalf,~ c_2 > \onehalf .
\end{cases}
\label{ElectronUpmassratio}
\end{align}
{It will be seen below that in a typical example for the third generation
$c_2= - 0.7$ and ${\mu_{\bf 11}'} C_L^2 / S_R^2 \sim 2.6$.}

\ignore{
$\mu_{\bf 11}' / \, \tilde{\mu}_2 \sim 20$.  {\checkpoint} 
In this case the observed charged lepton 
masses are obtained by adopting $c_2 < \onehalf < c_0$ for the first and second generations
and $c_2 < c_0 < \onehalf$ for the third generation.
}

One comment is in order about the masses of exotic charged leptons 
$\hat e, \hat e', \hat E_\pm'$ with charge $Q_{\rm EM} = +1$.
No zero modes exist for $\hat e$ and $\hat e'$, and all modes have masses larger than
$\onehalf m_{\KK_6}$.  On the other hand $\hat E_\pm'$ have zero modes.  They
acquire masses through the Hosotani mechanism and brane interactions.
We suppose that $c_2 < \onehalf$ and/or $\mu_{\bf 11}'$ is sufficiently large
so that their lightest masses are $O(m_{\KK_5})$.

\subsection{Neutrino}
\label{Sec:Neutrino-mass}

\subsubsection*{(iv) $Q_{\rm EM}=0$: 
$\nu,\nu',N_\pm', \hat N_\pm', S_\pm' , \eta_-$
$(\Psi_{\bf 32}, \Psi_{\bf 11}^{\prime}, \chi_{\bf 1})$}

6D $SO(11)$ spinor and vector bulk fermion $\Psi_{\bf 32}$ and 
$\Psi_{\bf 11}'$ and a 5D $SO(11)$ singlet
symplectic Majorana brane fermion $\chi$ appear in this sector.
From the action in Eqs.~(\ref{Eq:Action-bulk-fermion-check})
and ${\cal L}_5^m,  {\cal L}_6^m, {\cal L}_7^m$ in (\ref{Eq:Action-brane-fermion-mass}),
the equations of motion for the bulk fermions become
\begin{align}
-i\delta
\begin{pmatrix} \nu_{+L}^{\dag}\\ \nu_{+L}^{\prime\dag} \end{pmatrix} :\ &
\big( -k\hat{D}_-(c_0)+i\partial_v \big)
\begin{pmatrix} \check{\nu}_{+R}\\ \check{\nu}_{+R}^{\prime} \end{pmatrix}
+\sigma^\mu\partial_\mu
\begin{pmatrix} \check{\nu}_{+L}\\ \check{\nu}_{+L}^{\prime} \end{pmatrix}
=\delta(y) 
\begin{pmatrix} 2i\widetilde{\mu}_2\check{N}_{-R}^{\prime}\cr
(m_B/\sqrt{k}) \, \xi_- \end{pmatrix} , \cr
\noalign{\kern 5pt}
i\delta
\begin{pmatrix} \nu_{+R}^{\dag}\\ \nu_{+R}^{\prime\dag} \end{pmatrix} :\ &
\overline{\sigma}^\mu\partial_\mu
\begin{pmatrix} \check{\nu}_{+R}\\ \check{\nu}_{+R}^{\prime} \end{pmatrix}
+ \big( -k\hat{D}_+(c_{0})+i\partial_v\big)
\begin{pmatrix} \check{\nu}_{+L}\\ \check{\nu}_{+L}^{\prime} \end{pmatrix} \cr
&\hskip 5.cm
=\delta(y)
\begin{pmatrix}  0\\  -\sqrt{2}\widetilde{\mu}_2\check{S}_{-L}^{\prime}
+ (m_B/\sqrt{k}) \,  \eta_-  \end{pmatrix},  \cr
\noalign{\kern 5pt}
-i\delta
\begin{pmatrix} \hat{N}_{+L}^{\prime\dag}\\  N_{+L}^{\prime\dag}\\ 
S_{+L}^{\prime\dag} \end{pmatrix}   :\ &
\big(-k\hat{D}_-(c_{2})+i\partial_v\big)
\begin{pmatrix}  \check{\hat{N}}_{+R}^{\prime}\\ \check{N}_{+R}^{\prime}\\
\check{S}_{+R}^{\prime} \end{pmatrix}
+\sigma^\mu\partial_\mu
\begin{pmatrix}  \check{\hat{N}}_{+L}^{\prime}\\ \check{N}_{+L}^{\prime}\\
\check{S}_{+L}^{\prime} \end{pmatrix}
=\delta(y)
\begin{pmatrix} 2{\mu}_{\bf 11}^{\prime}\check{\hat{N}}_{-R}^{\prime}\\
2{\mu}_{\bf 11}^{\prime}\check{N}_{-R}^{\prime} \cr 0 \end{pmatrix},\cr
\noalign{\kern 5pt}
i\delta
\begin{pmatrix} \hat{N}_{+R}^{\prime\dag}\\  N_{+R}^{\prime\dag}\\ 
S_{+R}^{\prime\dag} \end{pmatrix}   :\ &
\overline{\sigma}^\mu\partial_\mu
\begin{pmatrix}  \check{\hat{N}}_{+R}^{\prime}\\ \check{N}_{+R}^{\prime}\\
\check{S}_{+R}^{\prime} \end{pmatrix}
+\big(-k\hat{D}_+(c_{2})+i\partial_v\big)
\begin{pmatrix}  \check{\hat{N}}_{+L}^{\prime}\\ \check{N}_{+L}^{\prime}\\
\check{S}_{+L}^{\prime} \end{pmatrix}
=\delta(y)
\begin{pmatrix} 0 \cr 0 \cr 2{\mu}_{\bf 11}^{\prime}\check{S}_{-L}^{\prime}\end{pmatrix},\cr
\noalign{\kern 5pt}
-i\delta
\begin{pmatrix} \hat{N}_{-L}^{\prime\dag}\\  N_{-L}^{\prime\dag}\\ 
S_{-L}^{\prime\dag} \end{pmatrix}   :\ &
\big( k\hat{D}_+ (c_{2}) - i\partial_v\big)
\begin{pmatrix}  \check{\hat{N}}_{-R}^{\prime}\\ \check{N}_{-R}^{\prime}\\
\check{S}_{-R}^{\prime} \end{pmatrix}
+\sigma^\mu\partial_\mu
\begin{pmatrix}  \check{\hat{N}}_{-L}^{\prime}\\ \check{N}_{-L}^{\prime}\\
\check{S}_{-L}^{\prime} \end{pmatrix} \cr
&\hskip 5.cm
=\delta(y)
\begin{pmatrix} 0\\ 0\\
-\sqrt{2}\widetilde{\mu}_{2} \check{\nu}_{+R}^{\prime}
+2\mu_{\bf 11}^{\prime}\check{S}_{+R}^{\prime} \end{pmatrix},\cr
\noalign{\kern 5pt}
i\delta
\begin{pmatrix} \hat{N}_{-R}^{\prime\dag}\\  N_{-R}^{\prime\dag}\\ 
S_{-R}^{\prime\dag} \end{pmatrix}   :\ &
\overline{\sigma}^\mu\partial_\mu
\begin{pmatrix}  \check{\hat{N}}_{-R}^{\prime}\\ \check{N}_{-R}^{\prime}\\
\check{S}_{-R}^{\prime} \end{pmatrix}
+\big( k\hat{D}_- (c_{2}) - i\partial_v\big)
\begin{pmatrix}  \check{\hat{N}}_{-L}^{\prime}\\ \check{N}_{-L}^{\prime}\\
\check{S}_{-L}^{\prime} \end{pmatrix} \cr
&\hskip 5.cm
=\delta(y)
\begin{pmatrix}
2\mu_{\bf 11}^{\prime}\check{\hat{N}}_{+L}^{\prime}\\
-2i\widetilde{\mu}_{2} \check{\nu}_{+L}
+2\mu_{\bf 11}^{\prime}\check{N}_{+L}^{\prime}\\
0 \end{pmatrix}.
\label{EoMneutral1}
\end{align}
The equations for the  5D brane symplectic Majorana fermion $\chi_{\bf 1}$ are
\begin{align}
-i \delta \eta_-^\dag  :\ &
\sigma^\mu\partial_\mu\eta_--i\partial_v\xi_-
-  \frac{m_B}{\sqrt{k}} \,  \nu_{+R}^{\prime}-M\eta_-^C =0 ~, \cr
i \delta \xi_-^\dag :\ &
\overline{\sigma}^\mu\partial_\mu\xi_--i\partial_v\eta_-
- \frac{m_B}{\sqrt{k}} \, \nu_{+L}^{\prime}-M\xi_-^C =0 ~.
\label{EoMneutral2}
\end{align}
Note that  $\eta_-$ has a zero mode but $\xi_-$ has no zero mode in the 6th
dimensional direction. 

For 6th dimensional $n=0$ KK modes, the equations of motion become
\begin{align}
\begin{matrix} (a) \cr (b) \end{matrix} :\ &
-k\hat{D}_-(c_0) 
\begin{pmatrix} \check{\nu}_{+R}\\ \check{\nu}_{+R}^{\prime} \end{pmatrix}
+\sigma^\mu\partial_\mu
\begin{pmatrix} \check{\nu}_{+L}\\ \check{\nu}_{+L}^{\prime} \end{pmatrix}
=\delta(y) 
\begin{pmatrix} 2i\widetilde{\mu}_2\check{N}_{-R}^{\prime}\\ 0 \end{pmatrix} , \cr
\noalign{\kern 5pt}
\begin{matrix} (c) \cr (d) \end{matrix} :\ &
\overline{\sigma}^\mu\partial_\mu
\begin{pmatrix} \check{\nu}_{+R}\\ \check{\nu}_{+R}^{\prime} \end{pmatrix}
 -k\hat{D}_+(c_{0}) 
\begin{pmatrix} \check{\nu}_{+L}\\ \check{\nu}_{+L}^{\prime} \end{pmatrix} 
=\delta(y)
\begin{pmatrix}  0\\  -\sqrt{2}\widetilde{\mu}_2\check{S}_{-L}^{\prime}
+ (m_B/\sqrt{k}) \, \eta_- \end{pmatrix},  \cr
\noalign{\kern 5pt}
\begin{matrix} (e) \cr (f) \cr (g) \end{matrix} :\ &
-k\hat{D}_-(c_{2}) 
\begin{pmatrix}  \check{\hat{N}}_{+R}^{\prime}\\ \check{N}_{+R}^{\prime}\\
\check{S}_{+R}^{\prime} \end{pmatrix}
+\sigma^\mu\partial_\mu
\begin{pmatrix}  \check{\hat{N}}_{+L}^{\prime}\\ \check{N}_{+L}^{\prime}\\
\check{S}_{+L}^{\prime} \end{pmatrix}
=\delta(y)
\begin{pmatrix} 2{\mu}_{\bf 11}^{\prime}\check{\hat{N}}_{-R}^{\prime}\\
2{\mu}_{\bf 11}^{\prime}\check{N}_{-R}^{\prime} \cr 0 \end{pmatrix},\cr
\noalign{\kern 5pt}
\begin{matrix} (h) \cr (i) \cr (j) \end{matrix} :\ &
\overline{\sigma}^\mu\partial_\mu
\begin{pmatrix}  \check{\hat{N}}_{+R}^{\prime}\\ \check{N}_{+R}^{\prime}\\
\check{S}_{+R}^{\prime} \end{pmatrix}
-k\hat{D}_+(c_{2}) 
\begin{pmatrix}  \check{\hat{N}}_{+L}^{\prime}\\ \check{N}_{+L}^{\prime}\\
\check{S}_{+L}^{\prime} \end{pmatrix}
=\delta(y)
\begin{pmatrix} 0 \cr 0 \cr 2{\mu}_{\bf 11}^{\prime}\check{S}_{-L}^{\prime}\end{pmatrix},\cr
\noalign{\kern 5pt}
\begin{matrix} (k) \cr (\ell ) \cr (m) \end{matrix} :\ &
k\hat{D}_+ (c_{2}) 
\begin{pmatrix}  \check{\hat{N}}_{-R}^{\prime}\\ \check{N}_{-R}^{\prime}\\
\check{S}_{-R}^{\prime} \end{pmatrix}
+\sigma^\mu\partial_\mu
\begin{pmatrix}  \check{\hat{N}}_{-L}^{\prime}\\ \check{N}_{-L}^{\prime}\\
\check{S}_{-L}^{\prime} \end{pmatrix} 
=\delta(y)
\begin{pmatrix} 0\\ 0\\
-\sqrt{2}\widetilde{\mu}_{2} \check{\nu}_{+R}^{\prime}
+2\mu_{\bf 11}^{\prime}\check{S}_{+R}^{\prime} \end{pmatrix},\cr
\noalign{\kern 5pt}
\begin{matrix} (n) \cr (o) \cr (p) \end{matrix} :\ &
\overline{\sigma}^\mu\partial_\mu
\begin{pmatrix}  \check{\hat{N}}_{-R}^{\prime}\\ \check{N}_{-R}^{\prime}\\
\check{S}_{-R}^{\prime} \end{pmatrix}
+  k\hat{D}_- (c_{2}) 
\begin{pmatrix}  \check{\hat{N}}_{-L}^{\prime}\\ \check{N}_{-L}^{\prime}\\
\check{S}_{-L}^{\prime} \end{pmatrix} 
=\delta(y)
\begin{pmatrix}
2\mu_{\bf 11}^{\prime}\check{\hat{N}}_{+L}^{\prime}\\
-2i\widetilde{\mu}_{2} \check{\nu}_{+L}
+2\mu_{\bf 11}^{\prime}\check{N}_{+L}^{\prime}\\
0 \end{pmatrix} , \cr
\noalign{\kern 5pt}
(q)
:\ &
\Big\{ {\sigma}^\mu\partial_\mu\eta_- 
- \frac{m_B}{\sqrt{k}} \, \nu_{+R}'-M\eta_-^C \Big\}
\delta(y)=0 ~.
\label{EoMneutralF3}
\end{align}
Here $\hat D_\pm (c)$ is given by (\ref{Dhat1}).  
To find the form of $\hat D_\pm (c_2)$ for  $(\hat N', N', S')$ in $\Psi_{\bf 11}'$, let us recall
that the original and twisted gauges are related by 
$\Psi_{\bf 11}' = \Omega (z) \widetilde \Psi_{\bf 11}' $ where 
$\Omega (z) = e^{i\theta(z)T_{4,11}}$.  Hence
\begin{align}
\psi_3' ~ &= \widetilde \psi_3' ~, ~~
\begin{pmatrix} \psi_4' \cr \psi_{11}' \end{pmatrix} =
\begin{pmatrix} \cos\theta(z) &\sin\theta(z) \cr -\sin\theta(z) &\cos\theta(z) \end{pmatrix}
\begin{pmatrix} \widetilde \psi_4' \cr \widetilde \psi_{11}' \end{pmatrix} .
\end{align}
As
\begin{align}
&\psi_{4}^{\prime}=\frac{1}{\sqrt{2}}
(\hat{N}^{\prime}-N^{\prime} ),\ \ \
 \psi_{3}^{\prime}=\frac{i}{\sqrt{2}}
(\hat{N}^{\prime}+N^{\prime} ),\ \ \
 \psi_{11}^{\prime}=S^{\prime} , 
\end{align}
one finds
\begin{align}
&\begin{pmatrix} \widetilde{\check{\hat{N}}}{}' \cr \widetilde{\check N}' \cr \widetilde{\check S}' \end{pmatrix}
= \bar{\Omega} (z)
\begin{pmatrix}\check{ \hat{N}}{}' \cr  \check N' \cr  \check S' \end{pmatrix} , \cr
\noalign{\kern 5pt}
&\bar{\Omega}^{-1} (z)  =
\begin{pmatrix}
\mfrac{1+\cos\theta(z)}{2}&\mfrac{1-\cos\theta(z)}{2}&
\mfrac{\sin\theta(z)}{\sqrt{2}} \cr
\mfrac{1-\cos\theta(z)}{2}&\mfrac{1+\cos\theta(z)}{2}&
-\mfrac{\sin\theta(z)}{\sqrt{2}} \cr
-\mfrac{\sin\theta(z)}{\sqrt{2}}&\mfrac{\sin\theta(z)}{\sqrt{2}}& \cos\theta(z)
\end{pmatrix} .
\label{twist11a}
\end{align}
It follows that
\begin{align}
\hat D_- (c_2) 
\begin{pmatrix} \check{\hat{N}}_{+R}' \cr \check N_{+R}' \cr  \check S_{+R}' \end{pmatrix}
= \bar{\Omega}^{-1} (z) \, D_- (c_2)
\begin{pmatrix} \widetilde{\check{\hat{N}}}{}_{+R}' \cr \widetilde{\check N}{}_{+R}' \cr 
\widetilde{\check S}{}_{+R}' \end{pmatrix}
\label{twist11b}
\end{align}
and so on.

To obtain boundary conditions at $y=0$, we integrate the above
$(a)$, $(d)$, $(e)$, $(f)$, $(j)$, $(m)$, $(n)$, $(o)$ 
($\frac{1}{2}\int_{-\epsilon}^{+\epsilon}dy\cdots$) in the vicinity of
$y=0$ for parity-odd (in $x^5=y$) fields 
$\nu_{+R},\nu_{+L}',$
$\hat{N}_{+R}^{\prime},N_{+R}^{\prime},$
$S_{+L}^{\prime},$
$S_{-R}^{\prime},$
$\hat{N}_{-L}^{\prime},N_{-L}^{\prime}$:
\begin{align}
(a)\ \Rightarrow\ &
+\check{\nu}_{+R}(x,\epsilon)
=i\widetilde{\mu}_2\check{N}_{-R}^{\prime}(x,0),\cr
\noalign{\kern 5pt}
(d)\ \Rightarrow\ & 
-\check{\nu}_{+L}^{\prime}(x,\epsilon)
=-\frac{\widetilde{\mu}_2}{\sqrt{2}}\check{S}_{-L}^{\prime}(x,0)
+  \frac{m_B}{2 \sqrt{k}} \, \eta_-(x),\cr
\noalign{\kern 5pt}
(e)\ \Rightarrow\ & 
+\check{\hat{N}}_{+R}^{\prime}(x,\epsilon)
=\mu_{\bf 11}^{\prime}\check{\hat{N}}_{-R}^{\prime}(x,0),\cr
\noalign{\kern 5pt}
(f)\ \Rightarrow\ & 
+\check{N}_{+R}^{\prime}(x,\epsilon)
=\mu_{\bf 11}^{\prime}\check{N}_{-R}^{\prime}(x,0),\cr
\noalign{\kern 5pt}
(j)\ \Rightarrow\ & 
-\check{S}_{+L}^{\prime}(x,\epsilon)
=\mu_{\bf 11}^{\prime}\check{S}_{-L}^{\prime}(x,0),\cr
\noalign{\kern 5pt}
(m)\ \Rightarrow\ & 
+\check{S}_{-R}^{\prime}(x,\epsilon)
=-\frac{\widetilde{\mu}_{2}}{\sqrt{2}}\check{\nu}_{+R}^{\prime}(x,0)
+\mu_{\bf 11}^{\prime}\check{S}_{+R}^{\prime}(x,0),\cr
\noalign{\kern 5pt}
(n)\ \Rightarrow\ & 
-\check{\hat{N}}_{-L}^{\prime}(x,\epsilon)
=\mu_{\bf 11}^{\prime}\check{\hat{N}}_{+L}^{\prime}(x,0),\cr
\noalign{\kern 5pt}
(o)\ \Rightarrow\ & 
-\check{N}_{-L}^{\prime}(x,\epsilon)
=-i\widetilde{\mu}_{2} \check{\nu}_{+L}(x,0)
+\mu_{\bf 11}^{\prime}\check{N}_{+L}^{\prime}(x,0).
\label{BCneutralF1}
\end{align}
For parity-even (in $x^5=y$) fields 
$\nu_{+L},\nu_{+R}',$
$\hat{N}_{+L}^{\prime},N_{+L}^{\prime},$
$S_{+R}^{\prime},$
$S_{-L}^{\prime},$
$\hat{N}_{-R}^{\prime},N_{-R}^{\prime}$,
we evaluate the equations of motion at 
$y=+\epsilon$ by making use of the above conditions to find
\begin{align}
(b)\ \Rightarrow\ &
-\hat{D}_-(c_{0})\check{\nu}_{+R}^{\prime}
-\frac{\widetilde{\mu}_{2}}{\sqrt{2}}\hat{D}_{+}\check{S}_{-R}^{\prime}
- \frac{m_B^2}{2k^2}  \,\check{\nu}_{+R}^{\prime}
- \frac{m_B M}{2k^{3/2}}  \, \eta_-^C
=0,\cr
\noalign{\kern 5pt}
(c)\ \Rightarrow\ &
\hat{D}_+(c_{0})\check{\nu}_{+L}
+i\widetilde{\mu}_{2}\hat{D}_{-}\check{N}_{-L}^{\prime}
=0,\cr
\noalign{\kern 5pt}
(g)\ \Rightarrow\ &
-\hat{D}_-(c_{0})\check{S}_{+R}^{\prime}
+\mu_{\bf 11}^{\prime}\hat{D}_{+}\check{S}_{-R}^{\prime}
=0,\cr
\noalign{\kern 5pt}
(h)\ \Rightarrow\ &
\hat{D}_+(c_{0})\check{\hat{N}}_{+L}^{\prime}
+\mu_{\bf 11}^{\prime}\hat{D}_{-}\check{\hat{N}}_{-R}^{\prime}
=0,\cr
\noalign{\kern 5pt}
(i)\ \Rightarrow\ &
\hat{D}_+(c_{0})\check{N}_{+L}^{\prime}
+\mu_{\bf 11}^{\prime}\hat{D}_{-}\check{N}_{-R}^{\prime}
=0,\cr
\noalign{\kern 5pt}
(k)\ \Rightarrow\ &
\hat{D}_+(c_{2})\check{\hat{N}}_{-R}^{\prime}
-\mu_{\bf 11}^{\prime}\hat{D}_{-}\check{\hat{N}}_{+R}^{\prime}
=0,\cr
\noalign{\kern 5pt}
(\ell)\ \Rightarrow\ &
\hat{D}_+(c_{2})\check{N}_{-R}^{\prime}
+i\widetilde{\mu}_{2} \hat{D}_{-}\check{\nu}_{+R}
-\mu_{\bf 11}^{\prime}\hat{D}_{-}\check{N}_{+R}^{\prime}
=0,\cr
\noalign{\kern 5pt}
(p)\ \Rightarrow\ &
\hat{D}_-(c_{2})\check{S}_{-L}^{\prime}
-\frac{\widetilde{\mu}_{2}}{\sqrt{2}}\hat{D}_{+}\check{\nu}_{+L}'
+\mu_{\bf 11}^{\prime}\hat{D}_{+}\check{S}_{+L}^{\prime}
=0,
\label{BCneutralF2}
\end{align}
at $y=+\epsilon$. 

There are nine fields in the neutral fermion sector which intertwine with
each others.
To simplify the discussions, we consider the case in which 
$m_B^2/k^2,  m_B M/k^2,\mu_{\bf 11}^{\prime}\gg\widetilde{\mu}_2$. 
In  the {$\widetilde{\mu}_2=0$} limit,  the equations of motion and boundary 
conditions in the neutral sector,  
(\ref{EoMneutralF3}), (\ref{BCneutralF1}), and  (\ref{BCneutralF2}),
split into two parts.
The first sector, the neutrino sector-1,   contains $\check \nu_{+R}'$, 
$\check \nu_{+L}'$ in $\Psi_{\bf 32}$ and $\eta_-$ in $\chi_{\bf 1}$, 
while the second sector,  the neutrino sector-2,  contains other components in 
$\Psi_{\bf 32}$ and $\Psi_{\bf 11}^{\prime}$.
{$\widetilde{\mu}_2 \not= 0$ mixes the neutrino sector-1 and sector-2.}

In the twisted gauge, the mode functions are determined by the BCs on
the IR brane. The mass spectra can be fixed by the BCs on the UV brane.
For the neutrino sector-1, by using the boundary conditions at $z=z_L$,
their mode functions can be written as
\begin{align}
&\begin{pmatrix} \widetilde{\check{\nu}}_R \cr  \widetilde{\check{\nu}}_R^{\prime} \cr
\eta_-^C \end{pmatrix}
= \begin{pmatrix} \alpha_\nu S_R(z;\lambda,c_0 ) \cr 
i \alpha_{\nu'} C_R(z;\lambda,c_0 ) \cr - i \alpha_\eta^* / \sqrt{k} \end{pmatrix} 
f_R(x) ~, ~~
\overline{\sigma}^\mu\partial_\mu f_R(x)=k\lambda f_L(x)~, \cr
\noalign{\kern 5pt}
&\begin{pmatrix} \widetilde{\check{\nu}}_L \cr  \widetilde{\check{\nu}}_L^{\prime} \cr
\eta_- \end{pmatrix}
=\begin{pmatrix}  \alpha_\nu C_L(z;\lambda,c_0) \cr 
i \alpha_{\nu'} S_L(z;\lambda,c_0) \cr i \alpha_\eta / \sqrt{k} \end{pmatrix}
f_L(x) ~, ~~
{\sigma}^\mu\partial_\mu f_L(x)=k\lambda f_R(x) ~.
\end{align}
Here $f_{R/L}(x)$ are chosen in the neutrino sector-1 such that
$f_L^C(x)=-e^{i\delta_C}\sigma^2f_L(x)^*=f_R(x)$ is satisfied where $\delta_C$ is defined
in (\ref{sympMajorana1}).
One can take $\alpha_\nu , \alpha_{\nu'}, \alpha_\eta$ to be real.
In this case  $\sigma^\mu \dd_\mu \eta_-  = -k\lambda \eta_-^C$ is satisfied
so that the equation $(q)$  in (\ref{EoMneutralF3}) implies that
\begin{align}
\frac{m_B}{\sqrt{k}} \, \check \nu_{+R}' \big|_{y=0} + (k\lambda + M) \eta_-^C =0 ~.
\label{BCneutralF3}
\end{align}
With this identity the first relation in (\ref{BCneutralF2}) can be rewritten as
\begin{align}
&
\hat{D}_-(c_{0})\check{\nu}_{+R}^{\prime}
+\frac{\widetilde{\mu}_{2}}{\sqrt{2}}\hat{D}_{+}\check{S}_{-R}^{\prime}
- \frac{m_B \lambda }{ 2\sqrt{k} } \, \eta_-^C =0.
\label{BCneutralF4}
\end{align}

Setting $\widetilde{\mu}_{2} =0$, one find that 
the BCs at $z=1^+$ in the twisted gauge can be written as
\begin{align}
{K_{\nu_1}}
\begin{pmatrix} \alpha_\nu \cr \noalign{\kern 10pt}  
\alpha_{\nu'} \cr \noalign{\kern 10pt}  \alpha_\eta \end{pmatrix}
=
\begin{pmatrix}
\cos \mfrac{\theta_H}{2}S_R^0 &\sin \mfrac{\theta_H}{2}C_R^0 &0 \cr
\noalign{\kern 5pt}
-\sin \mfrac{\theta_H}{2}C_L^0 &\cos \mfrac{\theta_H}{2}S_L^0 
&\mfrac{m_B}{2k}  \cr
\noalign{\kern 5pt}
-\sin \mfrac{\theta_H}{2}S_R^0 &\cos\mfrac{\theta_H}{2}C_R^0
& - \mfrac{k\lambda + M}{m_B} \end{pmatrix} 
\begin{pmatrix} \alpha_\nu \cr \noalign{\kern 10pt}  
\alpha_{\nu'} \cr \noalign{\kern 10pt}  \alpha_\eta \end{pmatrix} = 0 ~.
\label{SpectrumNeutral1}
\end{align}
From $\mbox{det} K_{\nu_1}=0$, we find the mass spectrum formula for the  neutrino
sector-1:
\begin{align}
{\det K_{\nu_1} = - \frac{k\lambda+M}{m_B} 
\left\{S_L^0 S_R^0+\sin^2\frac{\theta_H}{2}\right\}
- \frac{m_B}{2k } \, S_R^0 C_R^0=0 } ~.
\label{Eq:Mass-Spectrum-Neutrino-Sector-1}
\end{align}
This gives the gauge-Higgs 
seesaw mechanism for the small neutrino masses.
For $\lambda z_L\ll 1$ and $k\lambda\ll |M|$, the neutrino mass  
is given by
\begin{align}
m_\nu \simeq -\frac{2m_u^2 Mz_L^{2c_0+1}}{(2c_0+1)m_B^{2}}.
\label{Neutrino-mass}
\end{align}
where $m_u$ is given by (\ref{Eq:Up-type-quark-mass-approximate}).
{
As shown in Ref.\ \cite{Hosotani:2017ghg},
the gauge-Higgs seesaw mechanism is characterized by
a $3 \times 3$ mass matrix
\begin{align}
\begin{pmatrix} & m_D &\cr m_D && \tilde m_B \cr & \tilde m_B & M \end{pmatrix}
\end{align}
in the 4d effective theory in which $m_D = m_u$ and 
$\tilde m_B \sim m_B z_L^{- c_0 -1/2}$. 
The Majorana mass $M$ may take a moderate value.}

Next, we consider the neutrino sector-2.
The BCs at the IR brane determine the mode functions in the twisted gauge to be
\begin{align}
&\begin{pmatrix}
\tilde{\check{\hat{N}}}_{+R}^{\prime}\\
\tilde{\check{N}}_{+R}^{\prime}\\
\tilde{\check{S}}_{+R}^{\prime}\\
\tilde{\check{\hat{N}}}_{-R}^{\prime}\\
\tilde{\check{N}}_{-R}^{\prime}\\
\tilde{\check{S}}_{-R}^{\prime}\\
\end{pmatrix}
=\begin{pmatrix}
\alpha_{\hat{N}_+^{\prime}}S_R(z;\lambda,c_2)\\
\alpha_{N_+^{\prime}}S_R(z;\lambda,c_2)\\
\alpha_{S_+^{\prime}}C_R(z;\lambda,c_2)\\
\alpha_{\hat{N}_-^{\prime}}C_L(z;\lambda,c_2)\\
\alpha_{N_-^{\prime}}C_L(z;\lambda,c_2)\\
\alpha_{S_-^{\prime}}S_L(z;\lambda,c_2)\\
\end{pmatrix} f_R(x) ~, \cr
\noalign{\kern 5pt}
&\begin{pmatrix}
\tilde{\check{\hat{N}}}_{+L}^{\prime}\\
\tilde{\check{N}}_{+L}^{\prime}\\
\tilde{\check{S}}_{+L}^{\prime}\\
\tilde{\check{\hat{N}}}_{-L}^{\prime}\\
\tilde{\check{N}}_{-L}^{\prime}\\
\tilde{\check{S}}_{-L}^{\prime}\\
\end{pmatrix}
= \begin{pmatrix}
\alpha_{\hat{N}_+^{\prime}}C_L(z;\lambda,c_2)\\
\alpha_{N_+^{\prime}}C_L(z;\lambda,c_2)\\
\alpha_{S_+^{\prime}}S_L(z;\lambda,c_2)\\
-\alpha_{\hat{N}_-^{\prime}}S_R(z;\lambda,c_2)\\
-\alpha_{N_-^{\prime}}S_R(z;\lambda,c_2)\\
-\alpha_{S_-^{\prime}}C_R(z;\lambda,c_2)\\
\end{pmatrix} f_L(x) ~.
\end{align}
By making use of (\ref{twist11a}) and (\ref{twist11b}), 
the BCs at the UV brane are expressed as
\begin{align}
K_{\nu_2}
\left(
\alpha_{\hat{N}_+^{\prime}}\ 
\alpha_{N_+^{\prime}}\ 
\alpha_{S_+^{\prime}}\
\alpha_{\hat{N}_-^{\prime}}\
\alpha_{N_-^{\prime}}\
\alpha_{S_-^{\prime}}\
\right)^T
=0 ~,
\end{align}
where 
\begin{align}
K_{\nu_2}&= \begin{pmatrix} A & -\mu_{\bf 11}' B \cr 
\noalign{\kern 5pt}
-\mu_{\bf 11}' B & A \end{pmatrix} , \cr
\noalign{\kern 10pt}
A &= \begin{pmatrix}
\mfrac{1+\cos\theta_H}{2}S_R^2
&\mfrac{1-\cos\theta_H}{2}S_R^2
&\mfrac{\sin\theta_H}{\sqrt{2}} C_R^2 \cr
\noalign{\kern 5pt}
\mfrac{1-\cos\theta_H}{2}S_R^2
&\mfrac{1+\cos\theta_H}{2}S_R^2
&-\mfrac{\sin\theta_H}{\sqrt{2}} C_R^2 \cr
\noalign{\kern 5pt}
-\mfrac{\sin\theta_H}{\sqrt{2}}C_L^2
&\mfrac{\sin\theta_H}{\sqrt{2}}C_L^2
&\cos\theta_H S_L^2  \end{pmatrix} , \cr
\noalign{\kern 10pt}
B &= \begin{pmatrix}
\mfrac{1+\cos\theta_H}{2}C_L^2
&\mfrac{1-\cos\theta_H}{2}C_L^2
&\mfrac{\sin\theta_H}{\sqrt{2}} S_L^2 \cr
\noalign{\kern 5pt}
\mfrac{1-\cos\theta_H}{2}C_L^2
&\mfrac{1+\cos\theta_H}{2}C_L^2
&-\mfrac{\sin\theta_H}{\sqrt{2}} S_L^2 \cr
\noalign{\kern 5pt}
-\mfrac{\sin\theta_H}{\sqrt{2}}S_R^2
&\mfrac{\sin\theta_H}{\sqrt{2}}S_R^2
&\cos\theta_H C_R^2  \end{pmatrix} . 
\label{Eq:K-neutrino-2-matrix}
\end{align}

Note that
\begin{align}
&K_{\nu_2}' = \begin{pmatrix} V & 0 \cr \noalign{\kern 5pt} 0 &V \end{pmatrix}  K_{\nu_2}
\begin{pmatrix} V^{-1} & 0 \cr \noalign{\kern 5pt} 0 &V^{-1} \end{pmatrix}
= \begin{pmatrix} A' & -\mu_{\bf 11}' B' \cr 
\noalign{\kern 5pt}
-\mu_{\bf 11}' B' & A' \end{pmatrix} , \cr
\noalign{\kern 5pt}
&V = V^{-1} = \begin{pmatrix} 2^{-1/2} & 2^{-1/2} & 0 \cr
2^{-1/2} & - 2^{-1/2} & 0 \cr 0 &0 &1 \end{pmatrix} , \cr
\noalign{\kern 5pt}
&A' = \begin{pmatrix} S_R^2 & 0 &0 \cr 
\noalign{\kern 3pt}
0 & \cos \theta_H S_R^2 & \sin \theta_H C_R^2 \cr
\noalign{\kern 3pt}
0 & - \sin\theta_H C_L^2 & \cos\theta_H S_L^2 \end{pmatrix} , \cr
\noalign{\kern 5pt}
&B' = \begin{pmatrix} C_L^2 & 0 &0 \cr 
\noalign{\kern 3pt}
0 & \cos \theta_H C_L^2 & \sin \theta_H S_L^2 \cr
\noalign{\kern 3pt}
0 & - \sin\theta_H S_R^2 & \cos\theta_H C_R^2 \end{pmatrix} .
\end{align}
From $\mbox{det}K_{\nu_2}= \mbox{det}K_{\nu_2}'= 0$   the mass spectra for the neutrino
sector-2 are found to be
\begin{align}
{ \det K_{\nu_2} =}
&\left(S_R^2 S_R^2-\mu_{\bf 11}^{\prime 2}C_L^2C_L^2\right)
\Big\{
\left(S_L^2S_R^2 +\sin^2\theta_H\right)^2 \cr
\noalign{\kern 5pt}
&
+\mu_{\bf 11}^{\prime 2}
\left(2\sin^2\theta_H\cos^2\theta_H 
-C_R^2C_R^2S_R^2S_R^2
-C_L^2C_L^2S_L^2S_L^2\right) \cr
\noalign{\kern 5pt}
&
+\mu_{\bf 11}^{\prime 4}
\left(S_L^2 S_R^2+\cos^2\theta_H\right)^2
\Big\} = 0 ~.
\label{SpectrumNeutralF2a}
\end{align}
In the $\mu_{\bf 11}^{\prime}=0$ limit, or in the absence of the brane interactions, 
(\ref{SpectrumNeutralF2a}) becomes 
\begin{align}
\left(S_R^2\right)^2\left(S_L^2 S_R^2+\sin^2\theta_H\right)^2=0 ~.
\end{align}
For  large $\mu_{\bf 11}^{\prime}$, it becomes
\begin{align}
\left(C_L^2\right)^2\left(S_L^2 S_R^2+\cos^2\theta_H\right)^2
\simeq 0 ~.
\label{Eq:Mass-Spectrum-Neutrino-Sector-2}
\end{align}

{
For ${\widetilde{\mu}}_2 \not= 0$ the neutrino sector-1 and sector-2 mix through 
the boundary  conditions.  One needs to solve
\begin{align}
&K \, 
\big( \alpha_\nu \  \alpha_{\nu'} \  \alpha_\eta \ 
\alpha_{\hat{N}_+^{\prime}}\  \alpha_{N_+^{\prime}}\  \alpha_{S_+^{\prime}}\
\alpha_{\hat{N}_-^{\prime}}\ \alpha_{N_-^{\prime}}\ \alpha_{S_-^{\prime}}   \big)^T
=0 ~, \cr
\noalign{\kern 10pt}
&K= \begin{pmatrix} K_{\nu_1} & 0 & -i {\widetilde \mu}_2\,  D \cr
\noalign{\kern 5pt}
0 &A & -\mu_{\bf 11}' B \cr 
\noalign{\kern 5pt}
i {\widetilde \mu}_2 \, C & -\mu_{\bf 11}' B & A \end{pmatrix} , \cr
\noalign{\kern 10pt}
&C =  \begin{pmatrix} 0 & 0 & 0 \cr
\noalign{\kern 5pt}
 \cos \onehalf \theta_H C_L^0 &  \sin \onehalf \theta_H S_L^0 & 0 \cr
\noalign{\kern 5pt}
- \mfrac{\sin \onehalf \theta_H}{\sqrt{2}} S_R^0  
&  \mfrac{\cos \onehalf \theta_H}{\sqrt{2}} C_R^0 & 0 \end{pmatrix} , \cr
\noalign{\kern 10pt}
&D = \begin{pmatrix}
\mfrac{1-\cos\theta_H}{2}C_L^2
&\mfrac{1+\cos\theta_H}{2}C_L^2
&-\mfrac{\sin\theta_H}{\sqrt{2}} S_L^2 \cr
\noalign{\kern 5pt}
-\mfrac{\sin\theta_H}{2}S_R^2
&\mfrac{\sin\theta_H}{2}S_R^2
&\mfrac{\cos\theta_H}{\sqrt{2}}  C_R^2  \cr
\noalign{\kern 5pt}
0 & 0 & 0 \end{pmatrix} . 
\label{K-neutrino-matrix}
\end{align}
$\det K$ is evaluated to be
\begin{align}
&\det K = F_0 + F_2 \,  ({\widetilde \mu}_2)^2 + F_4\,  ({\widetilde \mu}_2)^4 ~, \cr
\noalign{\kern 5pt}
&F_0 = \det K_{\nu_1} \cdot \det K_{\nu_2} ~, \cr
\noalign{\kern 5pt}
&F_2 = \bigg\{ \frac{m_B}{4k} (S_L^0 S_R^0 + \cos^2 \onehalf \theta_H )
+ \frac{k\lambda +M}{2 m_B} C_L^0 S_L^0 \bigg\} \cr
\noalign{\kern 5pt}
&\qquad
\times \bigg\{ C_L^2 S_R^2 (2 X + \sin^2 \theta_H ) (X + \sin^2 \theta_H ) \cr
\noalign{\kern 5pt}
&\qquad 
- \mu_{11}^{\prime \, 2} \Big[ (C_L^2)^3 S_L^2 ( 2X + \sin^2 \theta_H ) 
+ C_R^2 (S_R^2)^3 (2X + 1 + \cos^2 \theta_H)\cr
\noalign{\kern 5pt}
&\qquad 
+ C_L^2 S_R^2 ( X^2 - 2) \sin^2 \theta_H \cos^2 \theta_H \Big]
+ \mu_{11}^{\prime \, 4} C_L^2 S_R^2 (X + \cos^2 \theta_H) (2X + 1 + \cos^2\theta_H) \bigg\}
\cr
\noalign{\kern 5pt}
&\quad
+\frac{k\lambda + M}{2 m_B} \big\{ (S_R^2)^2 - \mu_{11}^{\prime \, 2} (C_L^2)^2 \big\}
\bigg\{ \Big[ (1 + \mu_{11}^{\prime \, 2} ) X + \sin^2 \theta_H 
+  \mu_{11}^{\prime \, 2} \cos^2 \theta_H \Big] \cos \theta_H \sin^2 \theta_H 
\cr
\noalign{\kern 5pt}
&\qquad
+ C_R^0 S_R^0 \Big[ C_R^2 S_R^2 (X + \sin^2\theta_H ) - 
\mu_{11}^{\prime \, 2} C_L^2 S_L^2 (X + \cos^2 \theta_H ) \Big] \bigg\}, 
\cr
\noalign{\kern 5pt}
&F_4 = - \frac{k \lambda + M}{32 m_B} (S_L^0 S_R^0 + \cos^2 \onehalf \theta_H )  \cr
\noalign{\kern 5pt}
&\quad
\times \Big[ (S_R^2 )^2 \big\{ 16   (S_L^2 S_R^2)^2 + (20 - 4 \cos 2\theta_H)  S_L^2 S_R^2
+ 5 - 4 \cos 2\theta_H - \cos 4\theta_H \big\} \cr
\noalign{\kern 5pt}
&\qquad
- \mu_{11}^{\prime  \, 2} (C_L^2)^2 \big\{ 16   (S_L^2 S_R^2)^2 
+ (12 + 4 \cos 2\theta_H)  S_L^2 S_R^2 - \cos 4\theta_H \big\} \Big],
\end{align}
where $X = S_L^2 S_R^2$.  In evaluating the effective potential $V_\eff (\theta_H)$, 
we shall use the approximate formula $\det K \sim \det K_{\nu_1} \det K_{\nu_2}$
for small $\widetilde{\mu}_2^{2}$.  
}

\subsection{Dark fermion}

\subsubsection*{(v) $\Psi_{\bf 32}^{\prime} = \Psi_{\bf 32}^{\alpha = 4}$}

The 6D $SO(11)$ ${\bf 32}$ Weyl fermion $\Psi_{\bf 32}^{\prime}$, 
which may be called a dark fermion multiplet,  has  6th dimensional $n=0$ KK modes. 
For $\Psi_{\bf 32}'$ one needs not introduce brane interactions.
We  consider the action (\ref{Eq:Action-bulk-fermion-check}) for the dark fermion sector
$\Psi_{\bf 32}^{\prime}$. From the parity assignment shown
in Table~\ref{Tab:BC-fermion-spinor},  the mass formula for the
dark fermions is found to be
\begin{align}
S_L^{0'}S_R^{0'}+\cos^2\frac{\theta_H}{2}=0 ~.
\label{SpectrumDarkF}
\end{align}
For $\lambda z_L\ll 1$, the dark fermion mass  is given by
\begin{align}
m_{\rm Dark}\simeq
\begin{cases}
k z_L^{-1}\sqrt{1-4c_{0}^{\prime 2}} \, \cos\mfrac{\theta_H}{2}&
\mbox{for}\ c_{0}'<1/2 ~, \cr
\noalign{\kern 5pt}
k z_L^{-1/2-c_{0}'}\sqrt{4c_{0}^{\prime 2}-1} \, \cos\mfrac{\theta_H}{2}&
\mbox{for}\ c_{0}' >1/2 ~. 
\end{cases}
\end{align}
As in the case of the 5D $SO(11)$ GHGUT discussed in
Ref.~\cite{Furui:2016owe}, the bulk mass parameter of the dark fermions,
$c_{0}'$, must be  relatively small  not to become light exotic particles.

\section{Effective potential}
\label{Sec:Effective-potential}

We evaluate the Higgs effective potential $V_{\rm eff}(\theta_H)$ by
using the mass spectrum formulas of the 6th-dimensional $n=0$ KK modes of
the $SO(11)$ gauge bosons and fermions. 
The evaluation is done in the same manner as in Ref.~\cite{Furui:2016owe}.

One-loop effective potential from each KK tower is given by
\cite{Falkowski:2006vi,Hosotani:2008tx,Hatanaka:2011ru}
\begin{align}
V_{\rm eff}(\theta_H)&=
\pm\frac{1}{2}\int\frac{d^4p}{(2\pi)^4}
\sum_n \mbox{ln}\left(p^2+m_n(\theta_H)^2\right)
\label{Eq:Effective-potential}
\end{align}
where $m_n(\theta_H)$ is the mass spectrum of the KK tower and 
we take $+$ and $-$ sign for bosons and fermions, respectively.
When the mass spectrum $\{ m_n = k \lambda_n \}$ is determined by 
$1 + \tilde Q (\lambda_n) f(\theta_H) = 0$, (\ref{Eq:Effective-potential}) 
can be written as 
\begin{align}
V_{\rm eff}(\theta_H) &= \pm I\left[Q(q);f(\theta_H)\right] \cr
\noalign{\kern 5pt}
I\left[Q(q);f(\theta_H)\right] &=  \frac{(k z_L^{-1} )^4}{(4\pi)^2}
\int_0^\infty dq \, q^3 \ln [ 1 + Q(q) f (\theta_H) ] ~,
\label{Eq:Effective-potential2}
\end{align}
where $Q(q) = \tilde Q (iq z_L^{-1} )$.
We utilize the mass spectra $m_n(\theta_H)$ of
the $SO(11)$ bulk gauge and fermions obtained in
Sec.~\ref{Sec:Spectrum-Bosons} and \ref{Sec:Spectrum-Fermions}. 
Note that only $\theta_H$-dependent mass spectra contribute to the 
$\theta_H$-dependent part of $V_\eff (\theta_H)$.
It contains the $SO(11)$ bulk gauge field, $SO(11)$ spinor and vector
fermion fields. 

We summarize those mass spectra.
The $SO(11)$ gauge boson contribution is almost the same as in the 5D $SO(11)$ GHGUT. 
The difference from the 5D case is that the $Y$ boson
contributions can be neglected in the current scheme as they have
masses of $O(m_{\KK_6})$.
The relevant mass spectra of $SO(11)$ gauge fields $A_\mu$ and $A_z$
are given,
from (\ref{spectrumW}),  (\ref{spectrumZ2}) and (\ref{spectrumAz}),   by  
\begin{align}
&\mbox{$W^{\pm}$ tower}:\
1+\frac{\lambda}{2C'(1;\lambda)S(1;\lambda)}\sin^2\theta_H=0 ~, \cr
\noalign{\kern 5pt}
&\mbox{$Z$ tower}:\
1+\frac{\lambda}{2 \cos^2 \theta_W \, C'(1;\lambda)S(1;\lambda)}
\sin^2\theta_H=0  ~, \cr
\noalign{\kern 5pt}
&A_z^{a4},A_z^{a\, 11} (a=1,2,3) :\
1+\frac{\lambda}{C'(1;\lambda)S(1;\lambda)}\sin^2\theta_H=0 ~.
\label{Eq:Mass-Spectrum-gauge}
\end{align}
The mass spectra of up- and down-type quarks, charged leptons, neutrinos,
and dark fermions are given,
from (\ref{Up-type-quark-mass}), (\ref{Down-type-quark-mass}), 
(\ref{Charged-lepton-mass}), (\ref{Eq:Mass-Spectrum-Neutrino-Sector-1}),
(\ref{SpectrumNeutralF2a}) and (\ref{SpectrumDarkF}), 
 by 
\begin{align}
&\mbox{(i)}:\  Q_{\rm EM}=+\frac{2}{3}:\ \ 
1+\frac{\sin^2\onehalf \theta_H}{S_L^{0}S_R^{0}}=0 ~, \cr
\noalign{\kern 5pt}
&\mbox{(ii)}:\  Q_{\rm EM}=-\frac{1}{3}:\ \ 
1+\frac{\sin^2\onehalf \theta_H}{S_L^0 S_R^0
+\mfrac{\mu_1^2 S_R^0 C_R^0 S_L^1 C_R^1}
{\mu_{\bf 11}^2(C_R^1)^2-(S_L^1)^2}} = 0 ~, \cr
\noalign{\kern 5pt}
&\mbox{(iii)}:\  Q_{\rm EM}=-1:\ \ 
1+ \frac{\sin^2\onehalf \theta_H}{S_L^0 S_R^0
+\mfrac{\widetilde{\mu}_2^2 S_L^0 C_L^0 S_R^2 C_L^2}
{\mu_{\bf 11}^{\prime 2}(C_L^2)^2-(S_R^2)^2}}  =0 ~, \cr
\noalign{\kern 5pt}
&\mbox{(iv-1)}:\  Q_{\rm EM}=0:\ \ 
1+\frac{\sin^2\onehalf \theta_H}{S_L^0 S_R^0
+\mfrac{m_B^2 /k }{2(k\lambda+M)}S_R^0 C_R^0} =0 ~, \cr
\noalign{\kern 5pt}
&\mbox{(iv-2)}:\  Q_{\rm EM}=0:\ \ 
1 + \frac{f_1 (\theta_H)}{f_2 } = 0 ~, \cr
\noalign{\kern 5pt}
&\hskip .5cm
f_1 (\theta_H) = (1 - \mu_{\bf 11}^{\prime \,  2} )
\big\{ 2 (1 + \mu_{\bf 11}^{\prime  \, 2}) S_L^2S_R^2
+2\mu_{\bf 11}^{\prime \, 2} 
+( 1 - \mu_{\bf 11}^{\prime \, 2}) \sin^2\theta_H \big\} \sin^2 \theta_H  ~, \cr
\noalign{\kern 5pt}
&\hskip .5cm
f_2 = \mu_{\bf 11}^{\prime 4}
+2\mu_{\bf 11}^{\prime 4}S_L^2S_R^2
+(1+\mu_{\bf 11}^{\prime 4})(S_L^2S_R^2)^2
-\mu_{\bf 11}^{\prime 2}\left\{(C_R^2S_R^2)^2+(C_L^2S_L^2)^2\right\} ~, \cr
\noalign{\kern 5pt}
& 
\mbox{(v)}:\  Q_{\rm EM}=+\frac{2}{3},-\frac{2}{3},-1,0:\ \ 
1+\frac{\cos^2 \onehalf \theta_H}{S_L^{0'}S_R^{0'}} =0 ~,
\label{SpectrumAllFermion}
\end{align}
where $m_B^2/k^2, m_B M/k^2, \mu_{\bf 11}' \gg \widetilde \mu_2$
has been asuumed  for (iv-1) and (iv-2).
(v) is the dark fermion mass spectrum.
$c_0'$ stands for the bulk mass of the 6D $SO(11)$ spinor bulk Weyl
fermion $\Psi_{\bf 32}^{\alpha=4}$.

We evaluate the  effective potential 
$V_{\rm eff}(\theta_H) = V_{\rm eff}^{\rm gauge}(\theta_H)
+V_{\rm eff}^{\rm fermion}(\theta_H)$.
In the $R_\xi$ gauge $V_{\rm eff}^{\rm gauge}(\theta_H)$ is
decomposed into 
\begin{align}
&V_{\rm eff}^{\rm gauge}(\theta_H)=
V_{\rm eff}^{W^{\pm}} +V_{\rm eff}^{Z}
+V_{\rm eff}^{A_z^{a4},A_z^{a,11}} ~, \cr
\noalign{\kern 5pt}
&V_{\rm eff}^{W^{\pm}}(\theta_H)
=2(3-\xi^2)
I\left[\onehalf Q_{0} \left(q, {1} \right);\sin^2\theta_H\right] ~, \cr
\noalign{\kern 5pt}
&V_{\rm eff}^{Z}(\theta_H)
=(3-\xi^2)
I \bigg[ \frac{Q_{0}\left(q, {1} \right)}{2 \cos^2 \theta_W} ; \sin^2\theta_H \bigg]  ~, \cr
\noalign{\kern 5pt}
&V_{\rm eff}^{A_z^{a4},A_z^{a\, 11}}(\theta_H)
 =3\xi^2
I\left[Q_{0}\left(q, {1} \right);\sin^2\theta_H\right] ~.
\label{Eq:Effective-potential-gauge}
\end{align}
Here
\begin{align}
&Q_0\left(q,c\right) =\frac{z_L}{q^2}
\frac{1}{\hat{F}_{c}^{++}(q)\hat{F}_{c}^{--}(q)} ~, \cr
\noalign{\kern 5pt}
&\hat{F}_{c}^{\pm \pm}(q)
 = \hat F_{c\pm \onehalf, c\pm \onehalf} (q z_L^{-1}, q) ~, \cr
\noalign{\kern 5pt} 
&\hat F_{\alpha, \beta} (u,v) = I_\alpha (u) K_\beta (v) - 
e^{-i (\alpha -\beta) \pi}  K_\alpha (u) I_\beta (v)  
\end{align}
where $I_\alpha (u)$ and $K_\beta (u)$ are modified Bessel functions.

The fermion part $V_{\rm eff}^{\rm fermion}(\theta_H)$ is evaluated in a
similar manner.   
Following the decomposition in (\ref{SpectrumAllFermion})
and taking into account four degrees of freedom for each Dirac fermion and 
the color factor 3 for quarks,  one can  write as
\begin{align}
V_{\rm eff}^{\rm fermion}(\theta_H)
& =V_{\rm eff}^{\rm (i)}+V_{\rm eff}^{\rm (ii)}
+V_{\rm eff}^{\rm (iii)}
+V_{\rm eff}^{\rm (iv-1)}
+V_{\rm eff}^{\rm (iv-2)}
+V_{\rm eff}^{\rm (v)}~, \cr
\noalign{\kern 10pt}
V_{\rm eff}^{\rm (i)}(\theta_H)
&=-12I \left[Q_0\left(q,c_0\right);\sin^2\onehalf \theta_H \right] ~, \cr
\noalign{\kern 5pt}
V_{\rm eff}^{\rm (ii)}(\theta_H)
&=-12I\left[Q_{\rm (ii)}
(q,c_0,c_1,\mu_1,\mu_{\bf 11});\sin^2 \onehalf \theta_H \right] ~,\cr
\noalign{\kern 5pt}
V_{\rm eff}^{\rm (iii)}(\theta_H)
&=-4I\left[Q_{\rm (iii)}
(q,c_0,c_2,\widetilde{\mu}_2,\mu_{\bf 11}^{\prime});\sin^2 \onehalf \theta_H \right] ~,\cr
\noalign{\kern 5pt}
V_{\rm eff}^{\rm (iv-1)}(\theta_H)
&=- 4 \cdot \frac{1}{2}  I\left[\widetilde{Q}_{\rm (iv-1)}
(q,c_0,m_B,M;\theta_H);1\right] ~,   \cr
\noalign{\kern 5pt}
V_{\rm eff}^{\rm (iv-2)}(\theta_H)
&=-4I\left[\widetilde{Q}_{\rm (iv-2)}
(q,c_2,\mu_{\bf 11}^{\prime};\theta_H);1\right] ~,\cr
\noalign{\kern 5pt}
V_{\rm eff}^{\rm (v)}(\theta_H)
&=-32I \left[Q_0\left(q,c_0'\right);\cos^2 \onehalf \theta_H \right] ~,
\label{Eq:Effective-potential-fermion}
\end{align}
where
\begin{align}
&Q_{\rm (ii)}(q) =
\frac{z_L}{q^2}  
\Bigg\{ \hat{F}_{c_0}^{++}\hat{F}_{c_0}^{--}
+\frac{\mu_1^2 \hat{F}_{c_0}^{--} \hat{F}_{c_0}^{-+} \hat{F}_{c_1}^{++} \hat{F}_{c_1}^{-+}}
{\mu_{\bf 11}^{2} (\hat{F}_{c_1}^{-+} )^2 + (\hat{F}_{c_1}^{++} )^2} \Bigg\}^{-1},  \cr
\noalign{\kern 10pt}
&Q_{\rm (iii)}(q)=
\frac{z_L}{q^2}
\Bigg\{ \hat{F}_{c_0}^{++}\hat{F}_{c_0}^{--}
+\frac{\widetilde{\mu}_2^2 \hat{F}_{c_0}^{++} \hat{F}_{c_0}^{+-} 
\hat{F}_{c_2}^{--} \hat{F}_{c_2}^{+-}}
{\mu_{\bf 11}^{\prime 2} (\hat{F}_{c_2}^{+-} )^2 + (\hat{F}_{c_2}^{--} )^2} \Bigg\}^{-1},   \cr
\noalign{\kern 10pt}
&Q_{\rm (iv-1)}(q)=
\frac{z_L}{q^2}
\bigg\{\hat{F}_{c_0}^{++}\hat{F}_{c_0}^{--}
-\frac{ i m_B^2 / k^2}{2(iqz_L^{-1}+M/k)}
\hat{F}_{c_0}^{-+}\hat{F}_{c_0}^{--}\bigg\}^{-1},  \cr
\noalign{\kern 10pt}
&\widetilde{Q}_{\rm (iv-1)}(q;\theta_H) =
(Q_{\rm (iv-1)}(q)+Q_{\rm (iv-1)}(q)^*)\sin^2 \onehalf \theta_H \cr
\noalign{\kern 5pt}
&\hskip 5.cm 
+( Q_{\rm (iv-1)}(q)Q_{\rm (iv-1)}(q)^* )\sin^4 \onehalf \theta_H ~, \cr
\noalign{\kern 10pt}
&\widetilde{Q}_{\rm (iv-2)}(q,\theta_H)= \frac{z_L^2}{q^4}
\frac{h_1 (q, \theta_H)}{h_2 (q)} ~,  \cr
\noalign{\kern 10pt}
&
h_1  = ( 1 - \mu_{\bf 11}^{\prime \, 2} )
\bigg\{ 2\frac{q^2}{z_L}
(1 + \mu_{\bf 11}^{\prime \, 2}) \hat{F}_{c_2}^{++}\hat{F}_{c_2}^{--}
+ 2 \mu_{\bf 11}^{\prime 2}
+(1 - \mu_{\bf 11}^{\prime \, 2}) \sin^2\theta_H \bigg\} \sin^2\theta_H  ~,  \cr
\noalign{\kern 10pt}
&
h_2 = \frac{z_L^2}{q^4} \mu_{\bf 11}^{\prime 4}
+2\mu_{\bf 11}^{\prime 4} \frac{z_L}{q^2} \hat{F}_{c_2}^{++}\hat{F}_{c_2}^{--}
+(1+\mu_{\bf 11}^{\prime 4})
 (\hat{F}_{c_2}^{++}\hat{F}_{c_2}^{--} )^2 \cr
\noalign{\kern 10pt}
&\hskip 5.cm
 +\mu_{\bf 11}^{\prime 2}
\Big\{ (\hat{F}_{c_2}^{-+}\hat{F}_{c_2}^{--} )^2
+ (\hat{F}_{c_2}^{+-}\hat{F}_{c_2}^{++} )^2 \Big\} ~.
\end{align}
As $Q_{\rm (iv-1)}(q)$ is not a real function, 
$V_{\rm eff}^{\rm (iv-1)}(\theta_H)$ has been converted to the integral 
involving $\widetilde{Q}_{\rm (iv-1)}(q;\theta_H)$.

The Higgs mass $m_H$ is determined,
at the minimum $\theta_H = \theta_H^{\rm min}$ of $V_\eff (\theta_H)$,  
by 
\begin{align}
&m_H^2 =  \frac{1}{f_H^2}
\frac{d^2V_{\rm eff}(\theta_H)}{d\theta_H^2}
\bigg|_{\theta_H = \theta_H^{\rm min}}~, \cr
\noalign{\kern 5pt}
&f_H = { \frac{\sqrt{6}}{g} \sqrt{\frac{2\pi R_6 \, k}{z_L^3 -1} } }
= { \frac{\sqrt{6}}{g_w} \frac{k}{\sqrt{(1-z_L^{-1})  (z_L^3 -1)}} }~, \cr
\noalign{\kern 5pt}
&{g_w = \frac{e}{\sin \theta_W} 
= g \sqrt{ \frac{k}{2\pi R_6 (1 - z_L^{-1})}} }~. 
\label{Eq:Higgs_mass}
\end{align}
Note that the 4D neutral Higgs field $H(x)$ is related { to  $\phi_H(x)$} in
(\ref{AzHiggs1}) by 
\begin{align}
\phi_H(x)=  \theta_H f_H + H(x)  ~.
\end{align}

We give results  for the effective potential and the Higgs mass
for typical sample values of the parameters.
In the following calculation, we take into account the 6th
dimensional $n=0$ modes of 6D bulk gauge bosons $A_M$, the third generation
$SO(11)$ spinor and vector fermions $\Psi_{\bf 32}^{\alpha=3}$,
$\Psi_{\bf 11}^{\beta=3}$, $\Psi_{\bf 11}^{\prime\beta=3}$, 
and dark fermions $\Psi_{\bf 32}^{\alpha=4}$.
Contributions coming from the first and second generations of quarks 
and leptons are numerically negligible.
Some of the parameters in the fermion sector are fixed by 
 the observed masses of quarks and
leptons in their approximate forms given in 
Eqs.~(\ref{Eq:Up-type-quark-mass-approximate}),
(\ref{Eq:Down-type-quark-mass-approximate}),
(\ref{Eq:Charged-lepton-mass-approximate}), and
(\ref{Neutrino-mass}).

The calculation algorithm  is almost the same as   in 5D $SO(11)$ GHGUT 
given in Sec.~5 of Ref.~\cite{Furui:2016owe}.
We are interested in the effective potential $V_\eff (\theta_H)$ at the 
electroweak scale.
Some of the relations derived in this paper are those at the GUT scale.
The value of $\sin^2 \theta_W$  evolves from $\frac{3}{8}$ 
at the GUT scale to the observed value at the electroweak scale by the 
renormalization group equation (RGE).  The RGE analysis of the coupling
constants is beyond the scope of the current paper.
In this paper we content ourselves to insert the observed value of  $\sin^2 \theta_W$ 
into the formulas of $m_Z$ and $V_\eff (\theta_H)$.
As input parameters we take
$\alpha_{\rm EM}^{-1}=127.916$,
$\sin^2\theta_W=0.2312$,
$m_Z=91.1876\ \mbox{GeV}$,
$m_t=171.17\ \mbox{GeV}$,
$m_b=4.18\ \mbox{GeV}$,
$m_\tau=1.776\ \mbox{GeV}$ and 
$m_{\nu_\tau}=0.1\ \mbox{eV}$.
The evaluation  is carried out in the following steps.\\
(1) We first pick the  values for $z_L$ and $\theta_H$.
{So far  consistent sets of parameters have been found only for $30 \lesssim z_L \lesssim 40$.}\\
(2) From $m_Z$, (\ref{spectrumZ2}), $k$ and $m_{\KK_5}$ are determined.  \\
(3) $c_0$ is fixed from $m_t$ by (\ref{Eq:Up-type-quark-mass-approximate}).\\
(4) 
Some of the parameters in the brane interactions on the UV brane remain free.
We take {$c_1 =0$,  $c_2 = -0.7$,  and $M = -10^{7}\,$GeV. 
These three parameters are kept fixed in the evaluation below.
We temporarily assign a value for $\mu_{\bf 11} = \mu_{\bf 11}' $.}
Then,  $\mu_1$, $\widetilde{\mu}_{2}$, and $m_B$ are determined by 
(\ref{DownUpmassratio}), (\ref{ElectronUpmassratio}) and (\ref{Neutrino-mass}),
or by
\begin{align}
&\mu_{1}\simeq
\sqrt{\frac{1+2c_1}{1+2c_0}}z_L^{c_0-c_1}\frac{m_t}{m_b}
\mu_{\bf 11} ~, \cr
\noalign{\kern 10pt}
&
\widetilde{\mu}_{2}\simeq
\sqrt{\frac{1-2c_2}{1-2c_0}} \, z_L^{c_2-c_0} \frac{m_t}{m_\tau}
\mu_{\bf 11}^{\prime} ~,   \cr
\noalign{\kern 10pt}
&m_B=\sqrt{- \frac{2Mm_t^2}{m_{\nu_\tau}}\frac{z_L^{2c_0+1}}{1+2c_0}} ~.
\label{Parameter-relation}
\end{align}
(5) 
Given the value of the bulk mass parameter $c_0'$ of the dark fermion multiplet,
the effective potential $V_\eff (\theta_H)$ can be evaluated.  We adjust the value
of $c_0'$ such that the minimum of $V_\eff (\theta_H)$ is located at the initial 
value of $\theta_H$.
In this manner a consistent  parameter set has been obtained.\\
(6) The Higgs boson mass $m_H$ is determined from (\ref{Eq:Higgs_mass}),
which can be viewed as a function of {$z_L$, $\theta_H $ and $\mu_{\bf 11}'$, namely
$m_H (\theta_H, z_L,\mu_{\bf 11}')$}.\\
(7) By demanding that the resulting $m_H$ is the observed value, 
namely $m_H (\theta_H, z_L,\mu_{\bf 11}') = 125.09 \pm 0.24 \,$GeV, 
{$\mu_{\bf 11}'$ is determined with the given $\theta_H$ and $z_L$.} \\
(8) Take a different $\theta_H$.  By repeating the above procedure one finds a new
value for {$\mu_{\bf 11}'$}.

The physical value of $\theta_H$ need to be determined, for instance, 
from the observation of the $Z'$ bosons (the 1st KK bosons $\gamma^{(1)}, 
Z^{(1)}, Z_R^{(1)}$) and their masses.   
In the $SO(5) \times U(1)$ gauge-Higgs electroweak unification
many of the physical quantities depend on the value of $\theta_H$, but not on the details
of the parameters in the theory.  It has been shown that 
$m_{Z^{(1)}} \gtrsim 7\,$TeV ($m_{\KK_5} \gtrsim 8.5\,$TeV) 
to be consistent with the current LHC data.\cite{Funatsu:2016uvi, Funatsu:2017nfm}

\begin{figure}[tb]
\begin{center}
\includegraphics[bb=0 0 494 357,height=5cm]{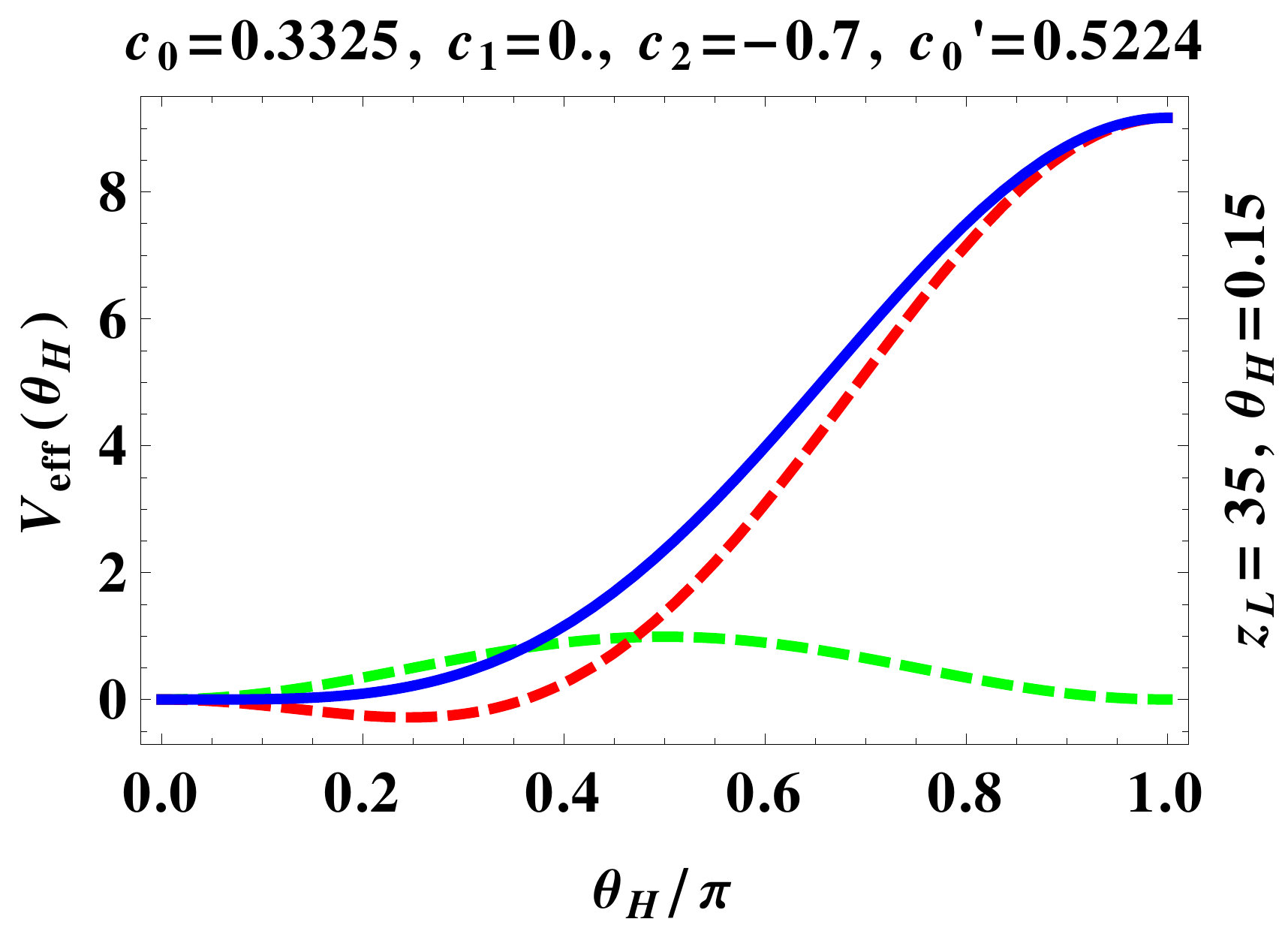}
\includegraphics[bb=0 0 556 352,height=5cm]{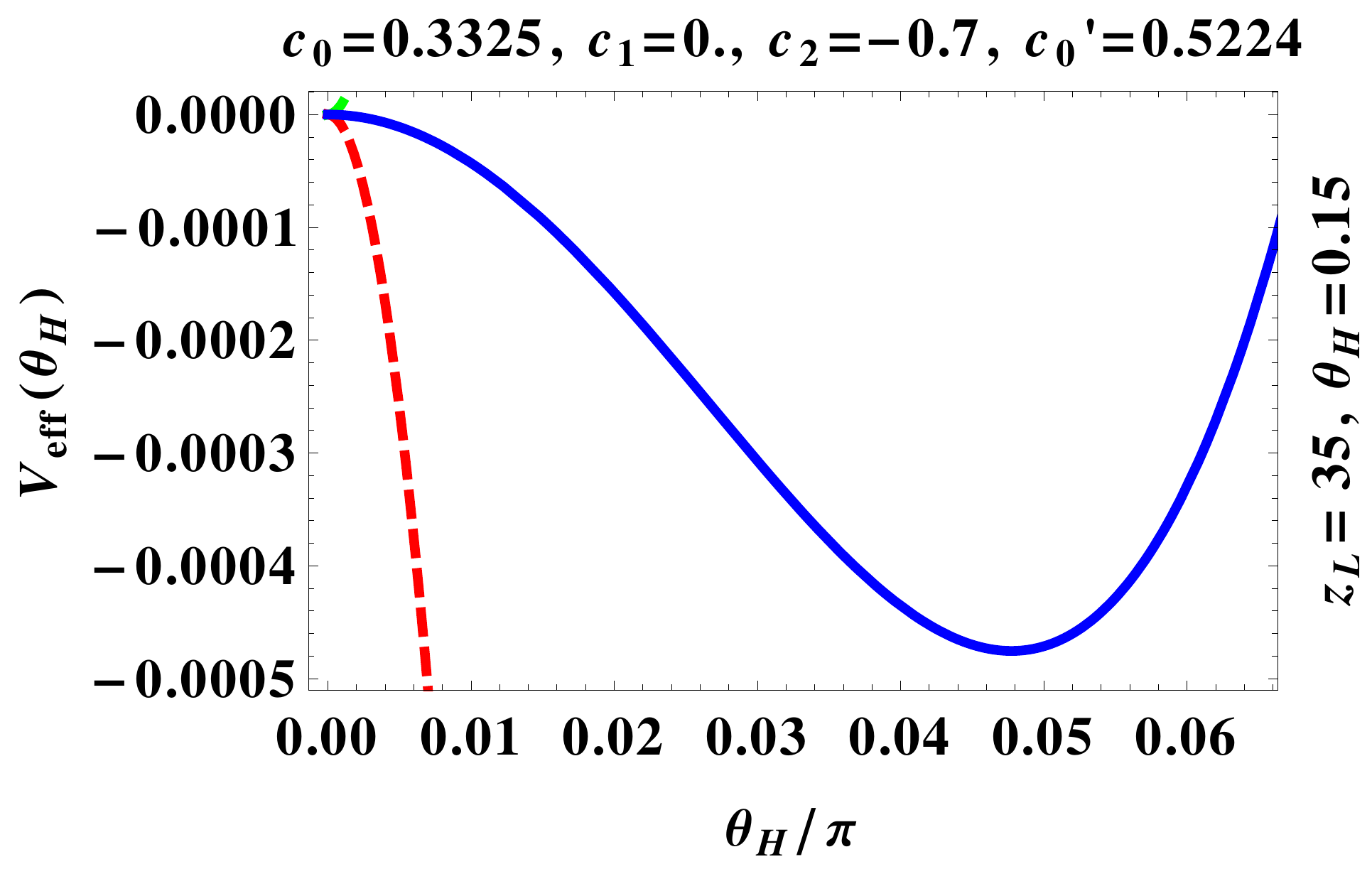}
\includegraphics[bb=0 0 484 330,height=5cm]{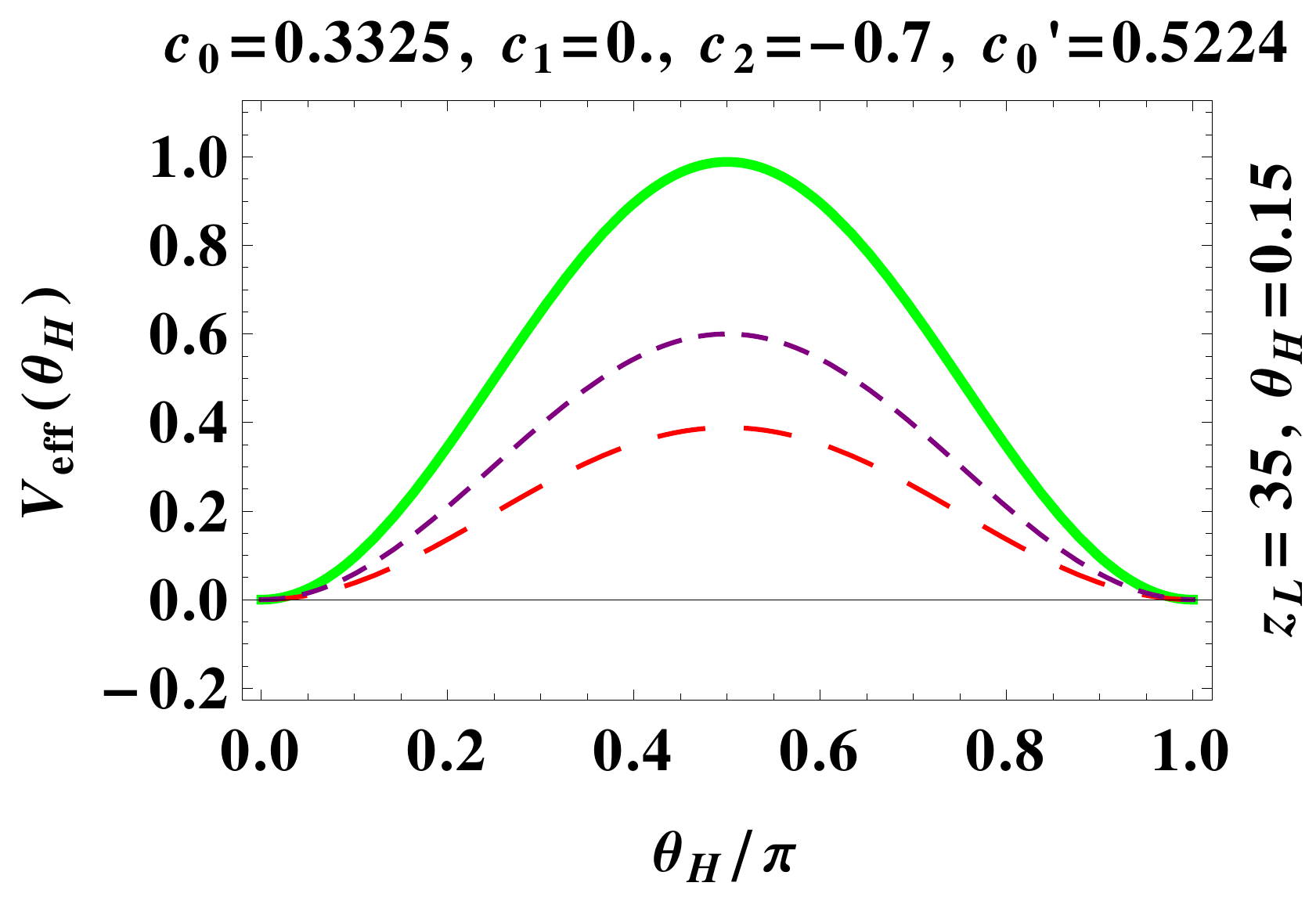}
\includegraphics[bb=0 0 494 340,height=5cm]{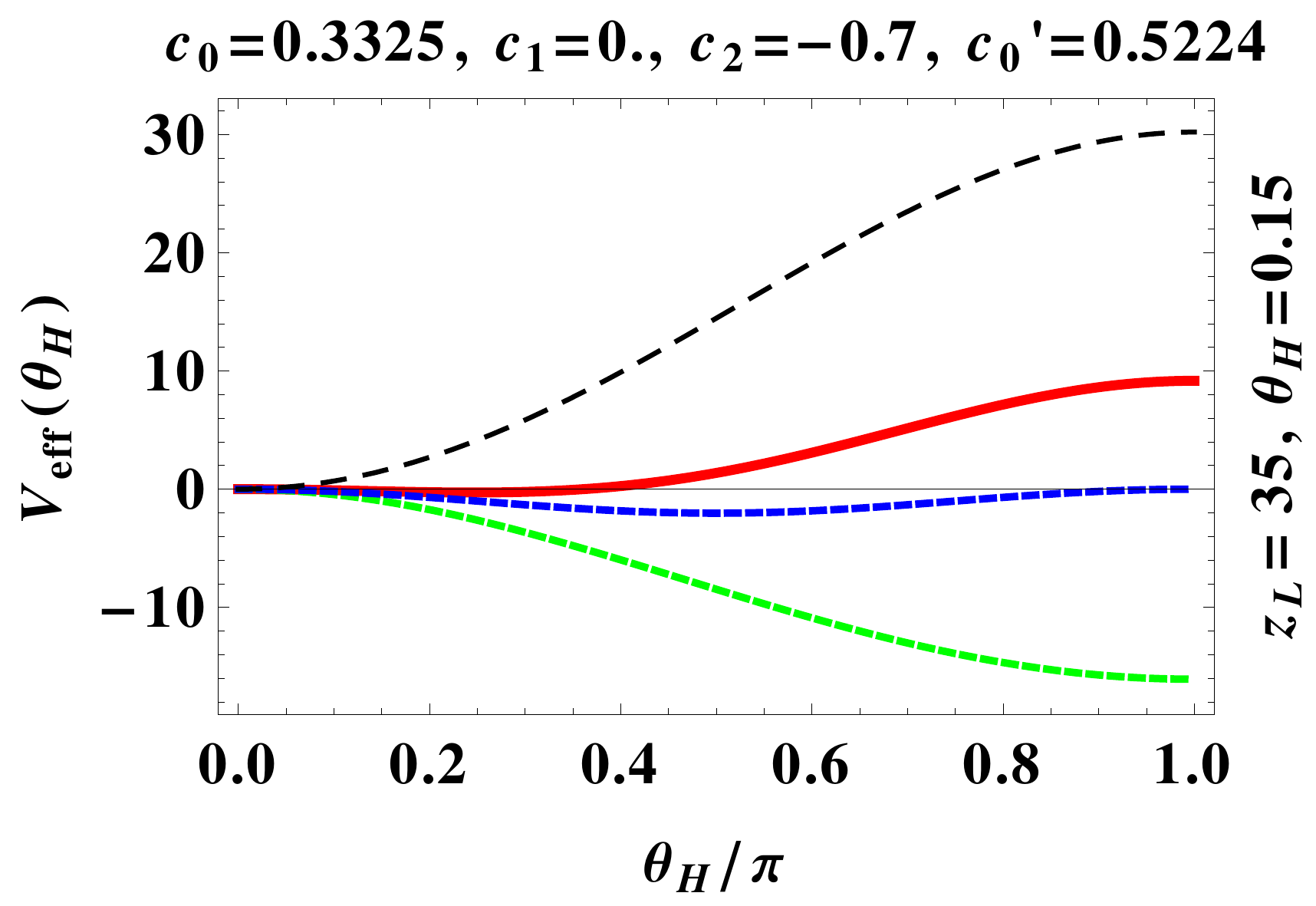}
\end{center}
\caption{
The effective potential $V_{\rm eff}(\theta_H)$ in the $R_{\xi=0}$ gauge
in which $z_L={35}$ 
and the minimum of $V_{\rm eff}(\theta_H)$ is located at 
$\theta_H={0.15}$.
For the top two figures, the blue solid line shows the effective
potential with both gauge and fermion effects. 
The red and green dashed lines represent the contributions of  fermions
and gauge fields, respectively.
For the bottom left figure, the green solid line represents the total
contributions of gauge fields. 
The red and purple dashed lines represent $Z$ and $W$ boson contributions.
In the bottom right figure the red solid line represents the total
contribution of all fermions.  
The green dashed lines represent contributions of top quark.
The blue dashed line represents that of the neutrino-2 sector.
The black dashed line represents the contribution of  dark fermions.
Contributions coming from other fermions are negligible.
}
\label{Figure:Effective-potential-xi=0}
\end{figure}

The results for the effective potential 
$V_{\rm eff}(\theta_H)$ in the $R_{\xi=0}$ gauge
are depicted  in Figure~\ref{Figure:Effective-potential-xi=0},
for which $z_L={35}$, $\theta_H={0.15}$, and $m_H = 125.1\,$GeV. 
The bulk and brane mass parameters are chosen as 
{
\begin{align}
&z_L=35,\ \ \theta_H=0.15,\ \ m_H=125.12\ \mbox{GeV},
\nonumber\\
&c_0=0.3325,\ \ c_1=0,\ \ c_2=-0.7 ,\ \ c_0'=0.5224,
\nonumber\\
&\mu_1\simeq 11.18,\ \
\widetilde{\mu}_2\simeq 0.7091,\ \
\mu_{\bf 11}=\mu_{\bf 11}^{\prime}=0.108,
\nonumber\\
&m_B=1.145\times 10^{12}\ \mbox{GeV},\ \
M=-10^{7}\ \mbox{GeV}.
\end{align}
}
The resultant $m_{\KK_5} = \pi k z_L^{-1}$, $k$, and $f_H$ are given by
\begin{align}
{
m_{\rm KK_5}\simeq 8.236\ \mbox{TeV},\ \
k\simeq 8.913\times 10^{4}\ \mbox{GeV},\ \
f_H\simeq 1.642\times 10^{3}\ \mbox{GeV}.
}
\end{align}

In Table \ref{zL-theta-mH-mKK},  we have tabulated various sets of the parameters
which yield $m_H=125.1 \pm 0.1$ GeV.  As $\theta_H$ increases, $m_\KK$ decreases.
One sees {$\theta_H \lesssim 0.15$ in order for $m_\KK > 8\,$TeV} to be consistent with
the current LHC data.
{
We have observed that
$\theta_H $ cannot be  too small  in order to reproduce $m_H \sim 125 \,$GeV.}
\begin{table}[htb]
{\small
\renewcommand{\arraystretch}{0.95}
\begin{center}
\begin{tabular}{|c|c|c|c|c|c|c|}
\hline
$\theta_H$&$\mu_{\bf 11}=\mu_{\bf 11}^\ell$&$m_{\KK_5}$ [TeV]&$c_0$&$c_0'$&$m_H$ [GeV]&
$\widetilde{\mu}_2$\\
\hline
$0.05$&$0.1730$&$24.63$&$0.3289$&$0.5960$&$125.17$&$1.138$\\
$0.06$&$0.1590$&$20.53$&$0.3292$&$0.5828$&$125.07$&$1.046$\\
$0.07$&$0.1480$&$17.60$&$0.3294$&$0.5714$&$125.13$&$0.9736$\\
$0.08$&$0.1400$&$15.40$&$0.3297$&$0.5625$&$125.03$&$0.9208$\\
$0.09$&$0.1330$&$13.70$&$0.3300$&$0.5544$&$125.08$&$0.8746$\\
\hline
$0.10$&$0.1270$&$12.33$&$0.3303$&$0.5471$&$125.17$&$0.8350$\\
$0.11$&$0.1225$&$11.21$&$0.3307$&$0.5414$&$125.05$&$0.8052$\\
$0.12$&$0.1180$&$10.28$&$0.3311$&$0.5356$&$125.14$&$0.7755$\\
$0.13$&$0.1145$&$9.495$&$0.3315$&$0.5311$&$125.06$&$0.7523$\\
$0.14$&$0.1110$&$8.821$&$0.3320$&$0.5264$&$125.12$&$0.7291$\\
\hline
$0.15$&$0.1080$&$8.236$&$0.3325$&$0.5224$&$125.12$&$0.7091$\\
$0.16$&$0.1055$&$7.725$&$0.3331$&$0.5192$&$125.03$&$0.6925$\\
$0.17$&$0.1030$&$7.275$&$0.3337$&$0.5158$&$125.02$&$0.6759$\\
$0.18$&$0.1005$&$6.875$&$0.3343$&$0.5125$&$125.11$&$0.6593$\\
$0.19$&$0.0985$&$6.517$&$0.3349$&$0.5099$&$125.05$&$0.6459$\\
\hline
$0.20$&$0.0965$&$6.195$&$0.3356$&$0.5073$&$125.05$&$0.6326$\\
$0.21$&$0.0945$&$5.903$&$0.3363$&$0.5047$&$125.21$&$0.6192$\\
$0.22$&$0.0930$&$5.639$&$0.3371$&$0.5030$&$125.00$&$0.6092$\\
$0.23$&$0.0910$&$5.398$&$0.3379$&$0.5003$&$125.17$&$0.5959$\\
$0.24$&$0.0895$&$5.177$&$0.3387$&$0.4986$&$125.14$&$0.5858$\\
\hline
$0.25$&$0.0880$&$4.974$&$0.3395$&$0.4969$&$125.16$&$0.5758$\\
$0.26$&$0.0867$&$4.786$&$0.3404$&$0.4956$&$125.11$&$0.5671$\\
$0.27$&$0.0855$&$4.613$&$0.3413$&$0.4945$&$125.04$&$0.5590$\\
$0.28$&$0.0840$&$4.452$&$0.3423$&$0.4928$&$125.16$&$0.5490$\\
$0.29$&$0.0830$&$4.302$&$0.3432$&$0.4921$&$125.05$&$0.5423$\\
$0.30$&$0.0818$&$4.163$&$0.3442$&$0.4911$&$125.07$&$0.5342$\\
\hline
\end{tabular}
\caption{Parameter sets which give dynamical EW symmetry breaking 
with $m_H=125.1 \pm 0.1$ GeV.
{Here $z_L=35$, $c_1=0.0, c_2=-0.7$, and $M = - 10^7\,$GeV.
$c_2$ can be varied without affecting $m_{\KK_5}$ and $c_0$.
For $\theta_H = 0.05$, for instance,  a set of $c_2 = - 0.9$ and
$\mu_{\bf 11}=\mu_{\bf 11}^\ell = 0.1174$  yields $m_H = 125.05\,$GeV, 
$c_0' = 0.6008$, and $\widetilde{\mu}_2 = 0.4084$.}
}
\label{zL-theta-mH-mKK}
\end{center}
}
\end{table}

\begin{figure}[tbh]
\begin{center}
\includegraphics[bb=0 0 512 349,height=6cm]{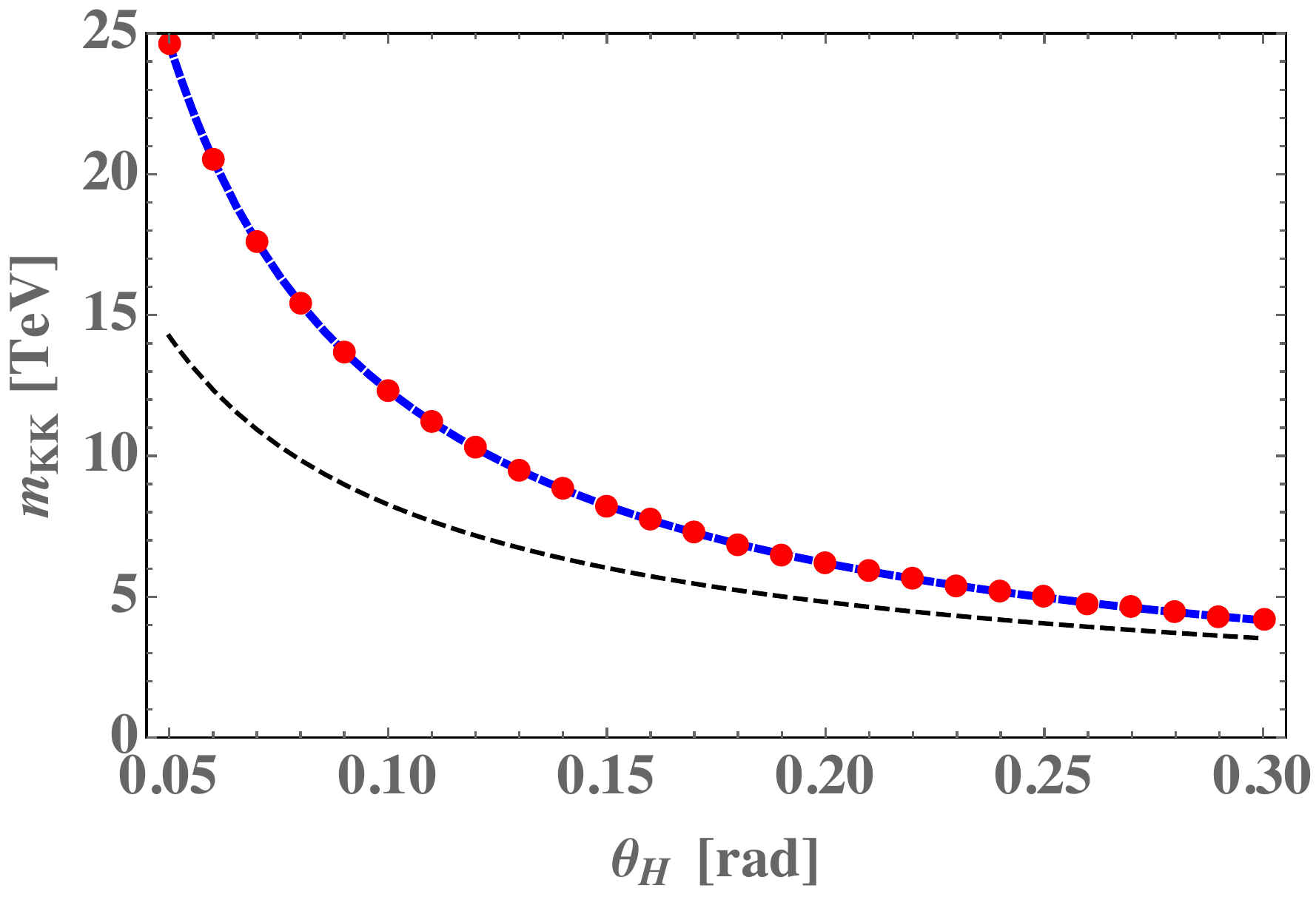}
\end{center}
\vskip -10pt
\caption{The Kaluza-Klein mass scale $m_\KK = m_{\KK_5}$ as a function of $\theta_H$.
The red dots represent  points in Table \ref{zL-theta-mH-mKK} which reproduce 
$m_H=125.1\pm 0.1$ GeV.
The blue thick dashed line represents the fitting curve given  in  (\ref{mKK-theta1}).
The black thin dashed line indicates $m_\KK (\theta_H)$ found in the 
$SO(5) \times U(1)$ gauge-Higgs EW unification.\cite{Funatsu:2013ni, Funatsu:2014fda}
}
\label{Figure:theta-mKK}
\end{figure}

In Figure \ref{Figure:theta-mKK},
$m_{\KK_5}$ is plotted as a function of $\theta_H$.  Data points in Table \ref{zL-theta-mH-mKK}
are fitted by
{
\begin{align}
m_{\KK_5} \sim \frac{\alpha}{(\sin \theta_H)^{\beta}}~,~~
\alpha = 1.230\,{\rm TeV} ~, ~~ \beta = 1.000
\label{mKK-theta1}
\end{align}
in the range $0.05 < \theta_H < 0.30$.}
A similar relation has been found in the $SO(5) \times U(1)$ gauge-Higgs EW unification
where $\alpha=1.352\,$TeV and $\beta = 0.786$.\cite{Funatsu:2013ni, Funatsu:2014fda}
The difference between the two is expected to originate from the different gauge group structure
of the models and the {different} matter content.
{So far we have found consistent parameter sets only for $c_1=0$ and $c_2 \le 0$.}

\section{Summary and discussions}
\label{Sec:Summary-Discussion}

In this paper, we discussed 6D $SO(11)$ GHGUT in the 6D hybrid warped space.
The GUT gauge symmetry $SO(11)$ is reduced to the Pati-Salam symmetry 
$G_{\rm PS}(=SU(2)_L\times SU(2)_R\times SU(4)_C)$ by the orbifold BCs, 
and the symmetry $G_{\rm PS}$ is spontaneously broken to 
the SM gauge symmetry $G_{\rm SM}(=SU(3)_C\times SU(2)_L\times U(1)_Y)$
by the nonvanishing VEV of the 5D brane scalar field $\Phi_{\bf 32}$ on
the UV brane. Finally,  the EW gauge symmetry $SU(2)_L\times U(1)_Y$ is broken 
to $U(1)_{\rm EM}$ by the Hosotani mechanism.
The SM Higgs boson appears as a zero mode of the 5th dimensional
components of the $SO(11)$ gauge field.

The SM fermions are contained in $\Psi_{\bf 32}$, 
$\Psi_{\bf 11}$, $\Psi_{\bf 11}^{\prime}$, $\chi_{{\bf 1}}$.
The SM fermions as well as the $W$ and $Z$ bosons acquire masses via the Hosotani mechanism.
In Sec.~\ref{Sec:Spectrum-Fermions}, we showed that the quark and lepton
mass spectra can be reproduced by choosing the parameters of the bulk
masses and the brane interactions.
In the 5D GHGUT in Ref.\ \cite{Furui:2016owe}, there necessarily have arisen
light exotic vectorlike fermions accompanied with SM fermions, 
which is  experimentally unacceptable.
In quite contrast to the case of the 5D GHGUT,
the non-SM fermions  have  masses  of either $O(m_{\rm KK_{6}})$ 
or $O(m_{\rm KK_{5}})$ in the current 6D model.
Thus the problem of light exotic fermions has been solved.

From the analysis of the effective potential $V_{\rm eff}(\theta_H)$ in
Sec.~\ref{Sec:Effective-potential}, we have seen that the EW symmetry is dynamically
broken  by the AB phase $\theta_H$.  
The Higgs boson mass $m_H = 125.1\,$GeV can be obtained
{for $0.05 \lesssim \theta_H  \lesssim  0.30$, namely for
$25\, {\rm TeV} \gtrsim  m_{\KK_5} \gtrsim 4.1 \,$TeV.}

In the present paper we have treated each generation of quarks and leptons
independently.  It is  necessary to incorporate the flavor mixing in the
gauge-Higgs grand unification.  The mixing arises from the brane interactions.
For quarks and charged leptons the mixing would result from
the matrix structure  of various $\kappa$'s and $\mu$'s in the brane
interactions (\ref{Eq:Action-brane-fermion}) of $\Psi_{\bf 32}$, $\Psi_{\bf 11}$
and $\Phi_{\bf 32}$.  
On the other hand
the dominant mixing in the neutrino sector would arise from the matrix
structure in the mass $M^{\beta \beta'}$ associated with the symplectic Majorana 
fields $\chi_{\bf 1}$ in (\ref{Eq:Action-brane-fermion-chi}).
The large mixing angles observed in the neutrino sector might be related to 
this special circumstance.

We would like to add a comment on the  effective potential  $V_{\rm eff}^6(\theta_H^6)$ 
of the AB phase $\theta_H^6$ along the 6th dimension.
From Table~\ref{Table:Venn-diagram-Symmetry},
not only the $SO(5)/SO(4)$ components of the 5th dimensional gauge field
$A_y$ but also those of the 6th dimensional gauge field $A_v$
have zero modes in the absence of brane interactions.
As in the  effective potential  $V_{\rm eff}(\theta_H)$ for the AB phase $\theta_H$ 
along the 5th dimension,  the $SO(11)$ bulk gauge bosons $A_M$, 
the $SO(11)$ bulk fermions $\Psi_{\bf 32}^{\alpha}$,
$\Psi_{\bf 11}^{\beta}$, and $\Psi_{\bf 11}^{\prime\beta}$,
contribute to   $V_{\rm eff}^6(\theta_H^6)$. 
If there were only gauge fields, it can be shown
that the minimum of  $V_{\rm eff}^6(\theta_H^6)$ would be  located at 
$\theta_H^6=\onehalf \pi \ ({\rm mod.}\pi)$.
In other words,  the EW symmetry would be broken
by the 6th dimensional AB phase $\theta_6$ and the $W$ and $Z$ bosons
would acquire masses of $O(m_{\rm KK_{6}})$.
The situation is similar to that in the  effective potential 
$V_{\rm eff}(\theta_H)$ in the 5D $SO(11)$ GHGUT.\cite{Hosotani:2015hoa,Furui:2016owe}
Further the contribution from the $SO(11)$ bulk spinor fermions $\Psi_{\bf 32}^{\alpha}$
to $V_{\rm eff}^6(\theta_H^6)$ has little  $\theta_H^6$ dependence as is 
inferred from Table~\ref{Tab:BC-fermion-spinor}.
The $\theta_H^6$-dependent contributions coming from a set of $SO(11)$ bulk vector fermions 
$\Psi_{\bf 11}^{\beta}$ and $\Psi_{\bf 11}^{\prime\beta}$ almost cancel 
each other as can be deduced  from Table~\ref{Tab:BC-fermion-vector}.
Hence, with the minimal fermion content presented in this paper 
the minimum of $V_{\rm eff}^6(\theta_H^6)$ would be located at around
$\theta_H^6=\pi/2\ ({\rm mod.}\pi)$. 
This undesirable result can be  avoided by introducing additional $SO(11)$ bulk
vector fermions $\Psi_{\bf 11}^{\beta}$ with appropriate boundary conditions.

The phase $\theta_H$ plays a crucial role in the $SO(11)$ gauge-Higgs grand unification. 
It is found that the 5D KK mass scale $m_{\KK_5}$ is determined by $\theta_H$ in a very good 
approximation by the formula (\ref{mKK-theta1}).  Similar relations are expected for
the 4D Higgs cubic and quartic couplings as in the $SO(5) \times U(1)$ gauge-Higgs EW
unification.  For $m_{\KK_5} (\theta_H)$ we have recognized {the difference between 
the gauge-Higgs EW and grand unification.   
We need   experimental data to find which one is prefered.}

We will  come back to these issues in the near future.

\section*{Acknowledgments}

{
We would like to thank Hisaki Hatanaka for critical reading of the manuscript and
valuable remarks.}
This work was supported in part by Japan Society for the Promotion of
 Science, Grants-in-Aid  for Scientific Research,  No.\ 15K05052.


\appendix

\section{Basics for 6D Kaluza-Klein expansion}
\label{Sec:6D-KK-expansion}

{
We present KK mode expansions  in 6D hybrid warped space with the metric given
in (\ref{Eq:6D-metric}).}

\subsection{{6D gauge fields}}
\label{Sec:6D-gauge-KK-modes}

{
The equations of motion for  free gauge fields in the 6D hybrid warped space 
are given by
\begin{align}
&\Big\{ \hat \eta^{\lambda \rho}  \big(\Box+k^2 {\cal P}_5  +\partial_v^2 \big)-
\Big( 1-\frac{1}{\xi} \Big) \partial^\lambda \partial^\rho \Big\} A_\rho = 0 ~, \cr
\noalign{\kern 5pt}
&
\left(\Box+\xi k^2{\cal P}_z+\partial_v^2\right)A_z = 0 ~,
\label{APgaugeEq1}
\end{align}
where ${\cal P}_5$ and ${\cal P}_z$ are defined in Eq.\ (\ref{gauge-free-action1}).
These ${\cal P}_5$ and ${\cal P}_z$ differ from ${\cal P}_4$ and ${\cal P}_z$ in the 5d case.
Basis mode functions in the $z$-coordinate in the 6d case are given, except for zero modes,  by
\begin{align}
&C(z;\lambda) 
= +\frac{\pi}{2}\lambda   z^{3/2}  z_L^{1/2}  F_{\frac{3}{2}, \frac{1}{2}}(\lambda z, \lambda z_L) ~, \cr
\noalign{\kern 5pt}
&S(z;\lambda) 
= -\frac{\pi}{2}\lambda   z^{3/2} z_L^{1/2} F_{\frac{3}{2}, \frac{3}{2}}(\lambda z, \lambda z_L) ~, \cr
\noalign{\kern 5pt}
&C'(z;\lambda)
= + \frac{\pi}{2}\lambda^2 z^{3/2} z_L^{1/2} F_{\frac{1}{2},\frac{1}{2}}(\lambda z, \lambda z_L) ~, \cr 
\noalign{\kern 5pt}
&S'(z;\lambda)
= -\frac{\pi}{2} \lambda^2 z^{3/2} z_L^{1/2} F_{\frac{1}{2},\frac{3}{2}}(\lambda z,  \lambda z_L) ~, 
\label{APgaugeF1}
\end{align}
where $F_{\alpha,\beta}(u,v) = J_\alpha (u) Y_\beta(v)-Y_\alpha(u) J_\beta(v)$.  
They can be expressed as
\begin{align}
&C(z;\lambda) 
= z \cos \lambda (z-z_L) - \frac{1}{\lambda} \sin \lambda (z-z_L) ~, \cr
\noalign{\kern 5pt}
&S(z;\lambda)
= \Big( z + \frac{1}{\lambda^2 z_L} \Big)  \sin \lambda (z-z_L) 
+ \Big( \frac{1}{\lambda} - \frac{z}{\lambda z_L} \Big) \cos \lambda (z-z_L)  ~, \cr
\noalign{\kern 5pt}
&C'(z;\lambda)
=  - \lambda z \sin \lambda (z-z_L) ~, \cr 
\noalign{\kern 5pt}
&S'(z;\lambda)
= \lambda z \cos \lambda (z-z_L) + \frac{z}{z_L} \sin \lambda (z-z_L)  ~, 
\label{APgaugeF2}
\end{align}
and satisfy
\begin{align}
&{\cal P}_5 \begin{pmatrix} C \cr S \end{pmatrix} = 
z^2 \frac{d}{dz} \frac{1}{z^2} \frac{d}{dz}
\begin{pmatrix} C \cr S \end{pmatrix} = 
- \lambda^2 \begin{pmatrix} C \cr S \end{pmatrix} , \cr
\noalign{\kern 5pt}
&{\cal P}_z \begin{pmatrix} C' \cr S' \end{pmatrix} = 
\frac{d}{dz} z^2 \frac{d}{dz} \frac{1}{z^2} 
\begin{pmatrix} C' \cr S' \end{pmatrix} = 
- \lambda^2 \begin{pmatrix} C' \cr S' \end{pmatrix} , \cr
\noalign{\kern 5pt}
&C(z_L; \lambda) = z_L ,~~ C' (z_L; \lambda) = 0 , ~~
S(z_L; \lambda) = 0 ,~~ S' (z_L; \lambda) =  \lambda z_L , \cr
\noalign{\kern 5pt}
&C S' - S C' = \lambda z^2 .
\label{APgaugeF3}
\end{align}
These basis functions have been employed in the text.  
We note that the basis functions for fermions, $C_{L/R} (z; \lambda, c)$ and 
$S_{L/R} (z; \lambda, c)$, remain the same as in the 5d case 
defined in Ref.~\cite{Furui:2016owe}.
}

\subsection{6D scalar fields}
\label{Sec:6D-scalar-KK-modes}

The  action of a massless 6D scalar field $\Phi(x,y,v)$ is given by
\begin{align}
S_{\rm bulk}^{\rm scalar}=
\int d^6x\sqrt{-\mbox{det}G} \, G^{MN}
 \partial_M\Phi(x,y,v)^\dag  \partial_N\Phi(x,y,v),
\end{align}
where $M,N=0,1,2,3,5,6$.
The equation of motion for the scalar field
$\Phi(x,z,v)$ is given by 
\begin{align}
z^2\left\{\Box_4+\partial_v^2
+k^2z^4\frac{\partial}{\partial z}\frac{1}{z^4}\frac{\partial}{\partial z}
\right\}\Phi(x,z,v)=0.
\label{APscalarEq1}
\end{align}
It follows that $\phi (x,z,v) = z^{-\nu} \Phi(x,z,v)$ ($\nu = \frac{5}{2}$) satisfies
\begin{align}
\left\{\Box_4+\partial_v^2
+k^2\left(
\frac{\partial^2}{\partial z^2}
+\frac{1}{z}\frac{\partial}{\partial z}
-\frac{\nu^2}{z^2}\right)
\right\}\phi(x,z,v)=0 ~.
\end{align}
The Bessel equation is given by 
\begin{align}
&\left(\frac{d^2}{dz^2}+\frac{1}{z}\frac{d}{dz}
+1-\frac{\nu^2}{z^2}\right)u(z)=0 ~, 
\nonumber
\end{align}
whose solutions are given by the Bessel functions $J_\nu(z)$ and $Y_\nu(z)$. 
Hence a mode function can be written as 
$\phi(x,z,v) = [\alpha J_\nu (\lambda z) + \beta Y_\nu (\lambda z)] \phi_\lambda(x,v)$
where $\phi_\lambda(x,v)$ satisfies
\begin{align}
\left\{\Box_4+\partial_v^2-k^2\lambda^2\right\}\phi_\lambda(x, v)=0 ~.
\end{align}

The orbifold boundary condition for the scalar field $\phi$ is given by 
\begin{align}
&\Phi(x,y_j-y,v_j-v)=P_j \, \Phi(x,y_j+y,v_j+v) ~.
\label{APscalarBC1}
\end{align}
In the $y$ coordinate $\phi(x,y,v) = e^{- \nu\sigma(y)} \Phi (x,y,v)$ satisfies
the same boundary condition as (\ref{APscalarBC1}) and Eq.\ (\ref{APscalarEq1})
becomes
\begin{align}
&
\left\{\Box_4+\partial_v^2+e^{\frac{1}{2}\sigma}
\left(\partial_y^2+\nu\sigma''-\nu^2\sigma'{}^2\right)
\right\}\phi=0 ~.
\label{APscalarEq2}
\end{align}
Due caution must be taken 
as $\sigma '' (y) = 2k \{ \delta_{2L_5} (y) -  \delta_{2L_5} (y-L_5) \}$.
If $\phi$ is parity even in the $y$ coordinate, the Neumann ($N$) condition becomes,
in the $z$ coordinate ($1 \le z \le z_L$), 
\begin{align}
N: ~ \Big( \frac{\dd}{\dd z} + \frac{\nu}{z} \Big) \phi = 0 \quad
{\rm at~} z = 1^+ ~{\rm or}~ z_L^- ~,
\label{APscalarBC2}
\end{align}
as can be confirmed by integrating (\ref{APscalarEq2}) over $y$ 
from $-\ep$ to $+\ep$ (or from $L_5 -\ep$ to $L_5 +\ep$).
If $\phi$ is parity odd in the $y$ coordinate, it satisfies 
the Dirichlet ($D$) condition $\phi = 0$ at $z=1$ or $z_L$.

\vskip 5pt
\noindent
(i) $(P_0=P_1,P_2=P_3)=(+,+)$

The 6D scalar field $\Phi(x,y,v)$ is expanded as 
\begin{align}
&\Phi(x,y,v)= e^{\nu \sigma (y) } \bigg\{ 
 \sum_{n=0}^{\infty}\widetilde{\phi}_n^C(x,y)f_n^C(v)
+\sum_{n=1}^{\infty}\widetilde{\phi}_n^S(x,y)f_n^S(v) \bigg\} ,
\end{align}
where $\widetilde{\phi}_n^C(x,y)$ and  $\widetilde{\phi}_n^S(x,y)$   satisfy
\begin{align}
&
\left\{
\begin{array}{l}
\widetilde{\phi}_n^C(x,-y)=\widetilde{\phi}_n^C(x,y) \cr
\noalign{\kern 5pt}
\widetilde{\phi}_n^C(x,L_5-y)=\widetilde{\phi}_n^C(x,L_5+y)\\
\end{array}
\right.  , \cr
\noalign{\kern 5pt}
&
\left\{
\begin{array}{l}
\widetilde{\phi}_n^S(x,-y)=-\widetilde{\phi}_n^S(x,y) \cr
\noalign{\kern 5pt}
\widetilde{\phi}_n^S(x,L_5-y)=-\widetilde{\phi}_n^S(x,L_5+y)\\
\end{array}
\right., 
\label{APPhi1}
\end{align}
and 
Fourier modes $f_n^C(v)$ and $f_n^S(v)$ are given by 
\begin{align}
&
f_n^C (v) = \begin{cases} \displaystyle \frac{1}{\sqrt{2\pi R_6}} & (n=0) \cr
\noalign{\kern 5pt}
\displaystyle
\frac{1}{\sqrt{\pi R_6}} \, \cos \frac{nv}{R_6} &(n\ge 1) \end{cases}~, \cr
\noalign{\kern 5pt}
&
f_n^S (v) =\frac{1}{\sqrt{\pi R_6}} \, \sin  \frac{nv}{R_6} ~~(n\ge 1) ~.
\label{APfnCS}
\end{align}
$\widetilde{\phi}_n^C$ and $\widetilde{\phi}_n^S$ satisfy the Neumann 
and Dirichlet conditions at $z=1, z_L$, respectively.
There is a  zero mode  for $\widetilde{\phi}_n^C$:
\begin{align}
\lambda_0^C = 0 ~,~~
\widetilde \phi_n^C(x, z)  = \phi_{n,0}^C(x) z^{- \nu} ~,
\end{align}
which corresponds to a mode constant in the 5th dimension.
Other modes   can be written as
\begin{align}
&\widetilde{\phi}_n^{C,S} (x, z) = Z_\nu(\lambda z)\phi_{m,\lambda}^{C,S}(x) ~, \cr
&Z_\nu(\lambda z)=
\alpha_{n,\lambda}^{C,S}J_\nu(\lambda z)
+\beta_{n,\lambda}^{C,S}Y_\nu(\lambda z) ~.
\end{align}
Eigenvalues $\lambda >0$ and the relative coefficient $\alpha/\beta$ are determined
by the boundary conditions.   Noting a recursion relation
\[
\bigg(\frac{d}{dz}+\frac{\nu}{z}\bigg) Z_\nu(\lambda z)
=\lambda Z_{\nu-1}(\lambda z) ~, 
\]
one finds eigenvalues $\big\{ \lambda_\ell^{C,S} \big\}$ ($\ell \ge 1$) from
\begin{align}
\widetilde{\phi}_n^C &: ~ 
Z_{\nu -1}  (\lambda_\ell^C)= Z_{\nu - 1} (\lambda_\ell^C z_L) = 0 ~,  \cr
\widetilde{\phi}_n^S &: ~ 
Z_{\nu }  (\lambda_\ell^S)= Z_{\nu} (\lambda_\ell^S z_L) = 0 ~,
\label{APscalarSpectrum0}
\end{align}
Thus $\Phi$ is expanded as
\begin{align}
&\Phi(x,z,v)= e^{\nu \sigma (y) } \bigg\{ 
\sum_{n=0}^{\infty} \sum_{\ell=0}^\infty 
{\phi}_{n, \ell}^C(x)  Z_{\nu} (\lambda_\ell^C z) f_n^C(v) \cr
\noalign{\kern 5pt}
&\hskip 3.2cm
+\sum_{n=1}^{\infty}  \sum_{\ell=1}^\infty 
{\phi}_{n, \ell} ^S(x)   Z_{\nu} (\lambda_\ell^S z)   f_n^S(v) \bigg\} ,
\label{APscalarExpansion1}
\end{align}
where $\widetilde{\phi}_{n, \ell}^{C,S} (x)$ satisfies
\begin{align}
&\Big\{\Box_4 - \big(m_{n,\ell}^{C,S} \big)^2 \Big\}  \, \phi_{n,\ell}^{C,S}(x)=0 ~, \cr
\noalign{\kern 5pt}
&\big(m_{n,\ell}^{C,S} \big)^2 = \frac{n^2}{R_6^2} + \big( k \lambda_\ell^{C,S} \big)^2 ~.
\label{APscalarSpectrum1}
\end{align}
There is one massless mode $\phi_{0,0}^C(x)$ with $m_{0,0}^C = 0$.

\noindent
(ii) $(P_0=P_1,P_2=P_3)=(-,-)$

The expansion is given by
\begin{align}
&\Phi(x,z,v)= e^{\nu \sigma (y) } \bigg\{ 
\sum_{n=0}^{\infty} \sum_{\ell=1}^\infty 
{\phi}_{n, \ell}^S(x)  Z_{\nu} (\lambda_\ell^S z) f_n^C(v) \cr
\noalign{\kern 5pt}
&\hskip 3.2cm
+\sum_{n=1}^{\infty}  \sum_{\ell=0}^\infty 
{\phi}_{n, \ell} ^C(x)   Z_{\nu} (\lambda_\ell^C z)   f_n^S(v) \bigg\} ,
\label{APscalarExpansion2}
\end{align}
There is no massless mode.

\noindent 
(iii) $(P_0=P_1,P_2=P_3)=(+,-)$

\begin{align}
&\Phi(x,z,v)= e^{\nu \sigma (y) } \bigg\{ 
\sum_{n=0}^{\infty} \sum_{\ell=0}^\infty 
{\phi}_{n, \ell}^C(x)  Z_{\nu} (\lambda_\ell^C  z)g_{n+\frac{1}{2}}^C(v) \cr
\noalign{\kern 5pt}
&\hskip 3.2cm
+\sum_{n=0}^{\infty}  \sum_{\ell=1}^\infty 
{\phi}_{n, \ell} ^S(x)   Z_{\nu} (\lambda_\ell^S z)   g_{n+\frac{1}{2}}^S(v) \bigg\} ,
\label{APscalarExpansion3}
\end{align}
where
\begin{align}
&g_{n+\frac{1}{2}}^C(v) = \frac{1}{\sqrt{\pi R_6}} \, \cos \frac{(n+\onehalf)v}{R_6} ~, ~~
g_{n+\frac{1}{2}}^S(v) = \frac{1}{\sqrt{\pi R_6}} \, \sin \frac{(n+\onehalf)v}{R_6} ~.
\ignore{
\noalign{\kern 5pt}
&\int_0^{2\pi R_6} dv \, g_{n+\frac{1}{2}}^C(v) g_{m+\frac{1}{2}}^C(v)
= \int_0^{2\pi R_6} dv \, g_{n+\frac{1}{2}}^S(v) g_{m+\frac{1}{2}}^S(v) = \delta_{nm} ~, \cr
\noalign{\kern 5pt}
&\int_0^{2\pi R_6} dv \, g_{n+\frac{1}{2}}^C(v) g_{m+\frac{1}{2}}^S(v) = 0 ~.}
\label{APgnCS}
\end{align}
Note that the mass spectrum is given by
\begin{align}
\big(m_{n,\ell}^{C,S} \big)^2 = \frac{(n + \onehalf)^2}{R_6^2} 
+ \big( k \lambda_\ell^{C,S} \big)^2 
\ge \Big( \frac{1}{2 R_6} \Big)^2  ~.
\end{align}

\noindent
(iv) $(P_0=P_1,P_2=P_3)=(-,+)$

\begin{align}
&\Phi(x,z,v)= e^{\nu \sigma (y) } \bigg\{ 
\sum_{n=0}^{\infty} \sum_{\ell=1}^\infty 
{\phi}_{n, \ell}^S(x)  Z_{\nu} (\lambda_\ell^S  z)g_{n+\frac{1}{2}}^C(v) \cr
\noalign{\kern 5pt}
&\hskip 3.2cm
+\sum_{n=0}^{\infty}  \sum_{\ell=0}^\infty 
{\phi}_{n, \ell} ^C(x)   Z_{\nu} (\lambda_\ell^C z)   g_{n+\frac{1}{2}}^S(v) \bigg\} .
\label{APscalarExpansion4}
\end{align}

\subsection{6D fermion field}
\label{Sec:6D-fermion-KK-modes}

Let us consider a free 6D Weyl fermion $\Psi(x,y,v)$ with $\gamma_{6D}^7=+1$:
\begin{align}
&\Psi(x,y,v)= e^{\frac{5}{2} \sigma (y)} \, \check \Psi ~, ~~  \check \Psi(x,z,v)=  
\left[ \begin{matrix} \xi(x,z,v) \cr  \eta(x,z,v) \cr 0 \cr 0 \end{matrix} \right] , \cr
\noalign{\kern 5pt}
&\gamma_{6D}^{7}= \begin{pmatrix} I_4& \cr  &-I_4 \end{pmatrix} 
=\gamma_{4D}^5\cdot (-i\gamma^5\gamma^6)~, ~~
\gamma_{4D}^5= \begin{pmatrix} I_2&&& \cr  &-I_2&& \cr  &&I_2&\cr &&&-I_2 \end{pmatrix}.
\end{align}
Its action is given by
\begin{align}
&I=\int d^6x\sqrt{-\mbox{det}G} \, 
\overline{\Psi}
\left\{\gamma^A E_A{}^{M}{\cal D}_M+ick\gamma^6\right\} \Psi \cr
&=\int d^4x \int_{0}^{2\pi R_6}dv \int_{1}^{z_L} \frac{dz}{k}
i [-\eta^\dag,\xi^\dag ]
\bigg[\begin{matrix} -kD_-(c)+i\dd_v &\sigma^\mu \dd_\mu \cr
\overline{\sigma}^\mu \dd_\mu & -kD_+(c)+i\dd_v \end{matrix} \bigg]
\bigg[\begin{matrix} \xi \cr  \eta  \end{matrix} \bigg]
\end{align}
where $D_\pm(c)$ is defined in (\ref{Dpm}).
As an example we consider the case in which 
the BCs are given by 
\begin{align}
\Psi(x,y_j-y,v_j-v) & =-i\gamma^5\gamma^6P_j\Psi(x,y_j+y,v_j+v) \cr
&=\gamma_{6D}^7\gamma_{4D}^5P_j\Psi(x,y_j+y,v_j+v) ~.
\end{align}
Equations of motion are
\begin{align}
&(-kD_- (c)  + i\partial_v) \xi  +\sigma^\mu\partial_\mu \eta=0 ~, \cr
&\overline{\sigma}^\mu\partial_\mu \xi  +(-kD_+(c) +i\partial_v)\eta =0 ~.  
\label{APeqF1}
\end{align}
To find eigenmodes, we separate 4D and 5, 6th dimensional parts as
\begin{align}
&\xi(x,z,v)=\widetilde{\xi}(z,v)f_R(x) ~,   ~~
\overline{\sigma}^\mu\partial_\mu f_R(x) =mf_L(x) ~, \cr
&\eta(x,z,v)=\widetilde{\eta}(z,v)f_L(x) ~, ~~
{\sigma}^\mu\partial_\mu f_L(x)=mf_R(x) ~. 
\end{align}
It follows that
\begin{align}
&(-kD_- (c) +i\dd_v) \widetilde{\xi} +m\widetilde{\eta}=0 ~, \cr
&m\widetilde{\xi}+ (-kD_+(c) +i\dd_v)\widetilde{\eta}=0 ~.
\end{align}
In the $y$ coordinate $\xi$ and $\eta$ satisfy the BCs 
\begin{align}
& 
\begin{pmatrix} \xi \cr  \eta  \end{pmatrix} (x,y_j-y,v_j-v)=
P_j  \begin{pmatrix} \xi \cr  - \eta  \end{pmatrix}  (x,y_j+y,v_j+v) ~.
\end{align}

\vskip 10pt

\noindent
(i) $(P_0=P_1,P_2=P_3)=(+,+)$

In this case
\begin{align}
&\xi(x,y_j-y,v_j-v)= + \xi(x,y_j+y,v_j+v) ~, \cr
&\eta(x,y_j-y,v_j-v)=-\eta(x,y_j+y,v_j+v) ~.
\end{align}
Mode functions  can be written  as 
\begin{align}
\xi(x,y,v)=&  \bigg\{
 \sum_{n=0}^{\infty}\widetilde{\xi}_n^C(y)f_n^C(v)
+\sum_{n=1}^{\infty}\widetilde{\xi}_n^S(y)f_n^S(v) \bigg\} f_R(x) ~, \cr
\noalign{\kern 5pt}
\eta(x,y,v)=&\bigg\{  
 \sum_{n=0}^{\infty}\widetilde{\eta}_n^S(y)f_n^C(v)
+\sum_{n=1}^{\infty}\widetilde{\eta}_n^C(y)f_n^S(v) \bigg\} f_L(x) ~,
\label{APexpansionF1}
\end{align}
where the $f_n^C(v), f_n^S(v)$ are defined in (\ref{APfnCS}).
$\widetilde{\xi}_n^{C,S}$ and $\widetilde{\eta}_n^{C,S}$ satisfy the BCs
\begin{align}
&\widetilde{\xi}_n^C (y_j - y) = + \widetilde{\xi}_n^C (y_j + y) ~,~~
\widetilde{\xi}_n^S (y_j - y) = - \widetilde{\xi}_n^S (y_j + y) ~, \cr
&\widetilde{\eta}_n^C (y_j - y) = + \widetilde{\eta}_n^C (y_j + y) ~,~~
\widetilde{\eta}_n^S (y_j - y) = - \widetilde{\eta}_n^S (y_j + y) ~.
\end{align}
By inserting (\ref{APexpansionF1}) into (\ref{APeqF1}), one finds
equations for $\widetilde \xi_n^{C,S},  \widetilde \eta_n^{C,S}$ 
in the $z$ coordinate ($1 \le z \le z_L$):
\begin{align}
&\begin{pmatrix} - k D_- & m \cr m & - kD_+ \end{pmatrix}
\begin{pmatrix} \widetilde{\xi}_0^C  \cr \widetilde{\eta}_0^S \end{pmatrix} = 0 ~, \cr
\noalign{\kern 5pt}
&\begin{pmatrix} - k D_-  & i \frac{n}{R_6} & m & 0 \cr
- i \frac{n}{R_6} & - k D_-  &0 & m \cr
m & 0 & - k D_+ & i \frac{n}{R_6} \cr  0 & m & - i \frac{n}{R_6} & - k D_+ \end{pmatrix}
\begin{pmatrix}  \widetilde{\xi}_n^C  \cr  \widetilde{\xi}_n^S  \cr
 \widetilde{\eta}_n^S \cr \widetilde{\eta}_n^C \end{pmatrix} = 0 \quad (n \ge 1) .
\end{align}
It follows that
\begin{align}
&( k^2 D_+ D_-  - m^2) \, \widetilde{\xi}_0^C =0 ~, \cr
\noalign{\kern 5pt}
&(k^2 D_- D_+ - m^2) \, \widetilde{\eta}_0^S = 0 ~, \cr
\noalign{\kern 5pt}
&\Big( k^2D_+ D_- \mp\frac{2kcn}{R_6}\frac{1}{z}+\frac{n^2}{R_6^2}-m^2 \Big)
\big(\widetilde{\xi}_n^C\pm i\widetilde{\xi}_n^S\big)=0 ~, \cr
\noalign{\kern 5pt}
&\Big( k^2D_- D_+ \mp\frac{2kcn}{R_6}\frac{1}{z}+\frac{n^2}{R_6^2}-m^2 \Big)
\big(\widetilde{\eta}_n^S \pm i \widetilde{\eta}_n^C\big)=0 \quad (n \ge 1) .
\end{align}

The equations for the zero ($n=0$) modes in the 6th dimension,  
$\widetilde{\xi}_0^C$  and $\widetilde{\eta}_0^S$, 
are reduced to the Bessel equation.
In terms of the basis functions $C_{R/L} (z; \lambda, c)$, $S_{R/L} (z; \lambda, c)$
given in Appendix B of Ref.~\cite{Furui:2016owe},  solutions satisfying the BCs 
are  given by
\begin{align}
&\widetilde{\xi}_{0,  0}^C =  a_0 z^c ~, ~~ \widetilde{\eta}_{0,  0}^S = 0 ~,~~
m_0 = 0 ~, \cr
\noalign{\kern 5pt}
&\widetilde{\xi}_{0,  \ell}^C = a_\ell C_R(z; \lambda_\ell, c) ~, ~~ 
\widetilde{\eta}_{0,  \ell}^S = a_\ell S_L(z; \lambda_\ell, c) ~, \cr
&S_L(1; \lambda_\ell, c) = 0 ~, ~~ m_\ell = k \lambda_\ell > 0
\end{align}
where $a_\ell$ is a normalization constant.  There is one massless mode
for the right-handed component.
For the $n \not= 0$ modes the solutions are more involved.

\vskip 5pt

\noindent
(ii) $(P_0=P_1,P_2=P_3)=(-,-)$

In this case mode functions of $\xi$ and $\eta$ can be written  as 
\begin{align}
\xi(x,y,v)=&  \bigg\{
 \sum_{n=0}^{\infty}\widetilde{\xi}_n^S(y)f_n^C(v)
+\sum_{n=1}^{\infty}\widetilde{\xi}_n^C(y)f_n^S(v) \bigg\} f_R(x) ~, \cr
\noalign{\kern 5pt}
\eta(x,y,v)=&\bigg\{  
 \sum_{n=0}^{\infty}\widetilde{\eta}_n^C(y)f_n^C(v)
+\sum_{n=1}^{\infty}\widetilde{\eta}_n^S(y)f_n^S(v) \bigg\} f_L(x) ~,
\label{APexpansionF2}
\end{align}
The equations for the zero ($n=0$) modes in the 6th dimension,  
$\widetilde{\xi}_0^S$  and $\widetilde{\eta}_0^C$,  are
\begin{align}
&\begin{pmatrix} - k D_- & m \cr m & - kD_+ \end{pmatrix}
\begin{pmatrix} \widetilde{\xi}_0^S  \cr \widetilde{\eta}_0^C \end{pmatrix} = 0 ~,
\end{align}
and  solutions satisfying the BCs 
are  given by
\begin{align}
&\widetilde{\xi}_{0,  0}^S =  0 ~, ~~ \widetilde{\eta}_{0,  0}^C = a_0 z^{-c}  ~,~~
m_0 = 0 ~, \cr
\noalign{\kern 5pt}
&\widetilde{\xi}_{0,  \ell}^S = a_\ell S_R(z; \lambda_\ell, c) ~, ~~ 
\widetilde{\eta}_{0,  \ell}^C= a_\ell C_L(z; \lambda_\ell, c) ~, \cr
&S_R(1; \lambda_\ell, c) = 0 ~, ~~ m_\ell = k \lambda_\ell > 0
\end{align}
where $a_\ell$ is a normalization constant.  There is one massless mode
for the left-handed component.

\vskip 5pt
\noindent
(iii) $(P_0=P_1,P_2=P_3)=(+,-)$

In this case mode functions  can be written  as 
\begin{align}
\xi(x,y,v)=&  \bigg\{
 \sum_{n=0}^{\infty}\widetilde{\xi}_n^C(y)g_{n + \frac{1}{2}}^C(v)
+\sum_{n=0}^{\infty}\widetilde{\xi}_n^S(y)g_{n +\frac{1}{2}}^S(v) \bigg\} f_R(x) ~, \cr
\noalign{\kern 5pt}
\eta(x,y,v)=&\bigg\{  
 \sum_{n=0}^{\infty}\widetilde{\eta}_n^S(y) g_{n +\frac{1}{2}}^C(v)
+\sum_{n=1}^{\infty}\widetilde{\eta}_n^C(y) g_{n +\frac{1}{2}}^S(v) \bigg\} f_L(x) ~,
\label{APexpansionF3}
\end{align}
where $g_{n +\frac{1}{2}}^{C,S} (v)$ are defined in (\ref{APgnCS}).  
There are no zero modes  in the 6th dimension.

\vskip 5pt

\noindent
(iv) $(P_0=P_1,P_2=P_3)=(-,+)$

In this case mode functions  can be written  as 
\begin{align}
\xi(x,y,v)=&  \bigg\{
 \sum_{n=0}^{\infty}\widetilde{\xi}_n^S(y)g_{n + \frac{1}{2}}^C(v)
+\sum_{n=0}^{\infty}\widetilde{\xi}_n^C(y)g_{n +\frac{1}{2}}^S(v) \bigg\} f_R(x) ~, \cr
\noalign{\kern 5pt}
\eta(x,y,v)=&\bigg\{  
 \sum_{n=0}^{\infty}\widetilde{\eta}_n^C(y) g_{n +\frac{1}{2}}^C(v)
+\sum_{n=1}^{\infty}\widetilde{\eta}_n^S(y) g_{n +\frac{1}{2}}^S(v) \bigg\} f_L(x) ~.
\label{APexpansionF3}
\end{align}
There are no zero modes  in the 6th dimension.

\bibliographystyle{utphys} 
\bibliography{../../arxiv/reference}

\end{document}